\documentclass[]{aastex631}   

\usepackage{amsmath}
\usepackage{amssymb}
\usepackage{amsfonts}

\newcommand{\fD}{\ensuremath{f_D}}
\newcommand{\fDi}{\ensuremath{f_{D1}}}  
\newcommand{\fDii}{\ensuremath{f_{D2}}}  
\newcommand{\tauscat}{\ensuremath{\tau_{\rm scatt}}}
\newcommand{\tauiss}{\ensuremath{\tau_{\rm scatt}}}
\newcommand{\tauissG}{\ensuremath{\tau_{\rm scatt,1GHz}}}

\newcommand{\etaiss}{\ensuremath{\eta_{\rm iss}}}
\newcommand{\etap}{\ensuremath{\eta_{\rm p}}}

\newcommand{\etacred}{\ensuremath{\eta_{\rm cred}}}
\newcommand{\etacreda}[2]{{\bf Arc credibility index} \etacred: #1 MHz (#2)}
\newcommand{\etacredb}[3]{{\bf Arc credibility indices} \etacred: #1 MHz, #2 MHz (#3)}
\newcommand{\etacredc}[4]{{\bf Arc credibility indices} \etacred: #1 MHz, #2 MHz, #3 MHz (#4)}

\newcommand \psrA {1.1} 
\newcommand \psrB {1.2} 
\newcommand \psrC {1.3} 
\newcommand \psrD {1.4} 
\newcommand \psrE {1.5} 
\newcommand \psrF {1.6} 
\newcommand \psrG {1.7} 
\newcommand \psrH {1.8} 
\newcommand \psrI {1.9} 
\newcommand \psrJ {1.10} 
\newcommand \psrK {1.11} 
\newcommand \psrL {1.12} 
\newcommand \psrM {1.13} 
\newcommand \psrN {1.14} 
\newcommand \psrO {1.15} 
\newcommand \psrP {1.16} 
\newcommand \psrQ {1.17} 
\newcommand \psrR {1.18} 
\newcommand \psrS {1.19} 
\newcommand \psrT {1.20} 
\newcommand \psrU {1.21} 
\newcommand \psrV {1.22} 

\newcommand{\nuiss}{\ensuremath{\Delta \nu_{\rm iss}}}
\newcommand{\tiss}{\ensuremath{ \Delta t_{\rm iss}}}
\newcommand{\dnuiss}{\ensuremath{\Delta \nu_{\rm iss}}}
\newcommand{\dtiss}{\ensuremath{ \Delta t_{\rm iss}}}

\newcommand{\eg}{e.g.\ }
\newcommand{\ie}{i.e.\ }

\newcommand{\veff}{\ensuremath{V_{\mathrm{eff}}}}
\newcommand{\Veff}{\ensuremath{V_{\mathrm{eff}}}}

\newcommand{\cucm}{\ensuremath{\textrm{ cm}^{-3}}}
\newcommand{\pc}{\ensuremath{\textrm{ pc}}}

\newcommand{\g}{\ensuremath{\textrm{ g}}}

\newcommand{\m}{\ensuremath{\textrm{ m}}}

\newcommand{\J}{\ensuremath{\textrm{ J}}}
\newcommand{\K}{\ensuremath{\textrm{ K}}}

\newcommand{\DM}{\hbox{DM }}                        

\newcommand{\danx}{\color{black}}

\def\Deff{D_{\rm eff}}
\def\Veff{V_{\rm eff}}

\newcommand\be{\begin{equation}}
\newcommand\ee{\end{equation}}
\newcommand{\bea}{\begin{eqnarray}}
\newcommand{\eea}{\end{eqnarray}}

\begin{document}

\title{A Scintillation Arc Survey of 22 Pulsars with Low to Moderate Dispersion Measures}
\shorttitle{Scintillation Arc Survey}
\shortauthors{Stinebring, D.\ R. , Rickett, B.\ J., {\em et al.}}

\author[0000-0002-1797-3277]{Dan~R.~Stinebring}
\affiliation{Department of Physics and Astronomy, 110 No. Professor St., Oberlin College, Oberlin, OH 44074 , USA}

\author[0000-0002-1138-2417]{Barney~J.~Rickett}
\affiliation{University of California, San Diego , USA}

\author[0000-0002-6555-312X]{Anthony H.\ Minter}
\affiliation{Green Bank Observatory, P.O.\ Box 2, Green Bank, WV 24944, USA}



\author[0000-0001-7301-5666]{Alex S.\ Hill}
\affiliation{Department of Computer Science, Math, Physics, and Statistics, University of British Columbia, Kelowna, BC V1V 1V7, Canada}
\affiliation{Dominion Radio Astrophysical Observatory, Herzberg Astronomy \& Astrophysics Research Centre, National Research Council, Penticton, BC, Canada}

\author [0000-0002-8380-6688]{Adam~P.~Jussila}
\affiliation{Department of Physics and Astronomy, 110 No. Professor St., Oberlin College, Oberlin, OH 44074 , USA}
\affiliation{Department of Bioinformatics and Systems Biology, 9500 Gilman Drive, Dept. 0419, University of California, San Diego, La Jolla, CA 92093-0419, USA}

\author[0000-0003-3491-1024]{Lele~Mathis}   
\affiliation{Department of Physics and Astronomy, 110 No. Professor St., Oberlin College, Oberlin, OH 44074 , USA}
\affiliation{Department of Chemistry, Northwestern University, 2145 Sheridan Rd, Evanston, IL 60208 , USA}

\author[0000-0001-7697-7422]{Maura A.\ McLaughlin}
\affiliation{Department of Physics and Astronomy, West Virginia University, P.O. Box 6315, Morgantown, WV 26506, USA}
\affiliation{Center for Gravitational Waves and Cosmology, West Virginia University, Chestnut Ridge Research Building, Morgantown, WV 26505, USA}

\author[0000-0002-4941-5333]{Stella~Koch~Ocker}
\affiliation{Department of Physics and Astronomy, 110 No. Professor St., Oberlin College, Oberlin, OH 44074 , USA}
\affiliation{Department of Astronomy and Cornell Center for Astrophysics and Planetary Science, Cornell University, Ithaca, NY 14853, USA}

\author[0000-0001-5799-9714]{Scott M.\ Ransom}
\affiliation{National Radio Astronomy Observatory, 520 Edgemont Road, Charlottesville, VA 22903, USA}

\begin{abstract}
{\em Context:} By providing  information about the location of scattering
material along the line of sight (LoS) to pulsars, scintillation arcs
are a powerful tool for exploring the distribution of ionized material
in the interstellar medium. Here, we present observations that probe the ionized ISM
on scales of $\sim$~0.001 -- 30~au.
{\em Aims:} We have surveyed pulsars for scintillation arcs in a relatively
unbiased sample with DM~$< 100\pc\cucm$. We present multi-frequency
observations of 22 low to moderate DM pulsars. Many of the 54
observations were also observed at another frequency within a few days.
{\em Methods:} For all observations we present dynamic spectra,
autocorrelation functions, and secondary spectra. We analyze these data
products to obtain scintillation bandwidths, pulse broadening times, and
arc curvatures.
{\em Results:} We detect definite or probable scintillation arcs in 19 of the
22 pulsars and 34 of the 54 observations, showing that scintillation
arcs are a prevalent phenomenon. The arcs are better defined in low DM
pulsars. We show that well-defined arcs do
not directly imply anisotropy of scattering. Only the presence of
reverse arclets and a deep valley along the delay axis, which occurs in
about 20\% of the pulsars in the sample,  indicates substantial
anisotropy of scattering.
{\em Conclusions:} The survey demonstrates substantial patchiness of the
ionized ISM on both au size scales transverse to the line of sight and
on $\sim$~100~pc scales along it. We see little evidence for distributed
scattering along most lines of sight in the survey.
\end{abstract}



\section{Introduction}
Nearly 55 years after their discovery \citep{hbp+68}, radio pulsars continue to be versatile probes of fundamental physics, plasma processes under extreme conditions, and the distribution of ionized gas in the Galaxy. 
Since early pioneering studies \citep{sch68,ric69,ric70} , the unique wideband, pulsed character of the signal has been employed to explore the ionized component of gas along the LoS to these sources. 
With more than 3300 pulsars  known \citep{mhth05}, they probe a wide range of distances and astrophysical conditions along sight lines and undergird the effort to develop a detailed model of the ionized gas distribution in the Milky Way \citep{cl02,ymw17}.

Classical studies of radio wave scintillation toward pulsars \citep[\eg][]{cwb85,cor86,grl94,lkm+01,bcc+04,kl07}, provided a broad-brush view of
the scattering along many lines of sight and the tools to interpret it.
In the study of interstellar scintillation (ISS) there has been an emphasis on measuring the characteristic bandwidth \nuiss\ and timescale \tiss\ of the scintillation structure in two-dimensional dynamic spectra (intensity as a
function of radio frequency and time).
This has yielded estimates of scattering angles toward pulsars, produced a better understanding of the distribution of scattering material along the LoS \citep{cr98}, and also allowed the estimate of pulsar proper space velocities through the estimation of scintillation speeds \citep{cor86,gup95}.

However, the discovery that pulsar dynamic spectra often have an underlying low-level modulation manifested as highly-organized parabolic structures in the power spectrum of the dynamic spectrum 
\citep{smc+01} 
 has provided a powerful new tool and uncovered several puzzles.  The position of features in scintillation arcs can move on $\sim$week timescales or shorter \citep{hsa+05,whh+18}, whereas the qualitative appearance of arcs can change on several month timescales \citep[][amongst others]{smc+01,msa+20,rcb+20}.

Scintillation arcs arise when the following conditions are met \citep{wmsz04,crsc06}: a) scattering occurs in a relatively small fractional portion of the LoS (thin screen condition)
\footnote{In general it is difficult to quantify this, but \citep{shr05} found a fractional thickness of less than 1\% of the 350~pc distance to B1929+10; also see \citep{mzsc22}},
b) the angular broadening function $B(\theta)$ has both a well-defined core and an outer halo; furthermore, the scintillation arc 
becomes narrower with a deeper valley, if the angular image on the sky is  {anisotropic and roughly aligned} along the effective velocity vector.

Most previous observational scintillation arc studies 
\citep[e.g.,][]{hsb+03, wmj+05, hsa+05, rscg11, bot+16, spg+17, whh+18, sro19, rcb+20, yzm+21, rszm21, mzsc22,cwy+22}
 have focused on a relatively small number of well-observed pulsars and have explored a range of diverse scintillation arc phenomena.
No prior study has explored the prevalence of scintillation arcs toward a sample of pulsars with {\danx fairly} uniformly applied selection criteria.
Since scintillation arcs often indicate the presence of highly organized linear scattering features toward pulsars --- and since the astrophysical origin of those features is not known --- it is of particular interest to characterize the frequency of occurrence of the arcs.

{\danx In addition to the work mentioned above, there has been a substantial amount of {\em precision} scintillation arc work -- both interferometric and single-dish -- in the past 10 years or so.
Much of this work was inspired by the remarkable interferometric study of PSR~B0834+06 by \citet{bmg+10},
 and work on interstellar holography \citep{ws05,wksv08} laid important groundwork, too.
Among other highlights in this scintillometry effort are several studies of binary pulsars \citep{rcn+14,msa+20,mma+22}, the detection of scattering from a supernova remnant around a pulsar \citep{yzm+21}, the location of multiple scattering screens toward nearby pulsars \citep{cwy+22,mzsc22}, and important new theoretical work \citep{spmb19a, spmb19b, swm+21, sx21, bbv+22}.

{Our approach in this paper is data-oriented with a minimum of model-fitting. After presenting necessary preliminaries, we start from a close inspection of features in the secondary spectra and proceed to more detailed analyses in the following sections. More specifically,}
in \S2 we present the observations and our data processing methods.
\S3 focuses on secondary spectra production and the extraction of basic scintillation parameters from the data set.
In  \S4, we closely inspect the secondary spectra from each of the 22 pulsars, highlighting salient arc features without much interpretation. We then group them into three sets based on the prominence of scintillation arcs.
\S5 develops an overall analysis of the observations and sets the results in a theoretical scattering context, with special attention to the DM and frequency dependence of arc parameters.
We discuss the key results in \S6 and summarize the paper in \S7.
All data used in this paper are being made available as described in \S\ref{sec:datashare}.

We have used the the ATNF Pulsar Catalogue ({\sc psrcat}) database\footnote{ http://www.atnf.csiro.au/research/pulsar/psrcat, V1.67} \citep{mhth05} extensively throughout this work.
}

\section{Observations and Data Processing}
In this Scintillation Arc Survey (SAS) we studied scintillation arcs in 22~pulsars.
The sources were chosen based on the following initial criteria: i) visible with either the {Green Bank Telescope (GBT)} or Arecibo, ii) {dispersion measure $\DM < 50 \pc\cucm$, and
iii) 400~MHz flux density} $> 25$~mJy. Later in the project we saw advantages to expanding \DM coverage out to $100 \pc\cucm$ for at least several sources. We then included 6 pulsars visible from Arecibo in order to accomplish this. Because of Arecibo's sensitivity, we relaxed the flux density limit somewhat. See Table~\ref{tab:psrparms}, which includes basic per-pulsar parameters such as source flux density, dispersion measure, transverse velocity, and previously measured scattering timescale. 





\nocite{cbb10, dgb+19}

\begin{deluxetable}{l c c c r r}




\tablecaption{Observational Parameters of the 22 Pulsars}

\tablenum{1}


\tablehead{\colhead{PSR} & 
\colhead{S$_{400}$} & 
\colhead{S$_{1400}$} & 
\colhead{DM} & \colhead{Dist.} & 
\colhead{V$_{\rm trans}$} \\ 
\colhead{} & 
\colhead{(mJy)} & 
\colhead{(mJy)} & 
\colhead{(pc cm$^{-3}$)} & 
\colhead{(kpc)} & 
\colhead{(km s$^{-1}$)} } 

\colnumbers
\startdata
B0138+59 & 49 & 4.5 & 34.93 & 2.30 & \nodata \\
B0450+55 & 59 & 13 & 14.59 & 1.18 & 314.2 \\
B0450--18 & 82 & 16.8 & 39.9 & 0.40 & 24.6 \\
B0523+11 & 19.5 & 1.94 & 79.42 & 1.84 & 270.7 \\
B0525+21 & 57 & 8.9 & 50.87 & 1.22 & 122.1 \\
B0540+23 & 29 & 10.7 & 77.7 & 2.06\tablenotemark{\footnotesize a} & 217.2\tablenotemark{\footnotesize a} \\
B0626+24 & 31 & 17.9\tablenotemark{\footnotesize b} & 84.18 & 3.00\tablenotemark{\footnotesize b} & 84.0\tablenotemark{\footnotesize b} \\
B0628--28 & 206 & 31.9 & 34.42 & 0.32 & 77.3 \\
B0809+74 & 79 & 10 & 5.75 & 0.43 & 102.7 \\
B0818--13 & 102 & 6 & 40.94 & 1.90 & 405.2 \\
B1508+55 & 114 & 8 & 19.62 & 2.10 & 962.6 \\
B1540--06 & 40 & 15.2\tablenotemark{\footnotesize b} & 18.4\tablenotemark{\footnotesize b} & 3.11\tablenotemark{\footnotesize b} & 247.4\tablenotemark{\footnotesize b} \\
B1706--16 & 47 & 14.5 & 24.89 & 0.56 & 125.3 \\
B1821+05 & 18 & 1.7 & 66.78 & 2.00 & 51.1 \\
B1857--26 & 131 & 15 & 37.99 & 0.70 & 170.3 \\
B1907+03 & 21 & 1.5 & 82.93 & 2.86 & \nodata \\
B2021+51 & 77 & 27 & 22.55 & 1.80 & 107.8 \\
B2045--16 & 116 & 22 & 11.46 & 0.95 & 510.1 \\
J2145--0750 & 46 & 10.3 & 9 & 0.71 & 44.5 \\
B2217+47 & 111 & 3 & 43.5 & 2.39 & 365.7 \\
B2310+42 & 89 & 15 & 17.28 & 1.06 & 125.0 \\
B2327--20 & 42 & 2.9 & 8.46 & 0.86 & 305.6 \\
\enddata

\tablenotetext{a}{{Chmyreva}, {Beskin},  and {Biryukov} (2010)}
\tablenotetext{b}{Deller et al.\ (2019)}

\tablecomments{All values from the ATNF {\sc psrcat}, v~1.67 unless indicated otherwise.}

\label{tab:psrparms}
\end{deluxetable}

In most cases we obtained multi-frequency dynamic spectra at two epochs separated by less than a week. However, the GBT~1400~MHz observations were made 14~years after those at the lower frequencies ( \S{\ref{sec:GBobs}}).

Three data sets are employed in this paper, two from the Green Bank Telescope and one from the Arecibo Observatory. They are described below.  Observational details such as epoch of observation, center frequency, and bandwidth are given in Table~\ref{tab:obsparms}.

\subsection{GBT Observations}
\label{sec:GBobs}
The observations in the first and largest portion of the dataset were made with the Robert C.\ Byrd Green Bank Telescope (GBT) between 2005 September 17 -- 24.
For each of the 16 GBT sources a 60-min dynamic spectrum was obtained with a 10-s dump time of the Spectral Processor spectrometer.
We used the Spectral Processor in a mode that produced $\rm N_{chans} = 1024$ across a bandwidth ranging from 5~MHz to 40~MHz in binary steps.
The center frequencies of the two bands were 340~MHz and 825~MHz.
Each front-end receiver was mounted at the prime focus, and only one front-end could be mounted at a time.
Initial observations were made at 340~MHz for two days followed, five days later, by observations for two days with the 825~MHz receiver in place.

\startlongtable
\begin{deluxetable}{c c c c c c c}
\tablecaption{Details of the Observations} 
\tablehead{
\colhead{    \#           } &
\colhead{     PSR      } &
\colhead{Telescope  } &
\colhead{Center Freq. } &
\colhead{Bandwidth } &
\colhead{Nchans\tablenotemark{\footnotesize a}     } &
\colhead{ MJD         } \\
 & & &
\colhead{(MHz)} &
\colhead{(MHz)} &
& 
}
\tablenum{2}
\colnumbers
\startdata
1 & B0138+59 & GBT &   340 &     5 &  1024 &    53630 \\
 2 & B0138+59 &  AO &   825 &    40 &  1024 &    53637 \\
 3 & B0450+55 & GBT &   340 &     5 &  1024 &    53632 \\
 4 & B0450+55 & GBT &   825 &    40 &  1024 &    53637 \\
 5 & B0450+55 & GBT &  1400 &   100 &   512 &    58920 \\
 6 & B0450-18 & GBT &   340 &     5 &  1024 &    53632 \\
 7 & B0450-18 & GBT &   824 &    40 &  1024 &    53637 \\
 8 & B0450-18 & GBT &  1400 &   100 &  1024 &    58913 \\
 9 & B0523+11 &  AO &   422 &     2 &  4096 &    58132 \\
10 & B0523+11 &  AO &  1450 &   160 &  4096 &    58131 \\
11 & B0525+21 &  AO &  1390 &    40 &  4096 &    58131 \\
12 & B0540+23 &  AO &   431 &    10 &  4096 &    58130 \\
13 & B0540+23 &  AO &  1450 &   160 &  4096 &    58130 \\
14 & B0626+24 &  AO &   431 &    10 &  4096 &    58132 \\
15 & B0626+24 &  AO &  1390 &    40 &  4096 &    58133 \\
16 & B0628-28 & GBT &   340 &     5 &  1024 &    53632 \\
17 & B0628-28 & GBT &   825 &    40 &  1024 &    53637 \\
18 & B0628-28 & GBT &  1400 &   100 &  1024 &    58904 \\
19 & B0809+74 & GBT &   340 &     5 &  1024 &    53630 \\
20 & B0809+74 & GBT &   832 &    24 &   251 &    53637 \\
21 & B0809+74 & GBT &  1400 &   100 &  1024 &    58877 \\
22 & B0818-13 & GBT &   340 &     5 &  1024 &    53632 \\
23 & B0818-13 & GBT &   824 &    40 &  1024 &    53637 \\
24 & B0818-13 & GBT &  1400 &   100 &  1024 &    58920 \\
25 & B1508+55 & GBT &   340 &     5 &  1024 &    53632 \\
26 & B1508+55 & GBT &   825 &    20 &  1024 &    53637 \\
27 & B1508+55 & GBT &  1400 &   100 &  1024 &    58933 \\
28 & B1540-06 & GBT &   340 &     5 &  1024 &    53630 \\
29 & B1540-06 & GBT &   825 &    40 &  1024 &    53634 \\
30 & B1540-06 & GBT &  1400 &   100 &  1024 &    58964 \\
31 & B1706-16 & GBT &   340 &     5 &  1024 &    53630 \\
32 & B1706-16 & GBT &   824 &    40 &  1024 &    53634 \\
33 & B1706-16 & GBT &  1400 &   100 &  1024 &    58965 \\
34 & B1821+05 &  AO &   431 &    10 &  4096 &    58132 \\
35 & B1857-26 & GBT &   340 &     5 &  1024 &    53631 \\
36 & B1857-26 & GBT &   824 &    40 &  1024 &    53635 \\
37 & B1907+03 &  AO &  1470 &    40 &  4096 &    58131 \\
38 & B2021+51 & GBT &   340 &     5 &  1024 &    53632 \\
39 & B2021+51 & GBT &   824 &    40 &  1024 &    53637 \\
40 & B2021+51 & GBT &  1400 &   100 &  1024 &    58922 \\
41 & B2045-16 & GBT &   340 &     5 &  1024 &    53631 \\
42 & B2045-16 & GBT &   824 &    40 &  1024 &    53635 \\
43 & B2045-16 & GBT &  1400 &   100 &  1024 &    58922 \\
44 & J2145-0750 & GBT &   340 &     5 &  1024 &    53631 \\
45 & J2145-0750 & GBT &   825 &    40 &  1024 &    53635 \\
46 & B2217+47 & GBT &   340 &     5 &  1024 &    53632 \\
47 & B2217+47 & GBT &   825 &    40 &  1024 &    53637 \\
48 & B2217+47 & GBT &  1400 &   100 &  1024 &    58920 \\
49 & B2310+42 & GBT &   340 &     5 &  1024 &    53632 \\
50 & B2310+42 & GBT &   825 &    40 &  1024 &    53637 \\
51 & B2310+42 & GBT &  1400 &   100 &  1024 &    58874 \\
52 & B2327-20 & GBT &   340 &     5 &  1024 &    53632 \\
53 & B2327-20 & GBT &   824 &    40 &  1024 &    53637 \\
54 & B2327-20 & GBT &  1400 &   100 &   256 &    58877 \\
\enddata

\tablenotetext{a}{Number of frequency channels. The GBT 1400~MHz data were taken with 8192 channels and downsampled.}

\label{tab:obsparms}%
\end{deluxetable}%

Another set of GBT observations, centered at 1400 MHz, was made during the period 2020 January -- April. 
Dynamic spectra were obtained for 13 of the pulsars in the sample at this frequency, with the frequency range chosen to minimize the RFI based on diagnostic scans performed with the GBT.
All observations used the VEGAS spectrometer with 8192 spectral channels across 100~MHz bandwidth, and
spectra were written out every 10~s.
These observations were made $\approx 15$~yr later than for the low-frequency GBT data.
Although the 1400 MHz data  shed important light on the scintillation arc structure seen at lower frequencies, care is needed in comparing features at widely separated epochs.

\subsection{Arecibo Observations} \label{AOobs}
Observations were made with the William E.\ Gordon Arecibo Telescope between 2018 January 12 -- 15.
Six pulsars were observed, three at the dual frequencies of 430~MHz and 1450~MHz. 
Successful observations of the other three were only possible at a single frequency, either 430~MHz or 1450~MHz. 
The Mock spectrometers were used for the observations, and bandwidths ranged between 2~MHz and 160~MHz, all with 4096 frequency channels and a 10~s interval between accumulated spectra.

\subsection{Data Processing}
Dynamic spectra were formed in the following fashion, similar to that done by Hill et al.\ (2003).
Data were binned into a 3D cube: pulse phase, radio frequency, and sub-integration number (or time; 10-s per time slice).
The cube was then collapsed along the pulse phase axis in order to locate the pulse.
An ON pulse window was established by eye that contained more than 95\% of the pulse energy.  For these pulsars, the ON window was typically about 5--10\% of the total pulse period.
An OFF pulse region of the same size was then identified from the cumulative pulse profile.
The dynamic spectrum was formed from each sub-integration by calculating
\be
S_i (\nu) = \frac{{\rm ON}_i(\nu) - {\rm OFF}_i (\nu)}{<{\rm OFF}(\nu)>},
\ee
where $\nu$ is the radio frequency, ${<{\rm OFF} (\nu)>}$ represents the average off-pulse spectrum (the bandpass), $i$ indexes the sub-integrations, and the division by this denominator partially corrects for varying sensitivity across the band.

When we substitute $t$ for time (0 -- 60~m in 10-s increments) in place of sub-integration number and $\nu$ is  also  discrete, the dynamic spectrum will be denoted as $S(t, \nu)$  as displayed, for example, in the upper panels of Figure~\ref{fig:SS0628}.
Because of the location of the GBT in the National Radio Quiet Zone and the differential (ON - OFF) nature of the spectrum formation, radio frequency interference (RFI) was not a major problem in the analysis.
Arecibo observations were more strongly affected by RFI, and the GBT/1400~MHz observations also had persistent RFI in several channels slightly below 1420~MHz.


{Some pulsars have deep amplitude modulations 
due to broadband intrinsic pulsar variability $p(t)$ such as nulling. Nulls appear as brief minima, which are often near zero amplitude. Across the spectrum these are detected at times offset by their relative dispersive delay \citep{ric70}. However, such offsets are smaller than our 10 sec time step for all the observed pulsars and so the nulls appear to be synchronous in the dynamic spectrum.  For example, see B0525+21 and B1706--16 in figure sets 1.11 and 1.31--1.33
We estimate the intrinsic modulation} by averaging $S(t, \nu)$ over frequency at each 10s time step to obtain the pulsed time series $p(t)$.    
As we describe below we subtract the estimated $p(t)$ from $S(t, \nu)$ at each time step, in order to minimize its effect on the secondary spectrum.

The secondary spectrum (SS) is the primary data product of interest in the study, computed from the power spectrum of the dynamic spectrum (DS).  Cordes et al.\ (2006) refer to this quantity as 
$S_2 (f_t, f_\nu) = |\tilde{S_c} (t, \nu)|^2$, where the tilde denotes a Fourier transform, and the
axes $f_t$ and $f_\nu$ are conjugate to the $t$ and $\nu$ axes, respectively. 
However, it has become more common in the literature to identify $f_t$ with differential Doppler frequency, $f_D$, and the conjugate frequency axis, $f_\nu$, with differential Doppler delay, $\tau$.
We use that notation in what follows.
An example of $S_2 (f_D, \tau)$ is shown in Figure~\ref{fig:SS0628} as well, where we follow the convention, standard in this field, of displaying $S_2$ using a logarithmic grayscale in order to encompass the large dynamic range often present in the data.
Note that the color table for the display of $S_2$ depends sensitively on the upper and lower power limits displayed 
as defined in \S\ref{sec:data}.

As noted above intrinsic pulsar variations $p(t)$ modulate $S(\nu,t)$ {synchronously} across the entire bandwidth. Hence they contribute power to the SS along the \fD\, axis 
$S_2(f_D, \tau=0)$, which is set to zero by subtracting $p(t)$ \citep{wvm+22}. 
$S_2$ is computed via a finite discrete Fourier transform and there is a corresponding spectral response function in delay and Doppler, whose sidelobes can allow leakage of power from isolated peaks to spread through the SS. By subtracting the intrinsic $p(t)$ from the DS we reduce the corresponding leakage to higher delays. We further reduce leakage of power from the peak near the origin of $S_2$ by applying a window function to the DS.  We use a cosine-squared window to taper the outer 20\% of the DS to zero in both time and frequency {in order to reduce leakage due to broadband RFI and to residual effects of intrinsic pulse variation.}

\subsection{Data Availability}
\label{sec:datashare}
{\bf Note for the arXiv version of this paper}, which has been accepted by ApJ and will appear in a forthcoming issue.
The zenodo link below will not be live until publication of the paper in ApJ.
Also, Figure sets 1 and 4, which will be included in the HTML version of the online ApJ paper are here attached as {\tt .png} files in the /anc directory.

 The dynamic spectra (DS) analyzed in this paper can be accessed at this DOI: \doi{10.5281/zenodo.6413233}.
Machine readable versions of the tables, including substantially more information in the case of Table~3, are available at the same DOI.
The software used to analyze and display the data is not in a public repository.
Any questions related to the details of the processing or access to the software employed should be directed to one of the first two authors.
{The Matlab code used to produce the theoretical secondary spectra displayed in Figures~8 and 9 can be obtained by contacting B.~Rickett.}

\section{Secondary Spectrum - Data Presentation}
\label{sec:data}

We have assembled the 54 observations of the 22 pulsars as a figure set visible online.
In each plot we show the dynamic spectra and associated secondary spectra together with two lower panels as described below. 
Of these, 13 pulsars were observed at three frequencies, 6 at two frequencies, and 3 were observed at one frequency only.

\begin{figure*}
\begin{center}
\begin{tabular}{lll}
\includegraphics[angle=0,width=5.5cm]{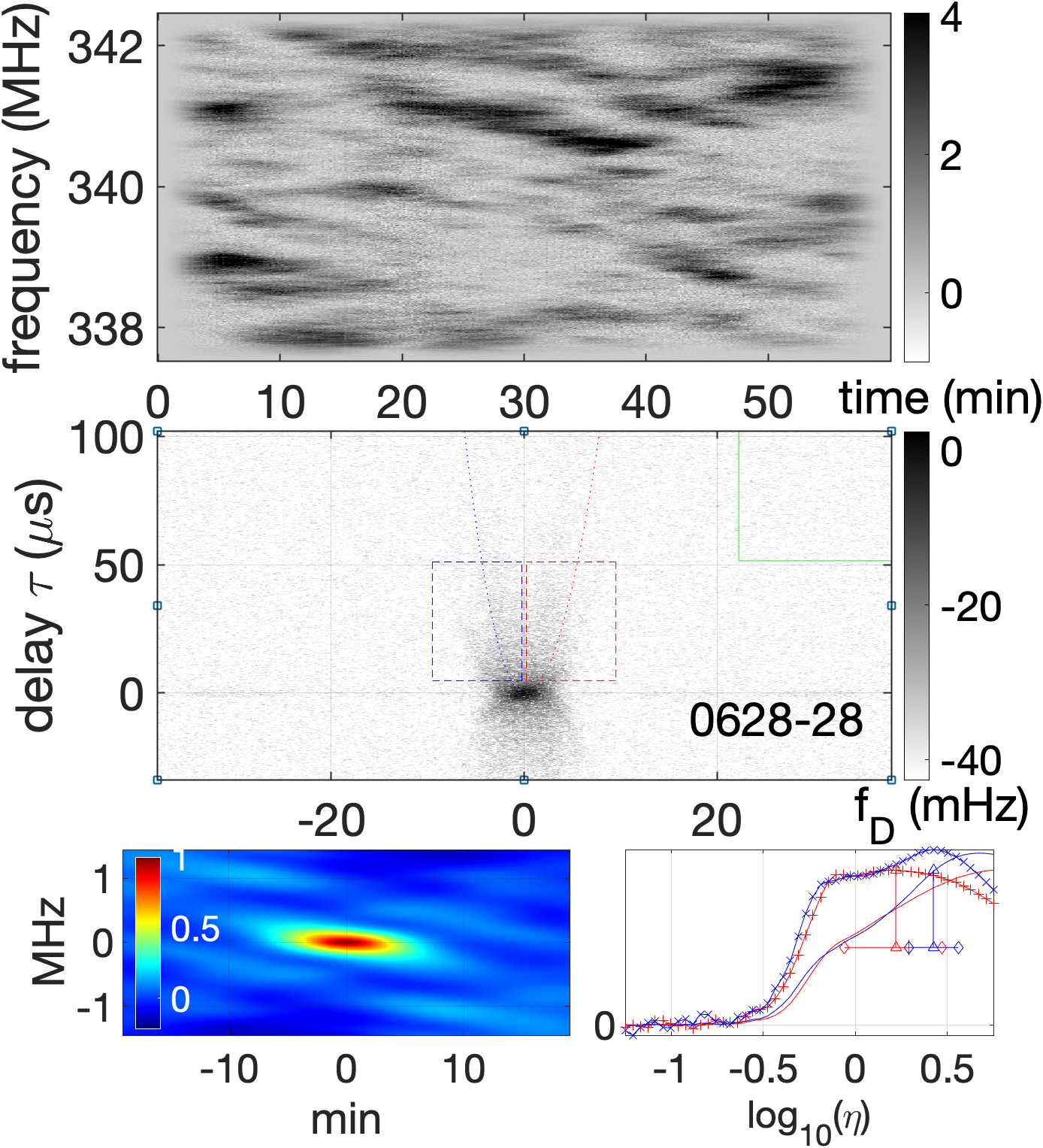} &
\includegraphics[angle=0,width=5.5cm]{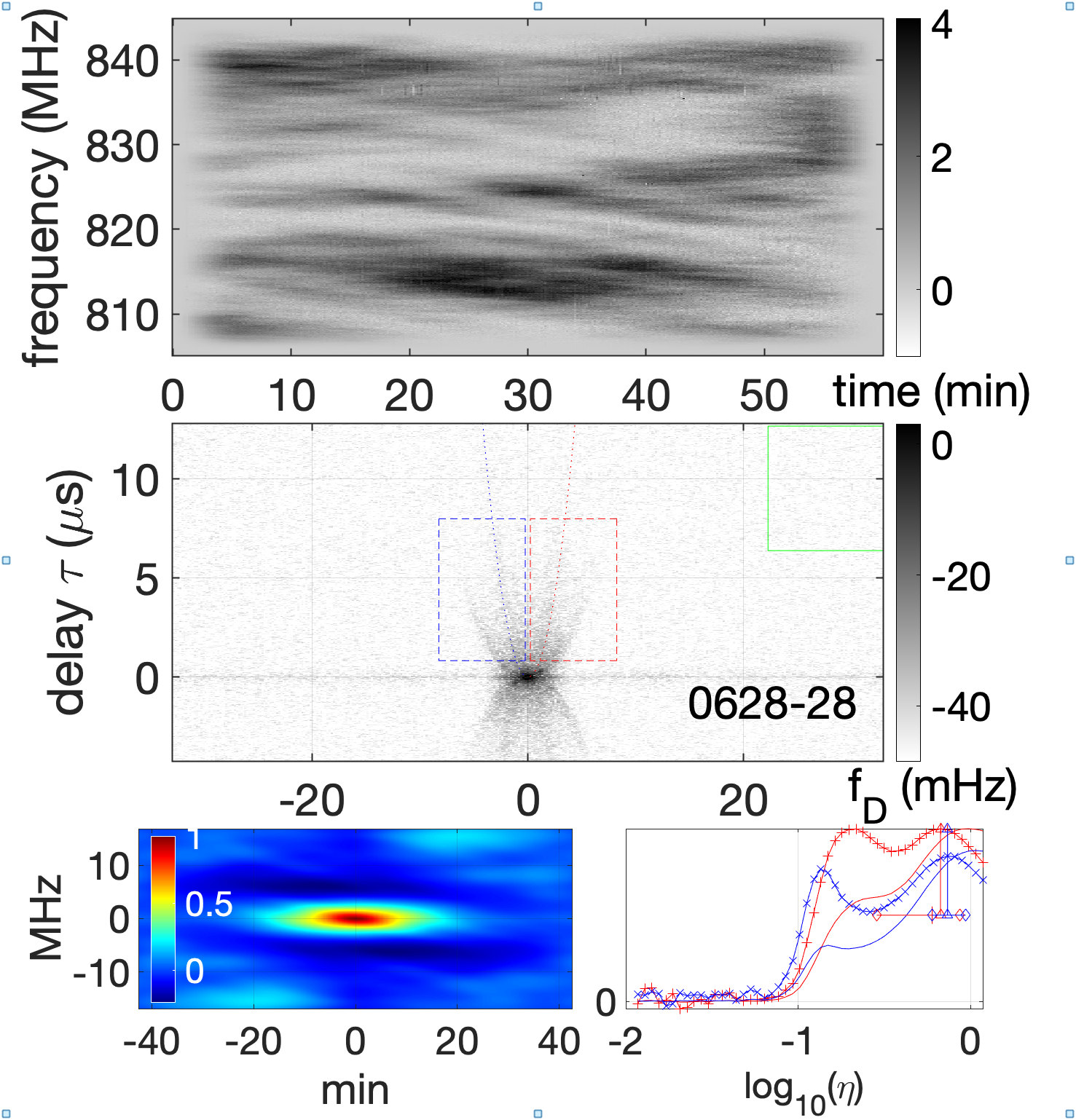} &
\includegraphics[angle=0,width=5.5cm]{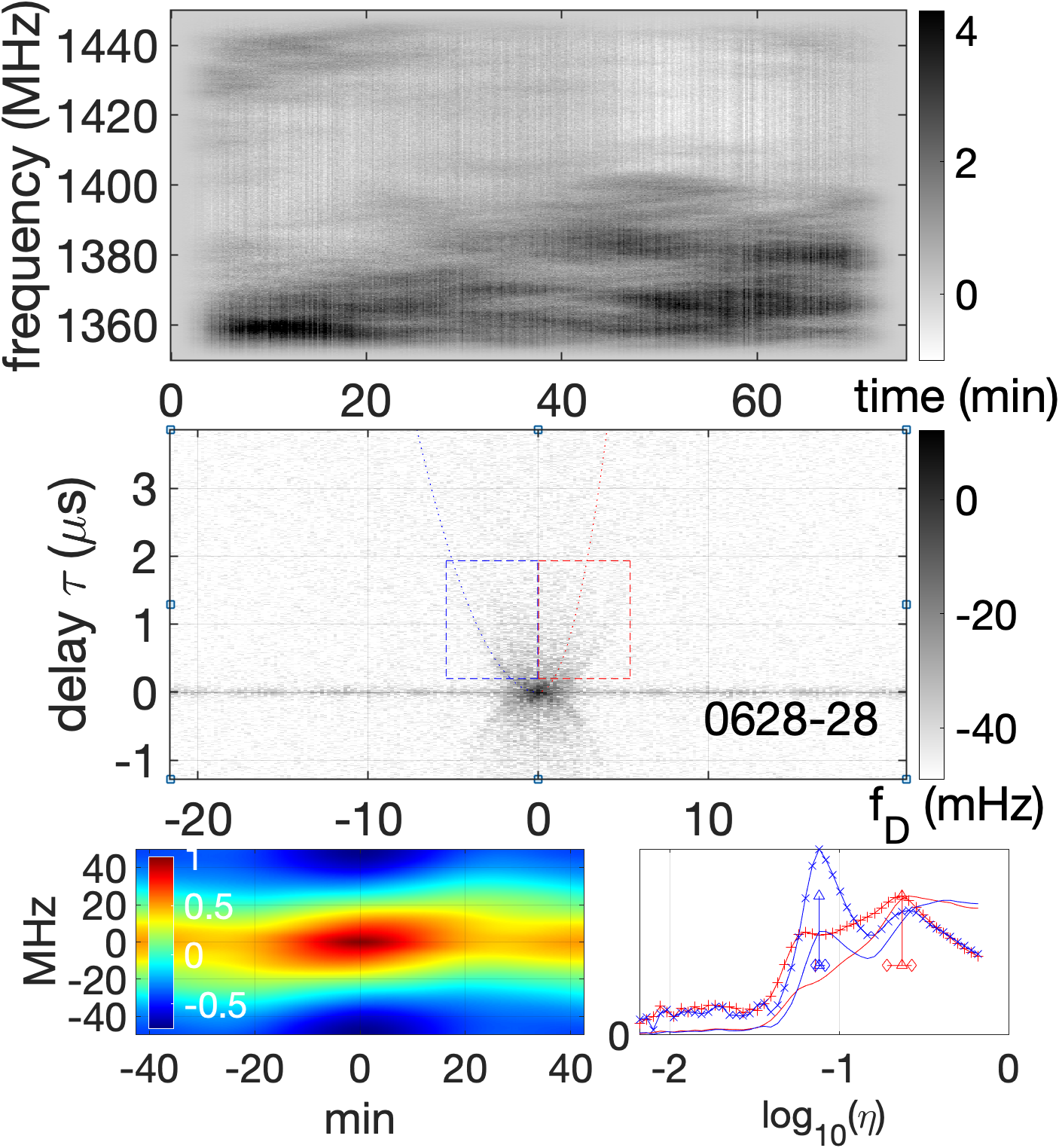} \\
\end{tabular}
\caption{Sample of main data display for pulsar B0628--28 (also Figure Set~\psrH).  \it Left: \rm 340~MHz on MJD 53632,  \it Middle: \rm 820~MHz on MJD 53637,  \it Right: \rm 1400 MHz on MJD 58904.   Upper panels are dynamic spectrum linearly scaled in units of the mean flux density;  middle panels are secondary spectra $S_2$ with decibel scaling. Lower left panels are auto-correlations of the dynamic spectra versus time lag (horizontal, minutes)  and frequency lag (vertical, MHz). The lower right panels show curvature estimation by parabolic summation of $S_2$ over a range in delay (defined by the blue and red rectangles in $S_2$).  The summation is plotted versus log of curvature $\eta$; all of the parabolas have apexes at the origin.  
As described in \S\ref{sec:scintarccurve} the plain red and blue curves plot the direct linear summations for negative and positive $\fD$, and the curves with x-marks are summations weighted by $| \fD |$.  The locations of the peaks in the weighted curves are flagged by vertical lines with a horizontal bar defining a width at 0.95 of the peak.
Faint dotted red and blue
parabolas are over-plotted on the SS at these estimated curvatures.
The complete figure set (54 images) is available in the online journal. } 
\label{fig:SS0628}
\label{fig:main}
\end{center}
\end{figure*}

Examples of the display format are shown in Figure \ref{fig:SS0628}  
as three plots for B0628--28 at frequencies 340, 825 and 1400 MHz.  In each case the upper panel is the dynamic spectrum (DS) in standard gray-scale format with linear scaling.  The lower panel shows the secondary spectrum (SS) with decibel scaling chosen to emphasize the structure at low levels (ranging from white at the mean noise level $S_{\rm noise}$ and saturating at black at a level 5dB below its global maximum).  
$S_{\rm noise}$ is estimated from the mean of $S_2$ in a rectangular region away from areas of ISS, which is outlined in green. {(Relative to the Nyquist limits $\tau_N, f_{D,N}$ in delay and Doppler, the noise region is
$0.49 \tau_N < \tau < 0.95 \tau_N$ and
$0.44 f_{D,N} < \fD < 0.93 f_{D,N}$ )}

Also shown at the lower left of each plot is the auto-correlation function (ACF) of the DS 
versus offsets in time and frequency. Cuts along the axes are used to estimate the decorrelation times and bandwidths, as described in more detail in \S\ref{sec:scintparam}.  The displayed range is set to be 5 times wider than the measured scales.  Lower right panels in each plot, discussed in \S\ref{sec:scintarccurve}, show how we estimate the curvature of any parabolic arc structure present.

\subsection{Overview}

The 54 plots provide a visual description of the ISS.  
As has long been known, the scintillation appears as a random distribution of peaks in the DS, whose widths in frequency and time can be characterized from its ACF.  Some such peaks (\it scintles\rm) can appear tilted causing tilts in the ACFs.  In traditional studies of ISS, the ACF widths are the main parameters extracted from an observation, and ISS was originally recognized by the narrowing in the ISS bandwidth for pulsars at increasing DM \citep{ric70}.

Scintillation arcs were discovered as systematic curved structures in $S_2(\tau,\fD)$, often  many decibels below the peak. 
The most common form of arc is a simple \it forward parabolic arc \rm $\tau = \eta\, \fD^2$, with its apex at or near the origin, as characterized by its curvature ($\eta$).   
In general we define arcs by secondary spectra that are peaked narrowly (or broadly) about such a parabola.  Such arcs can exhibit a dip or \it valley \rm along the delay axis near zero \fD;
for example, see B0450--18 at 340 MHz (Figure~\psrC).     

\begin{figure*}[t]
\begin{center}
\begin{tabular}{lll}
\includegraphics[angle=0,width=5.6cm]{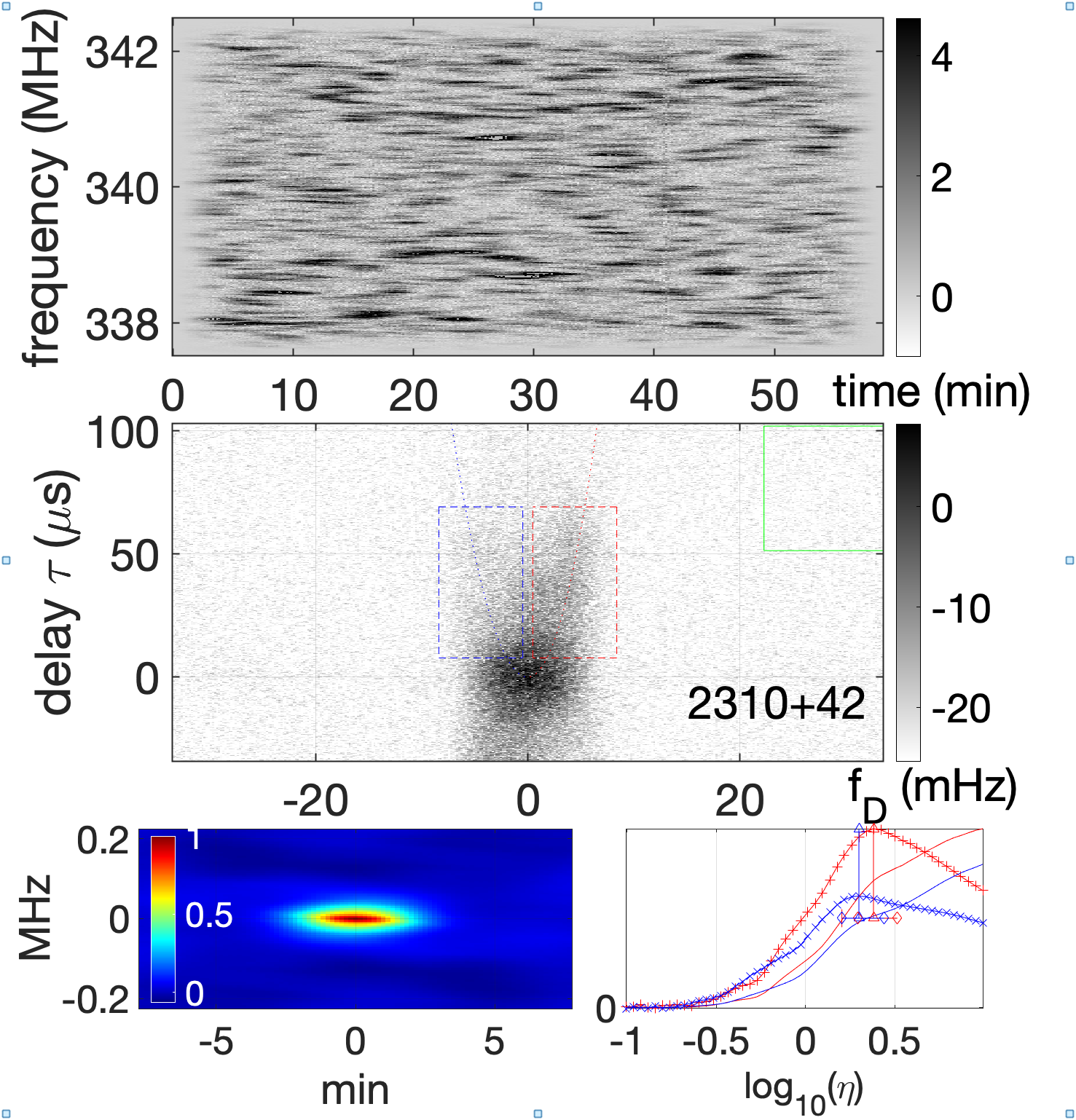}  &
\includegraphics[angle=0,width=5.6cm]{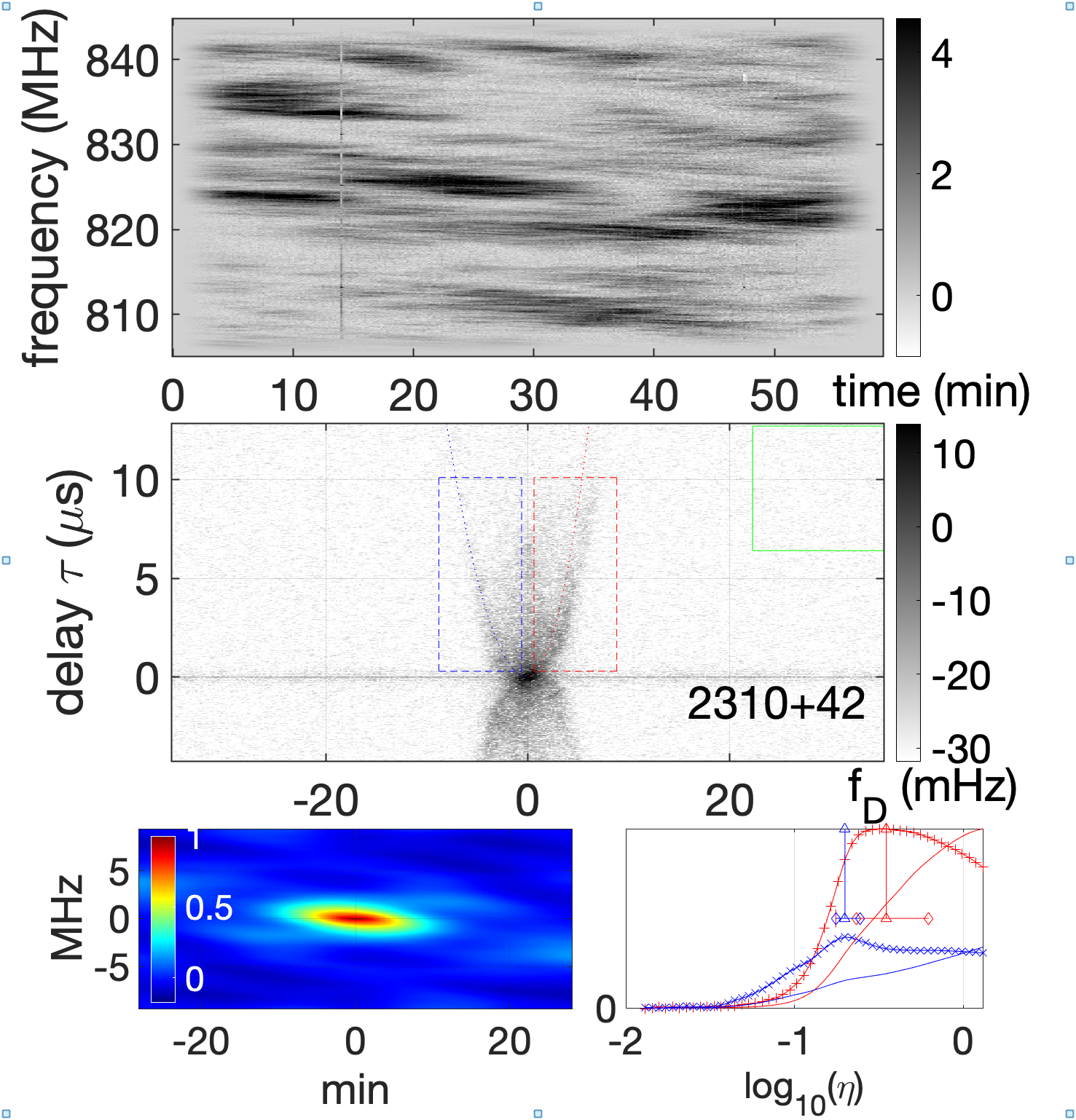} &
\includegraphics[angle=0,width=5.6cm]{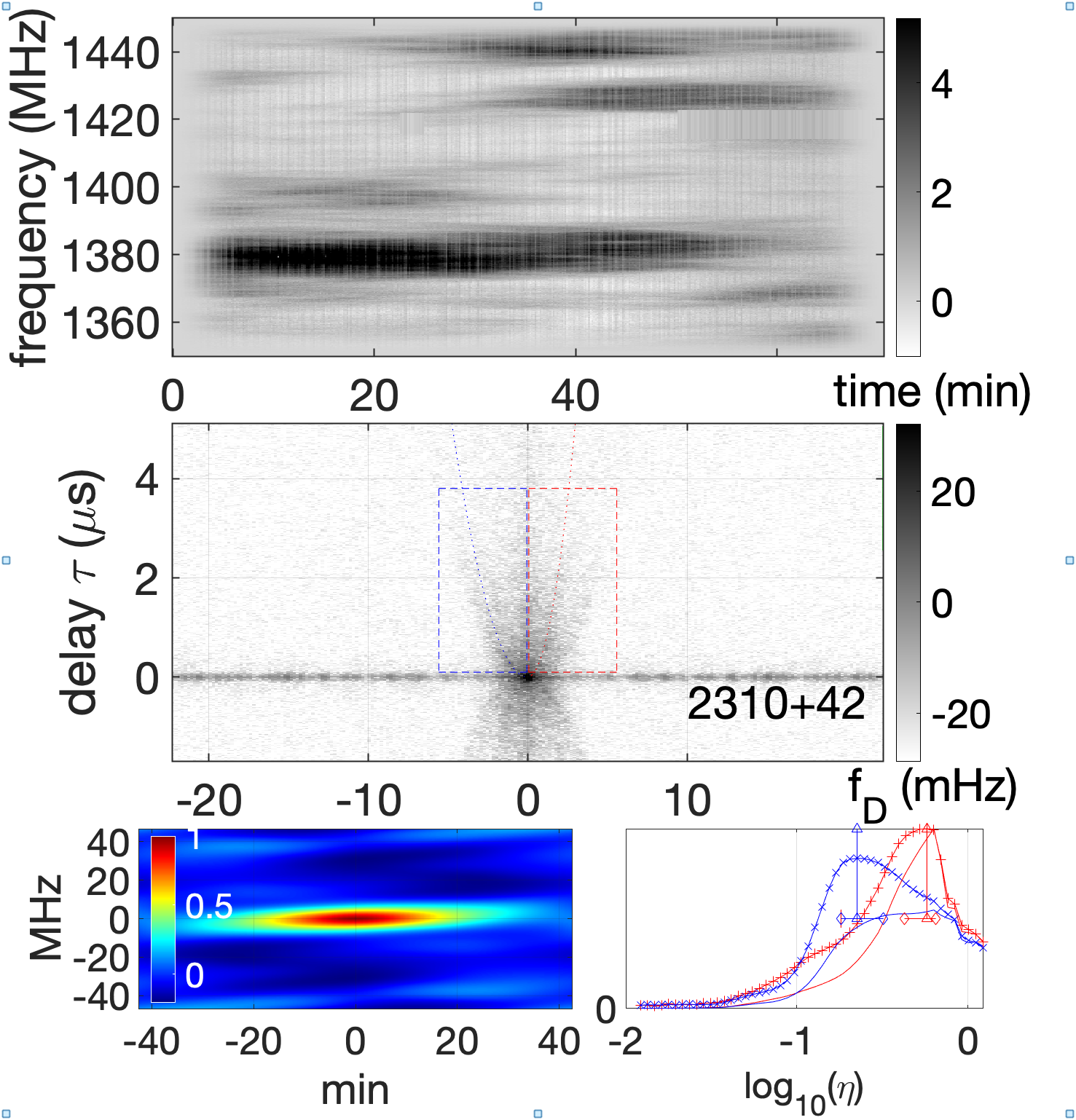}
\\
\end{tabular}
\caption{Dynamic and secondary spectra for B2310+42 (Figure Set \psrU). An example of a SS dominated by a centrally condensed core (CC) at the lowest frequency that becomes a scintillation arc structure at the two higher frequencies.
\it Left: \rm 340 MHz.  \it Center: \rm 825 MHz.  \it Right: \rm 1400 MHz. Same format as 
Figure~\ref{fig:SS0628}.}
\label{fig:SS2310} 
\end{center}
\end{figure*}

\it Multiple forward arcs \rm have been reported from some nearby pulsars: \eg {\danx Putney and Stinebring, (2006; pulsars B0329+54, B0823+26, B0919+06, B1133+16, B1642--03, and B1929+10); McKee et al.\ (2022; B1133+16);  and Reardon et al. (2020; J0437-4715).} 
\nocite{ps06a,rcb+20,mzsc22}
However, there are only a few examples in the observations reported here, which include many more distant pulsars. In Figure~\ref{fig:SS0628} for B0628--28 at 825 MHz  the outer arc provides a relatively sharp boundary with little SS power outside, while the inner arc is much less distinct.   A second example is B2021+51 at 1400 MHz in Figure~\psrQ.
A third example is B2310+42 at 825 MHz in Figure~\ref{fig:SS2310}.   It shows an outer arc which also acts as sharp boundary which we call a \it bounding arc\rm.  The inner structure is more like a broad ridge than an arc.  Note also that at 340 MHz the outer arc no longer acts as a sharp boundary.  Similar differences between low and high frequencies are common throughout the survey data and are discussed further in \S\ref{sec:comments}.   For B2310+42 we also show results at 1400 MHz, which we discuss in \S\ref{sec:comments}.

A number of pulsars exhibit isolated peaks {in their SS} which may or may not lie near a forward parabolic arc.  When such peaks follow a curved shape we refer to them as {\em reverse arclets}, which were discovered in pulsar B0834+06 \citep{hsb+03}.  Reverse arclets  have negative curvature and apexes that lie close to the underlying forward arc.   Figure~\psrC\ shows similar reverse arclets for B0450--18, which we have analyzed in detail elsewhere \citep{rszm21}.  Another example can be seen in Figure~\psrE\ for B0525+21.  
Reverse arclets can be understood as the interference of scattering from a discrete offset point with a central anisotropic scattered distribution \citep{wmsz04, crsc06}.  
Their apexes lie on the forward arc when the offset point lies along the axis of anisotropy.  Their curvature equals the reverse of the forward arc when the scattering is localized in the same screen as the main forward arc.  This situation can sometimes be recognized when a reverse arclet extends inwards as far as zero $\fD$ and passes through the origin.  Arclets with forward curvature are rare.\footnote{With hindsight, the fringe pattern identified by Rickett et al (1997) can be recognized as a one-sided forward arclet.  This can be caused by interference of the un-scattered wave with anisotropic scattering from a region that is displaced  in the perpendicular direction.}  
Pulsar B1508+55 exhibits an unusual variation of  \it flat arclets\rm\ at 825~MHz and 1400~MHz (Figure~\psrK).  
In another variation, Figure~\psrB\ for B0450+55 shows an isolated point in its SS, which is not extended in \fD.

Another common feature is \it asymmetry \rm in the intensity of $S_2$ versus differential Doppler \fD, which appears as {\em tilted} scintles in the DS and a tilted ACF.
 \footnote{The reciprocal tilt between the asymmetry in the SS and the ACF follows, of course, because they are a Fourier transform pair. Note, however, that the SS is normally displayed with a logarithmic color table -- highlighting low power values -- whereas a linear color table is usually used for an ACF.}   
 Asymmetry can also be seen between the height of the positive and negative peaks in the parabola summation curves.  There may also be asymmetry in that the apex of an arc may be slightly offset to positive or negative \fD, which can be due to refraction by a transverse gradient in the electron distribution somewhere along the LoS
 \citep{crsc06}. 

\begin{figure*}[t]
\begin{center}
\begin{tabular}{lll}
\includegraphics[angle=0,width=5.6cm]{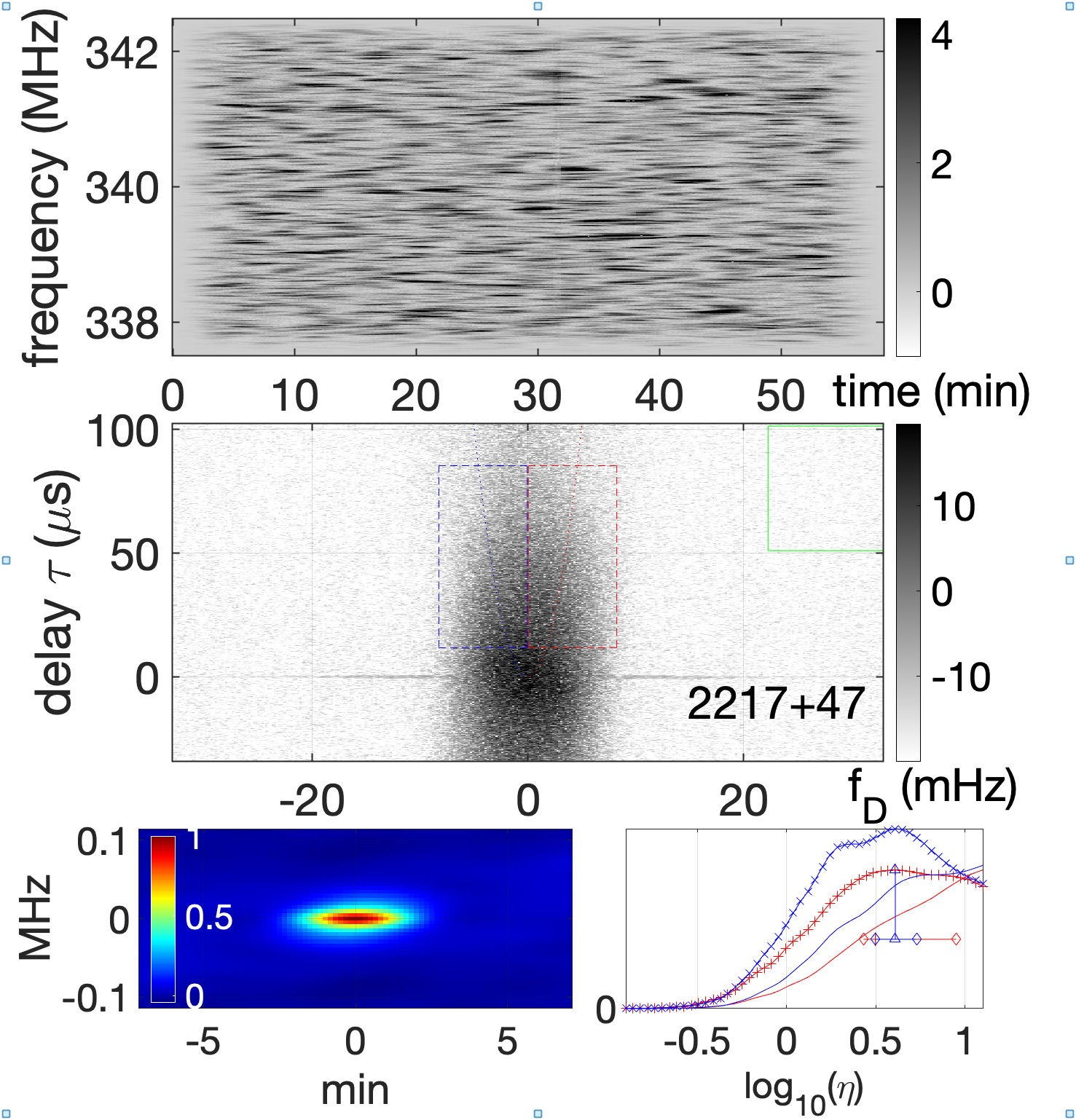}  &
\includegraphics[angle=0,width=5.6cm]{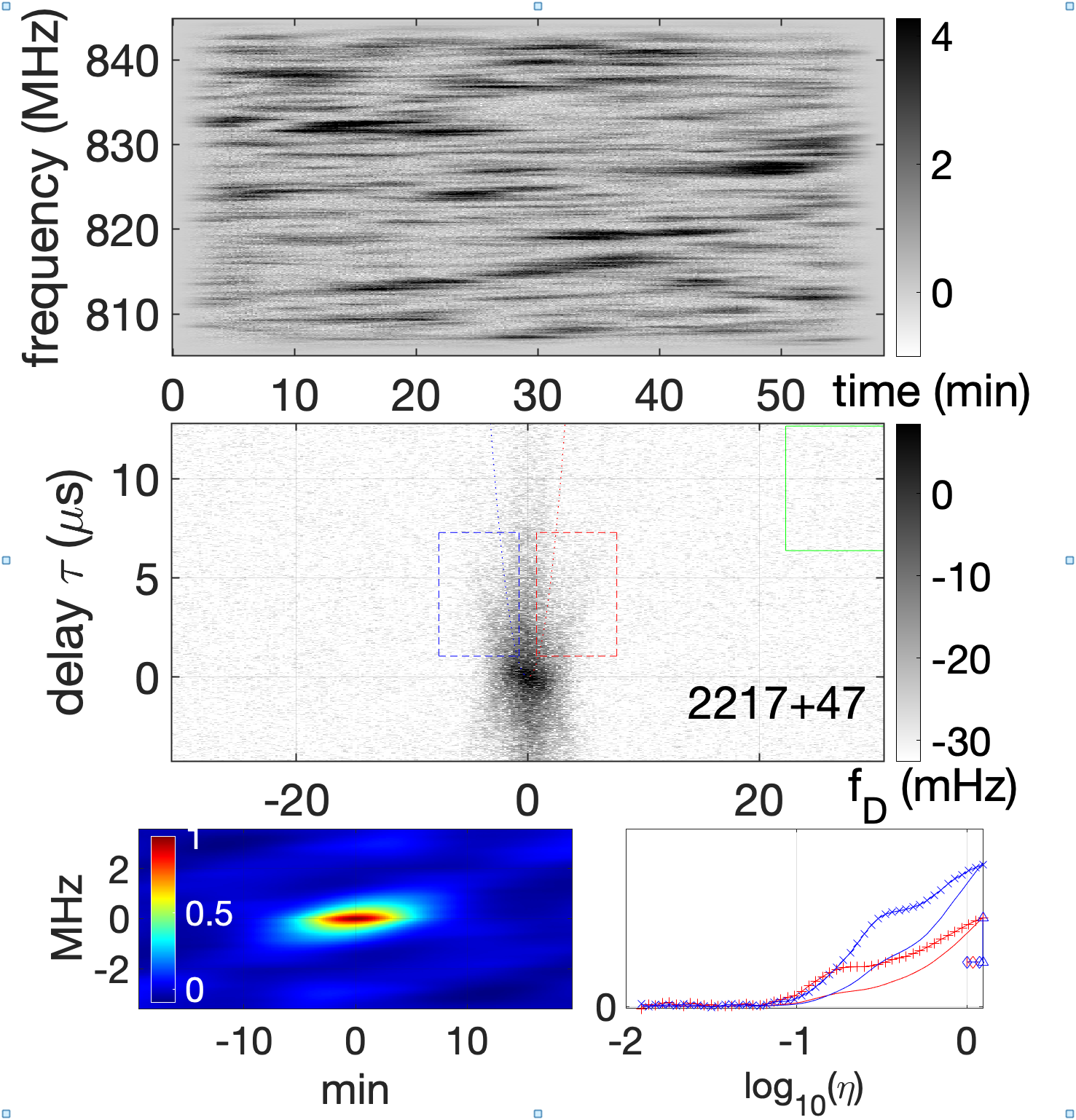}  &
\includegraphics[angle=0,width=5.6cm]{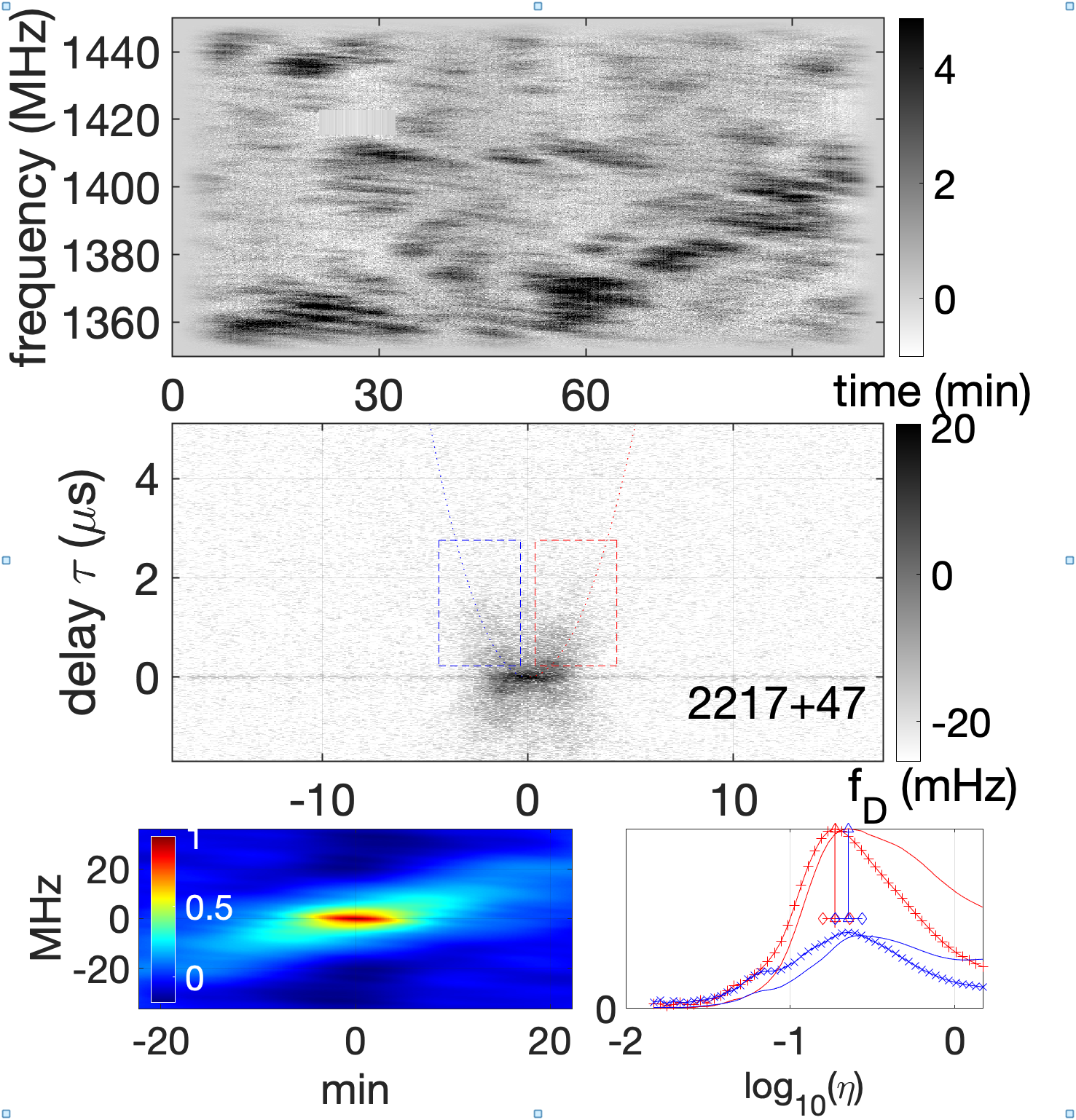}  \\
\end{tabular}
\end{center}
\caption{Dynamic and secondary spectra for B2217+47 (Figure Set \psrT). An example of observations that show a CC of power at the lowest frequency giving way to a more elongated distribution at the highest frequency.
\it Left: \rm 340 MHz at MJD 53632.    \it Middle: \rm 825 MHz at MJD 53637.   \it Right: \rm 1400 MHz at MJD 58920. Same format as Figure~\ref{fig:SS0628}. }
\label{fig:SS2217} 

\end{figure*}

{\danx Scintillation arc studies hold the promise of being able to locate scattering material along the LoS, at least in optimal cases. 
Dating back to the earliest days of ISS studies, the prevailing paradigm has been one of a pervasive turbulent medium punctuated by ``clouds" of increased turbulence along the LoS.
For example, see \citet{cwf+91}.
Although there is no thorough analysis of how distributed scattering along the LoS will show up in the SS,
several lines of argument indicate that it should produce a {\em centrally concentrated} (CC) region of power around the origin.
An example of a SS that displays this distribution is shown in the left panel of Figure~\ref{fig:SS2217}.
The DS in this case consists of a large number of scintles with no evidence for {fine structure} 
within a scintle.
Referring to the two higher frequency observations for this pulsar in the center and right panels, we see that the scintle structure broadens out with increased intra-scintle modulation in the DS and a corresponding tendency toward arc-like behavior in the SS.
We will discuss this generic frequency development further in \S\ref{sec:conditions}.
}

\subsection{Basic Scintillation Parameters}
\label{sec:arcparam}

\subsubsection{ISS Decorrelation Widths}

As is conventional procedure (\eg\ \citet{cwb85,grl94}), we estimated the frequency scale, \nuiss, and the time scale, \tiss, of the scintillation structure using the intensity autocorrelation function, $R(\nu,t)$, which we discuss in more detail in \S\ref{sec:scintparam}.
 These decorrelation scales are obtained, respectively, by determining where
$R(\nuiss,0)=0.5\, R(0,0)$ and $R(0,\tiss)= 0.5\, R(0,0)$ and are presented in Table~\ref{tab:analysis}. 
(We note that this definition differs from the convention of \citet{ric70} and \citet{cor86}, where the e$^{-1}$ point is used in the autocorrelation time lag.) 
The fractional errors in them are estimated as {the minimum of unity and} $N_{\rm iss}^{-1/2}$, where $N_{\rm iss}$ the number of independent ISS fluctuations over the observed bandwidth ($B$) and time span ($T$).  
We define $N_{\rm iss} \approx   (\epsilon B/2\nuiss)(\epsilon T/2\tiss)$, since $\nuiss$ and $\tiss$ are half-widths of the auto-correlations, and where $\epsilon = 0.2$ accounts for the fact that the exponentially distributed intensity of the scintles leads to  peaks in ISS that are sparsely distributed in time and frequency resulting in fewer independent  ISS fluctuations {(see equation (7) of \citet{cor86}). Although the value ($\epsilon = 0.2$) predicts conservatively large errors, we use it since it has been widely used in previous scintillation studies.  It needs to be validated by complete statistical modeling of scintillation, which is beyond the scope of the paper.} 

It is also important to note that there are a few observations in which the scintillations  have such a narrow frequency scale that they can be unresolved in the channel bandwidth of the spectrometer, {
We apply a simple quadrature correction for the resulting under-resolution in frequency.  $\Delta\nu_{\rm corr} = \sqrt{\nuiss^2 - \delta\nu^2)}$, where $\delta\nu$ is the channel bandwidth. A similar correction to the time scale $\tiss$ is applied with the 10~s integration time for the spectrometer, in place of the channel bandwidth.
There is one case in frequency and one case in time that this correction fails, giving an imaginary estimate; these cases are flagged in tabulating the results with an ellipsis indicating no valid data.
There are a few other cases where the scintillations are so slow or so wide in frequency that a single scintle may cover the entire observing range, \ie\ $N_{\rm iss} \lesssim 1$, with correspondingly large errors.}

\subsubsection{Scintillation Arc Parameters}
\label{sec:basicarc}

There is recognizable scintillation arc structure in more  than half of the 54 secondary spectra plots.   However, it is difficult to devise a simple yes/no criterion for the presence of parabolic arc structure.  Consequently, we have analyzed each observation to estimate a number of specific measurable quantities. We elaborate on this further in \S~\ref{sec:scintarccurve}, but briefly describe the fundamental quantities here.

We quantify forward arcs in the secondary spectrum  by estimating the curvature $\eta$ of the underlying parabola $\tau = \eta\, \fD^2$. 
We sum $S_2$ along parabolas that cover a range in curvature, and
in many cases we find a clear maximum in the summation curve and assign a value $\eta_p$ to this dominant parabola (Table~\ref{tab:analysis}).   As already noted, we occasionally also find a bounding arc outside this inner parabola; see \S\ref{sec:scintarccurve} for details.

We also include in the tabulated results a \it curvature credibility \rm criterion, \etacred, for each observation.  It is a subjective evaluation of the reliability of the curvature estimate obtained by examining the parabola summation curve for each case:  2, 1 or 0.  A compact maximum in the curve is rated 2; wide and double peaked curves are rated 1; cases where the peak is at the high or low limit in the search range or the secondary spectrum extends to the Nyquist delay are rated 0.  In the latter situation, the DS may be unresolved in frequency, and only an upper limit can be estimated for the decorrelation bandwidth \dnuiss.   
As detailed in \S\ref{sec:arcwidth}, we define and tabulate
a width measure $\Delta\eta$ to quantify how sharply $S_2$  is peaked about the forward parabolic arc. 

\section{Comments on Individual Sources}
\label{sec:comments}

Using terminology and results from the previous  sections, we  qualitatively discuss results 
from the  22 pulsars below.
We group them first by the prominence of the scintillation arc, followed by sorting them in RA.
The three groupings used below have a connection with the \etacred\ index.
However, the discussion here is on a per-pulsar basis, and a number of pulsars have
$\etacred = 2$ or 1 at one frequency with a lower index at one or more observing frequencies.
In some places, the signal to noise ratio (S/N) will be discussed qualitatively here. 
We treat it and parameters of the scintillation arcs  quantitatively in \S\ref{sec:analysis}.

\subsection{Pulsars with a Definite Scintillation Arc }

\subsubsection{B0450+55 (Figure Set \psrB)}
The 1400~MHz observation shows two fairly well-defined scintillation arcs.
 At 825~MHz, in data taken 14 years earlier, the scintillation arc structure is not well-defined, although there is a hint of an arc coincident with the dashed blue line in the 2nd quadrant of the SS.
 The CC of the 340~MHz data has a negative power asymmetry (\ie\ quadrant 2 power is greater than quadrant 1), as does the CC at 825~MHz, observed 5 days later.
\etacredc{340}{825}{1400}{0, 1, 1}
 
\subsubsection{B0450--18 (Figure Set \psrC)}  
We have reported on these observations in detail (Rickett et al.\ 2021).
\nocite{rszm21}
However, the 1400~MHz observation was not available at the time of that analysis.
In that paper we found  a 1D brightness distribution matched the data well, but  the overall brightness
function $B(\theta)$ was not consistent with simple Kolmogorov scattering in a thin screen. 
Instead, $B(\theta)$ scaled with frequency more slowly than Kolmogorov and various
local peaks in $B(\theta)$ were trackable across narrow frequency intervals, but not
between 340~MHz and 825~MHz. 
The 1400~MHz observation was made more than 14 years after the two at lower frequencies, so the LoS may be probing quite different ISM conditions. 
Two thin scintillation arcs are present. As discussed in \S\ref{sec:Discrepant}, neither of these is consistent with the curvature of the {\em heavily saturated} scintillation arc visible at the two lower frequencies.
Hence, this must be due to a different region of scattering along the LoS. 
\etacredc{340}{825}{1400}{2, 2, 2}

\subsubsection{B0525+21 (Figure Set \psrE)} 
The 3.7~s pulsar period and nulling causes modulation of the DS in time, particularly with our integration time of 10~s. But a criss-cross pattern is clear across the scintles,  seen as reverse arclets in the SS. 
As commented on in \S\ref{sec:pqAnalysis}, and widely in the scintillation arc literature since \citet{wmsz04}, such reverse arclets are indicative of a nearly 1D brightness distribution.
Our parabolic summing algorithm (lower right panel) reports a significantly wider arc on the RHS (positive Doppler) than on the LHS.
This is due to the influence of power near the origin; an algorithm that intercepted the apexes of the inverted arclets would produce more nearly equal values of $\eta$ for the two signs of Doppler frequency. 
\etacreda{1390}{1}

\subsubsection{B0628--28 (Figure Set \psrH)}   
See Figure~\ref{fig:SS0628}. 
These are  high S/N observations at all three frequencies. The low frequency data show a CC core with a clear bounding parabola for power further from the core of the SS. 
The 825~MHz  observation shows a compact CC core and two scintillation arcs or, alternatively, a boundary arc with
an interior arc due to anisotropic scattering and a velocity vector with significant tilt to the major axis of the scattered image \citep{rcb+20}.
The 1400~MHz data are consistent with the trend, seen elsewhere in this survey and in previously published data, for scintillation arcs to become substantially sharper at higher frequency.
\etacredc{340}{825}{1400}{1, 1, 1}

\subsubsection{1508+55 (Figure Set \psrK)}	
This pulsar, relatively distant (2.10 kpc) for the survey, has the highest transverse velocity (963 km~s$^{-1}$) in the sample and one of the highest of the entire pulsar population \footnote{Here and throughout the paper we refer to published measurements of pulsar dispersion measure, distance and velocity as listed in \citet{mhth05}, without discussion of their associated uncertainties.}.
The low frequency data show a clear CC core plus a very broad scintillation arc that also exhibits strong local maxima in the B($\theta$) distribution. This is even more pronounced at 825~MHz, where the highly unusual, flat-topped arclets are a prominent feature. 
All three frequencies show the presence of the same arc despite the fact that the pulsar has traveled approximately 2900~au transverse to the LoS during this time.
Similar flat arclets were recorded by \citet{msm+21}. 
Low curvature arclets could be due to localized scattering near the pulsar (i.e. small value of $s$) interfering with a core in brightness at small angles of deflection.
The scattering geometry is complex for this pulsar, as documented by 
\citet{btsd20}, who observed remarkable echoes of the pulse arriving 30~ms after the main pulse at 50 and 80~MHz that persisted over about 3 years.
Also see \citet{smw+22} who develop a two-screen explanation for the scattering from this pulsar.
\etacredc{340}{825}{1400}{2, 2, 2}
%
\subsubsection{B1540--06 (Figure Set \psrL)}
Similar to B0450--18 and B1508+55: a broad scintillation arc at 340~MHz that nevertheless shows signs of broadened arclet structure, confirmed at 825~MHz.
Consistent with highly anisotropic scattering in a plasma screen. Reverse arclets due to narrow peaks in a 1-dimensional scattered brightness profile.
\etacredc{340}{825}{1400}{2, 2, 0}
%
\subsubsection{B2021+51 (Figure Set \psrQ)} 
This is an excellent example of no scintillation arc at the lower frequency, but clear evidence for arcs at higher frequency. 
As discussed in \S\ref{sec:lowfreq}, if we only had the 340~MHz observation we would classify this as a pulsar without a scintillation arc.
Although the dynamic spectrum at 825~MHz shows about a dozen classical scintles, the secondary spectrum is remarkable in its sharpness and detail. 
The 1400~MHz observation, offset by 14 years, is fully consistent with the 825~MHz observation, showing a boundary arc with a filled interior that is similar to the signature expected for an anisotropic image with major axis not aligned with the effective velocity vector \citep{rcb+20}.
\etacredc{340}{825}{1400}{0, 1, 2}

\subsubsection{B2045--16 (Figure Set \psrR)}
Quite similar to B2021+51, this pulsar shows only a hint of a scintillation arc at 340~MHz, but that arc is fully developed -- and has a deep unfilled valley -- at 825~MHz and 1400~MHz.
Nulling of the pulsar causes vertical stripes in the DS that show up as power along the \fD\  axis in the SS.
\etacredc{340}{825}{1400}{1, 2, 2}

\subsubsection{B2217+47 (Figure Set \psrT)}
This pulsar, relatively distant ($D = 2.39$~kpc) for this survey,  is fairly heavily scattered. It also has a large transverse velocity.
It has been extensively studied for interesting propagation effects along the LoS.
\citet{mhd+18} observed pulse echoes delayed by about 10~ms near 150~MHz. 
They concluded that the echoes, which varied slowly over five years,  were scattered by a dense plasma concentration of 100~cm$^{-3}$.
Using LOFAR data, \citet{dvt+19} reported an episode of frequency dependent DM variation toward this pulsar.
Similar to B2021+51 and B2045--16, but even more heavily scattered, the SS at 340~MHz is completely dominated by a CC.
This is basically true at 825~MHz, too, although there is a hint of a scintillation arc developing.
At 1400~MHz, even though the CC is still the dominant feature in the SS, the parabolic summing algorithm shows clear evidence for a broad symmetric scintillation arc.
\etacredc{340}{825}{1400}{0, 0, 1}
%
\subsubsection{B2310+42 (Figure Set \psrU)}
 Yet another pulsar with a dominant CC at 340~MHz that displays a clear boundary arc at the two higher frequencies.
 At 340~MHz, the DS has well-resolved scintles with high S/N.
The SS  shows broad power centered on the origin, with slight positive asymmetry to larger delay.
There is a hint of a boundary arc but no valley.
At 825~MHz, the DS has about a dozen well-resolved scintles, with no obvious modulation of them. 
However, the SS has a clear boundary arc with a shallow valley partly filled by SS power extending along the delay axis.
The DS at 1400~MHz has three wide, moderately narrow scintles in it, with no obvious intra-scintle modulation.
The resulting SS has an outer bounding arc with an inner arc, neither very distinct.
The parabolic summing algorithm traces out the bounding arc on the LHS, but it appears to find a
faint interior arc on the RHS.
Taken together, these data illustrate the value of observing at multiple frequencies. 
They are consistent with strong plasma scattering with modest anisotropy \citep{crsc06,rcb+20}.  
 See \S\ref{sec:summary} under B2217+47 on the effect of orientation of the anisotropy axis with the \veff\ vector.
\etacredc{340}{825}{1400}{1, 1, 1}
%

\subsubsection{B2327--20 (Figure Set \psrV)}
This  set of observations shows the value of scintillation data even when the S/N is not large.
At 340~MHz, the DS has only moderate S/N with wide, tilted scintles crossed with finer modulation.
There appears to be some broadband pulse modulation, perhaps nulling, which puts power onto the \fD\  axis.
The SS has a clear narrow arc, predominantly one-sided (negative asymmetry) with several loci of higher
power along the arc.
 Although it is hard to tell from the low S/N observation at 825~MHz, the DS shows a loosely organized criss-cross pattern.
The SS has a sharply defined arc with a deep valley and a slight negative power asymmetry.
{\danx As can be seen in the quantitative results for $\eta_p$ in \S5, the boundary arc detected at 1400~MHz, observed 14 years after the low frequency observations, is not consistent with the curvature of the $\lambda^2$ scaled values at lower frequencies. $\eta_{p,1400}$ is approximately 7 times greater than $\eta_{p,825}$, when scaled by $\lambda^2$. This places the scattering material as close as 40~pc from the Earth.}
\etacredc{340}{825}{1400}{2, 2, 0}
%
\subsection{Pulsars with a Probable Scintillation Arc}
\subsubsection{B0523+11 (Figure Set \psrD)}
This is one of the most heavily scattered pulsars in the SAS.
With 4096 frequency channels across only 2~MHz of bandwidth, the frequency resolution
is barely adequate to resolve the scintles, and even Arecibo's sensitivity is not quite adequate to
display a clear secondary spectrum. 
However, with a careful choice of the color table, spanning only 10 dB in power, we are
able to see a tilted bar of power (negative asymmetry) extending out to more than 200~$\mu$s.
At 1450~MHz, the tilted SS shows a wide faint arc (negative asymmetry) with shallow valley.
It appears that the CC merges into a 
 broad scintillation arc as opposed to being a simple
tilted concentration with quasi-elliptical contours.
\etacredb{422}{1450}{0, 1}
%

\subsubsection{B0540+23 (Figure Set \psrF)}
This pulsar has very similar scintillation characteristics to B0523+11. 
They both show  narrow scintles at the lower frequency, consistent with strong scattering through the same region of plasma.  
At the higher frequency a similar asymmetrical CC emerges with arclike properties at higher delay.
Since a plasma wedge is one mechanism for an asymmetrical power distribution in arcs,
the same sense of asymmetry is, perhaps, a linkage in the source of dominant scattering for these two pulsars.
We note that they are relatively close on the sky ($12.7^\circ$) and that B0523+21 and B0540+23 have DMs of 79.4 and 77.7 pc cm$^{-3}$, respectively. 
At a screen location of $s = 0.5$, the angular separation would require a transverse screen extent of approximately 190~pc, however. 
Despite these similarities, the curvature of their scintillation arcs are significantly different, with the value for B0523+11 placing the scattering material about $\gtrsim 600$~pc from Earth and about twice that value for B0540+23 (see details in \S5 and Table~3).  
\etacredb{432}{1450}{0, 1}
%
%
\subsubsection{B0626+24 (Figure Set \psrG)}
Both the 432~MHz observation and that at 1390~MHz display a slightly tilted power distribution.
At the higher frequency there is a hint of a scintillation arc in the parabolic sum parameter.
Again, as for B0138+59, longer duration observations are necessary because of the particularly long timescale of the scintles.
\etacredb{432}{1390}{1, 1}
%
\subsubsection{B0809+74 (Figure Set \psrI)}  
There appears to be a bounding arc in the SS of the 340~MHz observations. 
Observations at the two higher frequencies were too low in S/N and contaminated
with RFI in order to show anything clearly in the respective SS.
\etacredc{340}{833}{1400}{1, 0, 0}
%
\subsubsection{B0818--13 (Figure Set \psrJ)}  
The observations at both 340~MHz and 825~MHz are high S/N, but have inadequate frequency resolution.
At 1400~MHz, the frequency and time resolution are adequate, and there is some hint of unorganized
wispiness around the CC core.
Close inspection of the 825~MHz SS shows a slight bifurcation of the power distribution (along the $\fD = 0$ axis) near the Nyquist frequency in delay.
Observations of higher frequency resolution and longer time duration would be needed to explore further the possibility of a scintillation arc in this pulsar.
\etacredc{340}{825}{1400}{0, 0, 1}
%
\subsubsection{B1706--16 (Figure Set \psrM)} 
 This is a nulling pulsar, which causes extra power along the \fD\  axis.
This is also a low-velocity pulsar, which causes problems because \tiss\ is long relative
to a typical observation length, particularly at the two higher frequencies.
The 825~MHz SS suggests a bounding arc with a slight negative power asymmetry.
As was the case with B2310+42, the $\eta_p$ value on the left looks more reliable than that on the right because of the 
more prominent boundary arc on the left.
\etacredc{340}{825}{1400}{0, 1, 0}
%
\subsubsection{B1907+03 (Figure Set \psrP)}
This single-frequency Arecibo observation has  low S/N in the dynamic spectrum. Hence, it is surprising to find a well-delineated patch of power at the origin that extends outward with parabolic wings. A parabola is found on both negative and positive \fD\  by the weighted summation algorithm.
\etacreda{1470}{1}
%
\subsubsection{J2145--0750 (Figure Set \psrS)}
This millisecond pulsar ($P=16$~ms) is lightly scattered in these observations, which is not surprising given its proximity and high Galactic latitude ($D = 0.61$~kpc and $b=-42^\circ$). The S/N of the observations is not large.
At 825~MHz there is a faint but definite scintillation arc.
The situation is reminiscent of B2327--20 at 825~MHz.
In both cases, low S/N scintles are organized in such a fashion that a scintillation arc
emerges in the SS.
\etacredb{340}{825}{0, 1}
%
\subsection{Pulsars with No Scintillation Arc or with Data Unsuitable to Determine}
\subsubsection{B0138+59 (Figure Set \psrA)} 
A 1~hour data span is not long enough to detect arcs, if they are present.  
Data are consistent with centrally concentrated (CC) distribution in the SS given the time-frequency span of the observation. 
Substantially longer observations than one hour are needed in order to spread out the distribution from the origin along the \fD\  axis.
\etacredb{340}{825}{0, 0}
\subsubsection{B1821+05 (Figure Set \psrN)} 
 A single frequency Arecibo observation. The scintles are short and narrow in frequency, but there is adequate resolution in both dimensions.
 The SS is CC with weak power extending upward in what might be a bounding arc. 
\etacreda{432}{0}
%
\subsubsection{B1857--26 (Figure Set \psrO)}
Although only 700~pc away, the LoS to this pulsar  ($l=10.3^\circ$, $b=-13.5^\circ$) is heavily scattered.
The extremely narrow and brief scintles at 340~MHz make this DS and SS unusable.
At 825~MHz, the scintles are easily visible in the the relatively high S/N DS.
(There is  a defect in the spectrum near 818~MHz that has only been partially corrected  in cleaning the data.)
The parabola traced out on the SS, determined by the parabolic summing algorithm, is
unlikely to be reliable except as a rough guide for the curvature of a scintillation arc that
would need to be explored at higher frequency.
\etacredb{340}{825}{0, 0}
\section{Analysis of the Secondary Spectra} 
\label{sec:analysis}  

We report on detailed analysis of the data in this section.
The first two subsections present results of a more classical scintillation analysis, focused on the DS.
The remainder of the subsections concentrate on detailed analyses of the SS.
Throughout \S5 we refer to parameters extracted from the data set in a uniform manner and presented in Table~\ref{tab:analysis} for all 54 observations.

\subsection{Scintillation Parameter Estimation}
\label{sec:scintparam}

The scintillation decorrelation time  \tiss~  and bandwidth \nuiss~  were already defined in \S\ref{sec:arcparam}, as the half-widths in time and frequency of the autocorrelation 
function (ACF) $R(\nu,t)$ of the DS. 
Here we provide additional information about the way we construct $R(\nu,t)$.
For many pulsar observations in the SAS the contribution of noise to the ACF is unimportant.  
However, noise is significant in some of our data  and needs to be corrected.  
In order to correct the ACF for system noise we started from $S_2$ and estimated the mean noise level in the SS from a rectangular region outlined in green in the SS figures away from the ISS. (See \S\ref{sec:scintarccurve} for how we identify a rectangle in delay-Doppler (DD), where the ISS signal is evident, which we refer to as the DD box.)

The inverse Fourier transform of $S_2$, after subtracting the mean noise level from every pixel, yields the autocorrelation function of the ISS, $R(\nu,t)$, versus lags in frequency and time.  
The result, 
as plotted in the lower left subpanels of Figure~\ref{fig:main},
is free from a spike at zero lag due to additive white noise, which would be present when the autocorrelation is computed directly from the dynamic spectrum.  Hence we use the value $R(0, 0)$ 
to estimate the scintillation variance, corrected for noise, in defining the decorrelation widths.  
Its square root gives the rms needed for estimating modulation index.  

It also allows us to define a signal-to-noise ratio as the ratio of the scintillation variance to the noise variance, found by summing the noise level $S_{\rm noise}$ over delay and Doppler. We include this ratio of signal variance to noise variance in Table~\ref{tab:analysis}.
Note that the signal-to-noise ratio for $S_2$ itself is typically higher, since the ISS only spans part of the observed domain, while the noise is uniform out to the Nyquist point in delay and Doppler.  
Consequently we also tabulate a signal-to-noise for the ISS, defined as the ratio of the variance in ISS summed over the DD box, divided by the variance of the noise, summed over the same DD box, which approximates a matched filter for the ISS.
We tested the S/N estimation process
against simulated data with known signal-to-noise ratio and found it reported accurate values within the statistical uncertainties.





\begin{longrotatetable}  
\begin{deluxetable}{ccccccccccccccccc}


\tabletypesize{\scriptsize}


\tablecaption{Analysis of Dynamic and Secondary Spectra}

\tablenum{3}

\tablehead{\colhead{   \#   } & \colhead{PSR} & \colhead{Freq.} & \colhead{MJD} & \colhead{\nuiss} & \colhead{$\pm \nuiss$} & \colhead{\tiss} & \colhead{$\pm \tiss$} & \colhead{$\eta_{\rm p,l}$} & \colhead{$\eta_{\rm p}$} & \colhead{$\eta_{\rm p,u}$} & \colhead{$\eta_{\rm edge}$} & \colhead{S/N} & \colhead{S/N$_{\rm iss}$} & \colhead{$\eta_{\rm cred}$} & \colhead{$s_{\rm max}$} & \colhead{Arc Power} \\ 
\colhead{} & \colhead{} & \colhead{(MHz)} & \colhead{} & \colhead{(MHz)} & \colhead{(MHz)} & \colhead{(s)} & \colhead{(s)} & 
\colhead{(s3)} & \colhead{(s3)} & 
\colhead{(s3)} & \colhead{(s3)} & 
\colhead{} & \colhead{} & \colhead{} & \colhead{} & 
\colhead{Asym., $\kappa$} } 

\colnumbers
\startdata
1 & B0138+59 & 340 & 53630 & 0.0253 & 0.00681 & 495 & 131 & \nodata & \nodata & \nodata & 8.27 & 0.325 & 17.6 & 0 & \nodata & 0.319 \\
2 & B0138+59 & 825 & 53637 & 0.98 & 0.737 & 847 & 637 & \nodata & \nodata & \nodata & 1 & 0.61 & 49.2 & 0 & \nodata & --0.186 \\
3 & B0450+55 & 340 & 53632 & 0.129 & 0.0383 & 123 & 36.7 & \nodata & \nodata & \nodata & 0.112 & 1.69 & 65 & 0 & 0.514 & --0.662 \\
4 & B0450+55 & 825 & 53637 & 2.85 & 2.04 & 257 & 183 & 0.214 & 0.259 & 0.296 & 0.0204 & 3.26 & 37.5 & 1 & 0.766 & --0.207 \\
5 & B0450+55 & 1400 & 58920 & 9.6 & 7.24 & 367 & 277 & 0.0489 & 0.0533 & 0.061 & 0.0384 & 10.8 & 3930 & 1 & 0.66 & 0.566 \\
6 & B0450--18 & 340 & 53632 & 0.00756 & 0.000536 & 70.9 & 4.27 & 2.77 & 3.55 & 5.41 & 1.79 & 4.99 & 30.8 & 2 & 0.136 & 0.0157 \\
7 & B0450--18 & 825 & 53637 & 0.151 & 0.0178 & 121 & 13.9 & 0.703 & 0.797 & 0.904 & 0.624 & 8.13 & 71.5 & 2 & 0.173 & --0.127 \\
8 & B0450--18 & 1400 & 58913 & 9.7 & 11.5 & 1140 & 1350 & 1.53 & 1.79 & 2.15 & 1.11 & 57.9 & 8830 & 2 & 0.575 & --0.34 \\
9 & B0523+11 & 422 & 58132 & 0.000588 & $7.56\times10^{-6}$ & \nodata & \nodata & \nodata & \nodata & \nodata & 0.264 & 0.071 & 1.06 & 0 & 0.697 & --0.286 \\
10 & B0523+11 & 1450 & 58131 & 0.0697 & 0.002 & 39.7 & 1.03 & 0.101 & 0.123 & 0.148 & 0.0302 & 0.385 & 2.54 & 1 & 0.695 & --0.468 \\
11 & B0525+21 & 1390 & 58131 & 0.105 & 0.00457 & 20.9 & 0.999 & 0.0414 & 0.0468 & 0.0533 & 0.0327 & 1.58 & 9.28 & 1 & 0.198 & 0.451 \\
12 & B0540+23 & 432 & 58130 & 0.0032 & $6.61\times 10^{-5}$& 13.6 & 0.278 & \nodata & \nodata & \nodata & 0.0945 & 0.563 & 2.94 & 0 & 0.202 & --0.336 \\
13 & B0540+23 & 1450 & 58130 & 0.234 & 0.0129 & 47.7 & 2.65 & 0.0349 & 0.0415 & 0.0493 & 0.0215 & 14.5 & 74.6 & 1 & 0.255 & --0.857 \\
14 & B0626+24 & 432 & 58132 & 0.00365 & 0.000144 & 81.9 & 2.71 & 7.83 & 10.2 & 14.4 & 3.36 & 1.39 & 11 & 1 & 0.506 & 0.507 \\
15 & B0626+24 & 1390 & 58133 & 0.288 & 0.0676 & 200 & 47 & 0.905 & 1.29 & 1.69 & 0.568 & 0.275 & 27.9 & 1 & 0.573 & 0.363 \\
16 & B0628--28 & 340 & 53632 & 0.145 & 0.0621 & 228 & 98.1 & 1.41 & 2.16 & 3.29 & 0.467 & 7.02 & 74.7 & 1 & 0.509 & --0.0623 \\
17 & B0628--28 & 825 & 53637 & 1.68 & 1.36 & 559 & 452 & 0.442 & 0.702 & 0.899 & 0.113 & 19 & 185 & 1 & 0.665 & 0.086 \\
18 & B0628--28 & 1400 & 58904 & 12.9 & 23.3 & 1150 & 2090 & 0.132 & 0.156 & 0.175 & 0.0504 & 30.8 & 6040 & 1 & 0.559 & --0.145 \\
19 & B0809+74 & 340 & 53630 & 0.413 & 0.532 & 691 & 889 & 1.34 & 1.45 & 1.64 & 0.0792 & 1.52 & 65.5 & 1 & 0.478 & --0.00856 \\
20 & B0809+74 & 833 & 53637 & 1.4 & 1.16 & 294 & 242 & \nodata & \nodata & \nodata & 0.0136 & 0.0765 & 6.58 & 0 & 0.501 & --0.257 \\
21 & B0809+74 & 1400 & 58877 & 9.28 & 11.7 & 762 & 961 & \nodata & \nodata & \nodata & 0.0244 & 0.417 & 856 & 0 & 0.523 & 0.137 \\
22 & B0818--13 & 340 & 53632 & \nodata & \nodata & 78.2 & 3.47 & \nodata & \nodata & \nodata & 2.67 & 2.86 & 22.1 & 0 & 0.937 & --0.0529 \\
23 & B0818--13 & 825 & 53637 & 0.0775 & 0.0101 & 224 & 25.9 & \nodata & \nodata & \nodata & 2.53 & 33.1 & 435 & 0 & 0.984 & --0.131 \\
24 & B0818--13 & 1400 & 58920 & 2.26 & 0.85 & 333 & 125 & 0.197 & 0.287 & 0.451 & 0.104 & 16.8 & 2060 & 1 & 0.915 & 0.151 \\
25 & B1508+55 & 340 & 53632 & 0.0193 & 0.00154 & 51 & 4.03 & 1.07 & 1.45 & 1.77 & 0.794 & 26.2 & 101 & 2 & 0.943 & --0.359 \\
26 & B1508+55 & 825 & 53637 & 0.449 & 0.1 & 92.2 & 20.7 & 0.185 & 0.207 & 0.229 & 0.145 & 3.3 & 13.9 & 2 & 0.932 & --0.84 \\
27 & B1508+55 & 1400 & 58933 & 8.77 & 4.99 & 234 & 133 & 0.0504 & 0.0579 & 0.0662 & 0.0353 & 7.79 & 736 & 2 & 0.918 & --0.139 \\
28 & B1540--06 & 340 & 53630 & 0.0199 & 0.00181 & 68.5 & 6.13 & 1.66 & 1.95 & 2.33 & 0.981 & 2.38 & 12 & 2 & 0.506 & --0.146 \\
29 & B1540--06 & 825 & 53634 & 0.35 & 0.0598 & 118 & 20.1 & 0.284 & 0.319 & 0.363 & 0.231 & 9.19 & 60.8 & 2 & 0.496 & --0.149 \\
30 & B1540--06 & 1400 & 58964 & 10.9 & 14.7 & 1010 & 1350 & \nodata & \nodata & \nodata & 0.239 & 0.92 & 1150 & 0 & 0.915 & 0.203 \\
31 & B1706--16 & 340 & 53630 & 0.0664 & 0.0235 & 342 & 121 & \nodata & \nodata & \nodata & 1.32 & 2.03 & 105 & 0 & 0.943 & 0.0385 \\
32 & B1706--16 & 825 & 53634 & 1.68 & 1.32 & 536 & 420 & 1.54 & 1.76 & 3.02 & 0.762 & 18.9 & 626 & 1 & 0.882 & --0.403 \\
33 & B1706--16 & 1400 & 58965 & 5.28 & 4.7 & 934 & 830 & \nodata & \nodata & \nodata & 0.46 & 14.6 & 19000 & 0 & 0.894 & 0.077 \\
34 & B1821+05 & 432 & 58132 & 0.0298 & 0.00288 & 103 & 9.99 & \nodata & \nodata & \nodata & 0.888 & 0.4 & 22 & 0 & 0.123 & 0.0697 \\
35 & B1857--26 & 340 & 53631 & 0.00152 & $9.58\times 10^{-5}$& 7.82 & 0.239 & \nodata & \nodata & \nodata & 0.0258 & 0.129 & 0.664 & 0 & 0.0721 & --0.0731 \\
36 & B1857--26 & 825 & 53635 & 0.0134 & 0.000873 & 12.5 & 0.341 & \nodata & \nodata & \nodata & 0.00522 & 2.52 & 12.9 & 0 & 0.167 & --0.235 \\
37 & B1907+03 & 1470 & 58131 & 0.046 & 0.00176 & 27.2 & 1.08 & 0.121 & 0.171 & 0.193 & 0.0394 & 0.0938 & 2.69 & 1 & \nodata & 0.124 \\
38 & B2021+51 & 340 & 53632 & 0.154 & 0.072 & 251 & 117 & \nodata & \nodata & \nodata & 1.03 & 3.29 & 80.2 & 0 & 0.353 & --0.00716 \\
39 & B2021+51 & 825 & 53637 & 1.32 & 0.762 & 363 & 210 & 0.825 & 1.19 & 1.45 & 0.21 & 36.8 & 1040 & 1 & 0.537 & --0.134 \\
40 & B2021+51 & 1400 & 58922 & 11.5 & 17.3 & 1230 & 1850 & 0.116 & 0.13 & 0.141 & 0.0941 & 69.1 & 8510 & 2 & 0.267 & --0.0643 \\
41 & B2045--16 & 340 & 53631 & 0.306 & 0.127 & 101 & 42.1 & 0.109 & 0.122 & 0.136 & 0.0519 & 6.75 & 96.5 & 1 & 0.461 & --0.0557 \\
42 & B2045--16 & 825 & 53635 & 4.05 & 3.12 & 215 & 166 & 0.0175 & 0.0192 & 0.0208 & 0.0145 & 3.72 & 66.3 & 2 & 0.442 & --0.292 \\
43 & B2045--16 & 1400 & 58922 & 10.8 & 8.1 & 312 & 234 & 0.00684 & 0.00752 & 0.00844 & 0.00558 & 13.4 & 2320 & 2 & 0.473 & 0.115 \\
44 & J2145--0750 & 340 & 53631 & 0.218 & 0.173 & 523 & 416 & \nodata & \nodata & \nodata & 1.51 & 0.0553 & 19.7 & 0 & 0.378 & --0.234 \\
45 & J2145--0750 & 825 & 53635 & 1.6 & 1.08 & 407 & 274 & 1.13 & 1.32 & 1.46 & 0.0539 & 0.0196 & 8.59 & 1 & 0.309 & --0.262 \\
46 & B2217+47 & 340 & 53632 & 0.0104 & 0.000859 & 84.8 & 6.39 & \nodata & \nodata & \nodata & 1.45 & 38.3 & 279 & 0 & 0.854 & --0.13 \\
47 & B2217+47 & 825 & 53637 & 0.357 & 0.0878 & 233 & 56.9 & 1.04 & 1.24 & 1.24 & 0.168 & 7.9 & 90.3 & 0 & 0.913 & --0.229 \\
48 & B2217+47 & 1400 & 58920 & 3.61 & 1.42 & 266 & 105 & \nodata & \nodata & \nodata & 0.078 & 3.09 & 531 & 1 & 0.834 & 0.404 \\
49 & B2310+42 & 340 & 53632 & 0.0219 & 0.00245 & 92.6 & 10.1 & 1.78 & 2.19 & 3.01 & 1.2 & 4.04 & 35.9 & 1 & 0.454 & 0.232 \\
50 & B2310+42 & 825 & 53637 & 0.941 & 0.447 & 337 & 160 & 0.204 & 0.274 & 0.434 & 0.151 & 10.9 & 79 & 1 & 0.379 & 0.433 \\
51 & B2310+42 & 1400 & 58874 & 4.66 & 5.02 & 1050 & 1140 & 0.303 & 0.4 & 0.485 & 0.228 & 104 & 10000 & 1 & 0.72 & 0.0898 \\
52 & B2327--20 & 340 & 53632 & 0.34 & 0.2 & 184 & 108 & 0.993 & 1.09 & 1.19 & 0.556 & 0.298 & 13.4 & 2 & 0.752 & --0.786 \\
53 & B2327--20 & 825 & 53637 & 1.75 & 0.805 & 177 & 81.5 & 0.18 & 0.186 & 0.194 & 0.0878 & 0.0481 & 4.78 & 2 & 0.753 & 0.116 \\
54 & B2327--20 & 1400 & 58877 & 7.93 & 7.02 & 431 & 382 & \nodata & \nodata & \nodata & 0.0551 & 0.707 & 1310 & 0 & 0.955 & 0.0452 \\
\enddata


\tablecomments{Three values are given for $\eta$ from the parabolic summing algorithm: a lower bound ($\eta_{\rm p,l}$), the most likely value ($\eta_{\rm p}$), and an upper bound ($\eta_{\rm p,u}$). $\eta_{\rm edge}$ is the point of maximum positive slope in the parabolic summing curve. S/N is, in the DS, the ratio of the signal variance to the noise variance. S/N$_{\rm iss}$ is a similar quantity, but with the signal variance calculated in the DD box (see text). $s_{\rm max}$ is the estimate of primary screen location (maximum distance from the pulsar; see Equation~(\ref{eq:smax})). The Arc Power Asymmetry is $\kappa \equiv$ (R -- L)/(R + L), where R is the peak parabolic summation on the RHS ($\fD > 0$) and L is the corresponding peak power on the LHS of the parabola. Columns (6) and (8) are finite scintle estimates of the uncertainty in \dnuiss and \dtiss, respectively. See text for details.}

\label{tab:analysis}
\end{deluxetable}
\end{longrotatetable} 

\subsection{Pulse Broadening Time}
\label{sec:tauscatt}

The pulse broadening time \tauiss\ is a useful measure of scattering along a LoS and an important parameter to know when planning a timing or scintillation observation.
Although many of the parameters in the ATNF {\sc psrcat} database \citep{mhth05} are extremely well-determined, others such as \tauissG, the pulse broadening time scaled to 1~GHz, are drawn from a wide range of disparate observational programs conducted over the last 50 years.
The heterogeneous nature of {\sc psrcat} \tauiss\ data is increased because values are typically determined by frequency domain techniques for relatively lightly scattered pulsars and time domain techniques for moderate to heavily scattered pulsars.
In this section we report 22 newly determined values of \tauissG\ and compare them with currently tabulated {\sc psrcat} values.

\begin{figure}[h]    
\gridline{\fig{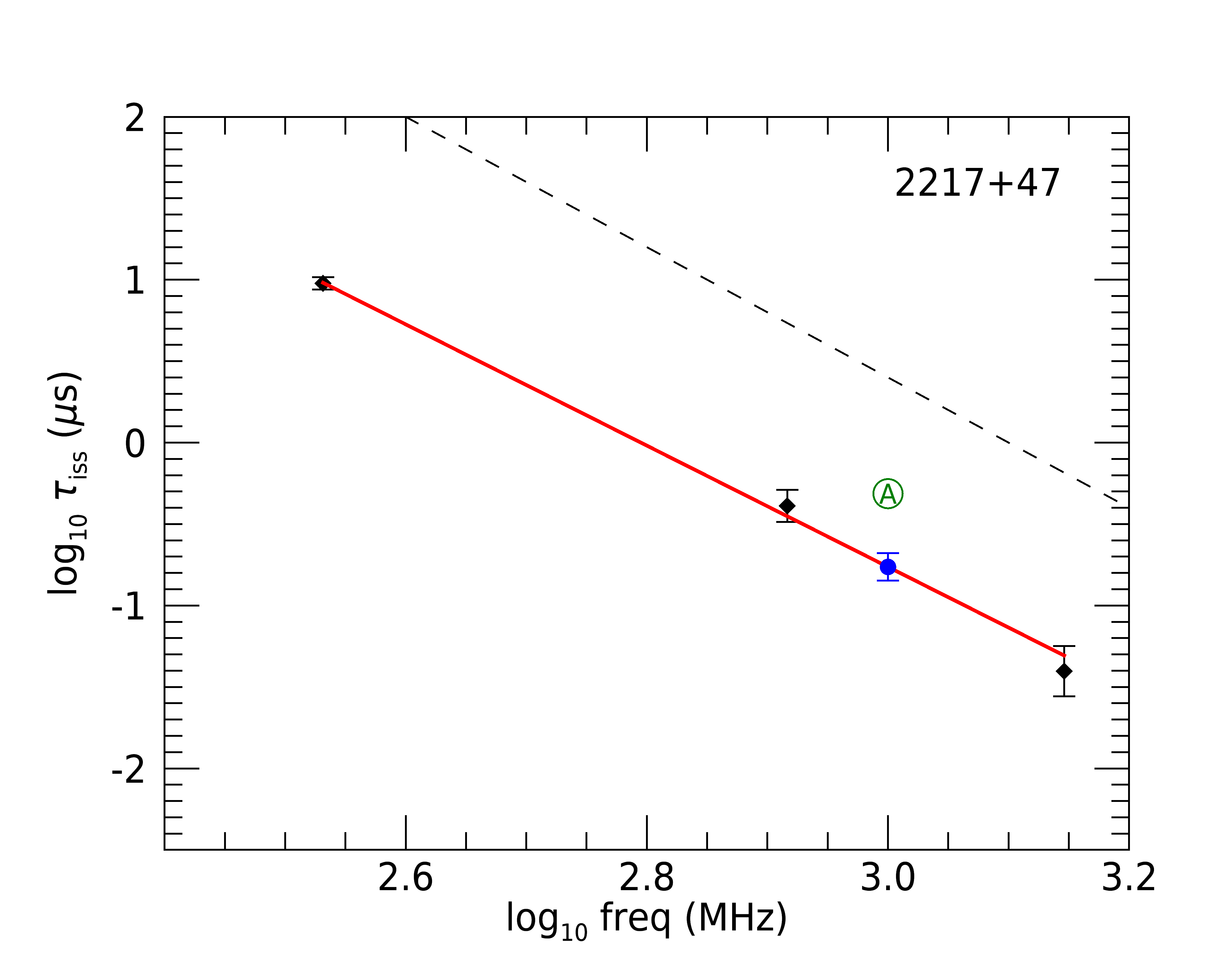}{0.47\textwidth}{(a)} 
             \fig{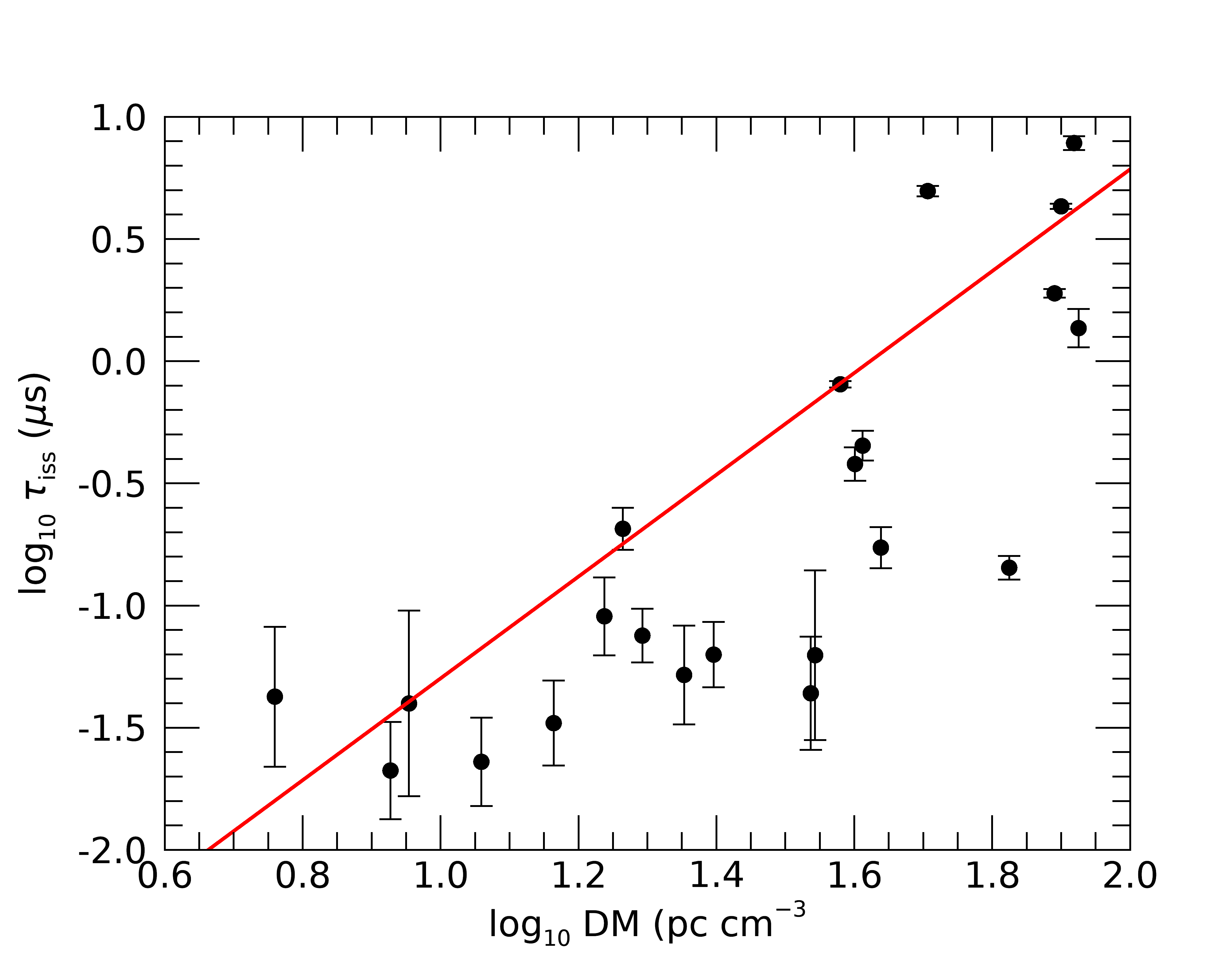}{0.47\textwidth}{(b)}
             }
 \caption{\footnotesize 
 {\em (a)} An example of how we calculated \tauissG and its uncertainty for each of the 22 pulsars in the survey.
A figure set is available online with a similar plot for each of the pulsars.
The black points are  estimates and 1-$\sigma$ uncertainties of $\tau_{\rm iss}$ at each of the available frequencies.
The red line represents the weighted linear least squares fit through these points. 
We report a value, marked by a filled blue circle, where that line crosses 1~GHz, and we assign an uncertainty to it from the least squares fit.
The {\sc psrcat} value is marked with a circled green A.
The dashed black line has a logarithmic slope of $-4$.
{\em (b)}
The $\tau_{\rm iss,1GHz}$ values from the SAS are plotted against the \DM\ values.
The red line is a weighted least squares fit. 
It has a logarithmic slope of 2.08, which is close to the $\tau \propto {\rm DM}^2$ behavior expected for scattering in a uniformly turbulent medium.
However, it is clear that the line is a very poor fit to the data if the individual errors are to be believed, and they are well-determined at the high DM part of the plot.
See the text for further comments.
 }
\label{fig:tausc1G}
\end{figure}
Figure~\ref{fig:tausc1G}(a) shows \tauiss\ values calculated from the observed \dnuiss\ values using $\tauiss = (2\pi\,\dnuiss )^{-1}$.  The error bars on the individual points are calculated from the formula for the number of independent scintles used by Cordes (1986). 
Since most of the uncertainties are not too large we use the approximation of a symmetric uncertainty in the log of the displayed value.
The best-fit line represents the weighted ($w_i = 1/\sigma_i^2$) least squares fit through these points in a log-log representation.
The interpolated or, in the case of some pulsars with only two observed frequencies, extrapolated values at 1~GHz are noted with a blue filled circle and a 1-$\sigma$ uncertainty from the linear least-squares fitting process.  This is the value and uncertainty reported in Table~\ref{tab:tauiss}.
In the case of  a single-frequency observation, $\tau_{\rm iss}$ was determined from the single \dnuiss\ value assuming  $\tau_{\rm iss} \propto \nu^{-4}$.



\begin{deluxetable}{l c c c c c c c}


\tabletypesize{\small}


\tablecaption{Scattering Delay and Related Quantities}

\tablenum{4}
\label{tab:tauiss}

\tablehead{\colhead{PSR} & 
\colhead{Nfreq} & 
\colhead{   \tauissG} & 
\colhead{   $\delta$\tauissG} & 
\colhead{    Slope} & 
\colhead{    $\delta$Slope} &
\colhead{$\tau_{\rm psrcat}$} &
\colhead{Ratio: } \\ 
\colhead{} & 
\colhead{} & 
\colhead{(ns)} & 
\colhead{(ns)} & 
\colhead{} & 
\colhead{} &
\colhead{(ns)} &
\colhead{$\tau_{\rm psrcat}/\tau_{\rm SAS}$}
} 

\colnumbers
\startdata
B0138+59 &2& \phn\phn60 &  \phn\phn70 & --4.08 & 0.80 & 607 & 9.70 \\
B0450+55 &3&  \phn\phn33 &  \phn\phn16 & --3.28 & 0.44 & 152 & 4.61 \\
B0450--18 &3& \phn380 &  \phn\phn70 & --2.91 & 0.18 & 835 & 2.20 \\
B0523+11 &2&4300& \phn110 &--3.33& 0.03 &3036&0.71\\
B0525+21 &1&5000& \phn250 & \nodata & \nodata &1518&0.31\\
B0540+23 &2&1900&  \phn\phn80 &--3.05& 0.05 &2277&1.20\\
B0626+24 &2&1400& \phn300 &--3.36& 0.22 &8349&6.12\\
B0628--28 &3&  \phn\phn40 &  \phn\phn30 &--3& 0.60 &13&0.30\\
B0809+74 &3&  \phn\phn40 &  \phn\phn40 &--2.09& 0.98 &10&0.24\\
B0818--13 &3& \phn450 &  \phn\phn70 &--3.12& 0.15 &650&1.44\\
B1508+55 &3&  \phn\phn80 &  \phn\phn20 &--3.9& 0.25 &56&0.74\\
B1540--06 &3& \phn210 & \phn\phn40 &--3.17& 0.21 &9&0.04\\
B1706--16 &3&  \phn\phn60 &  \phn\phn20 &--3.3& 0.37 &195&3.10\\
B1821+05 &1& \phn140 &  \phn\phn16 & \nodata & \nodata &5412&37.90\\
B1857--26 &1& \phn800 &  \phn\phn20 & \nodata & \nodata &\nodata&\nodata \\
B1907+03 &1&7800& \phn530 & \nodata & \nodata &\nodata&\nodata \\
B2021+51 &3&  \phn\phn50 &  \phn\phn30 &--2.82& 0.57 &197&3.79\\
B2045--16 &3&  \phn\phn22 &  \phn\phn11 &--2.85& 0.49 &5&0.22\\
J2145--0750 &2&  \phn\phn40 &  \phn\phn60 &--2.51& 1.09 &10&0.25\\
B2217+47 &3& \phn170 &  \phn\phn40 &--3.72& 0.20 &486&2.82\\
B2310+42 &3&  \phn\phn90 &  \phn\phn40 &--3.87& 0.36 &46&0.51\\
B2327--20 &3&  \phn\phn21 &  \phn\phn12 &--2.72& 0.64 &9&0.43\\
\enddata

\tablecomments{Column (2) is the number of frequencies available to estimate \tauscat.
Columns (3) and (4) give the value and uncertainty of \tauscat that we determine, referenced to $\nu = 1$~GHz using an assumed $\tauscat \propto \nu^{-4}$ relations. See text
for details. Columns (5) and (6) give information about the slope of the best fit line in the
equivalent of Figure~4(a) for each pulsar.
Column (7) gives the {\sc psrcat} value of \tauscat, also referenced to 1~GHz.
Column (8) gives the ratio between the SAS value and the {\sc psrcat} value.
}


\end{deluxetable}

 Figure~\ref{fig:tausc1G}(b) plots our determinations of \tauissG\ versus \DM. Although the best-fit line has a logarithmic slope close to the value of 2.0 expected for uniformly distributed scattering, we believe that this is coincidental as indicated by the poor match to points with small error bars for data with $\log_{10}$~DM $\ga 1.6$. See further comments below.

Plots such as this comprised the first observational evidence that showed how the strength of ISS increases with the interstellar column depth of electrons (Rickett, 1969).  
Many studies have shown that \tauscat\ typically increases more steeply than $\propto$~DM$^2$ \citep{sut71,bcc+04}, expected for uniformly distributed scattering, particularly for longer lines of sight through the Galaxy.
 Since pulsars are concentrated in the Galactic plane and toward the Galactic Center, the steeper \DM\  dependence is interpreted as increasing concentrations of turbulent plasma toward the inner Galaxy.  The survey observations extend only to about 3~kpc and so do not add new distance dependence. 

Inspection of individual plots in Figureset~4 or column 8 in Table~\ref{tab:tauiss} shows  substantial discrepancies between \tauissG\ from  {\sc psrcat} and those from the SAS.
The ratio of the two (column 8) is evenly split between ratio~$> 1$ and ratio~$< 1$ (10 instances of each; two comparisons missing).
To quantify the severity of the discrepancy, we took all the ratios less than 1 and found their reciprocals. Combining these with the ratios greater than 1, we found the median of the list to be 3.3, giving some indication of the difficulty of measuring this parameter.
It is well known \citep[\eg][]{grl94,rdb+06}
that \tauiss\ is time variable, sometimes by at least a factor of 3 in both directions, which no doubt accounts for some of the discrepancies in values, both between the SAS values and the  {\sc psrcat} values and probably within our own survey.

\subsection{Scintillation Arc Curvature}
\label{sec:scintarccurve}
The most fundamental parameter of a scintillation arc is its curvature, $\eta$.
In this section we explore many aspects, both theoretical and observational, of arc curvature as it occurs in the survey.

\subsubsection{Fundamental Relations}
\label{sec:theoryCurv}

Under the simple hypothesis of partial or fully 1-D scattering caused in a single screen located at some 
distance,
the predicted curvature depends on the distances and angles involved as follows \citep{crsc06}:
\bea
\eta  &=&   \frac{cD_{\rm eff}}{2\nu^2V_{\rm eff}^2} ,   \label{eq:eta} \\  
\mbox{where}  \;\;  D_{\rm eff} &=& D_{\rm psr}(1-s)/s \;\;      \\  
\Veff &=&  V_{\rm psr} \cos\psi ((1-s)/s + V_{\rm obs} - V_{\rm scr}
\eea
where the distance of the observer from the pulsar is $D_{\rm psr}$ and from the screen is $(1-s) D_{\rm psr}$;
$\psi$ is the angle between the effective velocity $\Veff$ and the long axis of the scattering. {The velocities of the observer $V_{\rm obs}$, the screen $V_{\rm scr}$ and the pulsar $V_{\rm psr}$ are summed as vectors. However, in what follows we assume that $\Veff$ is dominated by the motion of the pulsar.}
If the scattering is isotropic the same relations are obtained with $\cos\psi = 1$.
$V_{\rm eff}$ depends on the transverse velocity of the pulsar $V_{\rm psr}$ and also that of the observer and the screen, both of which we assume to be negligible relative to that of the pulsar (see \citealt{crsc06} for the full expressions).

\subsubsection{Methodology for Estimating Arc Curvature}
\label{sec:methodology}

As can be seen in the examples in Figures~1--3, $S_2$ typically peaks sharply near the origin, and arcs are only recognized at many decibels below the peak.  Hence we focus on rectangles in Delay-Doppler space away from the origin, selected visually where $S_2$ is significantly above the noise floor. The rectangles are defined by $f_{DA} > | \fD | > f_{DB}$,
$\tau_{A} > \tau > \tau_{B}$, which we refer to as the DD box.  The curvature is estimated in separate DD boxes for positive and negative \fD, with box coordinates listed in the online version of Table~\ref{tab:analysis}.

Elaborating on the discussion in \S\ref{sec:basicarc},
we use two methods to estimate $\eta$. 
In the first, we examine cross-cuts through $S_2$ over a range of fixed delays.  We tabulate the location in $\fD$ of the maximum in each cross-cut separately for both positive and negative $\fD$.  We then fit a parabola to the resulting set of peak locations in ($\fD, \tau$). The fitted curvature is the estimate $\eta_c$.  

In the second method we sum $S_2$ along each parabola from a search range in $\eta_p$.  Examples of the search, as parabola-summation versus curvature $\eta_p$ are plotted in the lower right hand sub-panels of Figures 1-3.
For each $\eta_p$ we compute the sum of $S_2(\fD,\tau) - S_{\rm noise}$ at each delay, interpolated in $\fD$ on each parabola.   The search range in $\eta_p$ is centered on the value $\eta_A = \tau_{A}/f_{DA}^2$ defined by the parabola that passes through the outer corner of the DD box, with 50 equal steps in log$(\eta)$ between $0.1 \eta_A$ and $10 \eta_A$.   
$S_2$ is summed in linear power over all delays between $\tau_B$ and $\tau_A$ and covers $\fD$ out to the Nyquist frequency, but excluding $\fD=0$.  
Separate summations over positive and negative $\fD$ are plotted in red and blue, respectively.   This method is similar to a Hough transform \citep{bot+16}.

The solid lines in Figures~1--3 plot the direct summations, while the lines with ``x'' markers are summations of $S_2$ weighted by $|\fD|$. Such a weighting is motivated by the theoretical relationship between $S_2$ and a one-dimensional model for scattered brightness.  In this model (\eg Stinebring, Rickett, \& Ocker 2019), \nocite{sro19} 
the $S_2$ contribution from interference between each pair of brightness components is divided by $|\fD|$, which is thus compensated by the weighting.   
Note that our weighting is the same as the Jacobian of the transformation to ``normalized Doppler profiles'' as used in curvature estimation by \citet{rcb+20}, which does not assume one dimensional scattering.  
It differs from the theta-theta mapping method of \citet{bbv+22}, which is based on one-dimensional scattering.

As a consequence of the weighting, however, an obvious broad peak in the weighted parabola summation, such as in B77+47 at 340 MHz (Figure~\ref{fig:SS2217}), does not necessarily correspond to visible parabolic arc structure in $S_2$.   For the same pulsar at 825 MHz, the summation curve only reaches its peak at the maximum curvature searched, and so only a lower limit on $\eta$ is given.  However, the curve does exhibit a sharp rise beyond which it flattens somewhat.   This behavior is characteristic of a parabolic boundary in $S_2$ outside of which $S_2$ drops off sharply.   In the survey there are several examples of this behavior, which is expected in the presence of a core of lightly scattered waves that interferes with a broadened distribution.

\begin{figure}
\plottwo{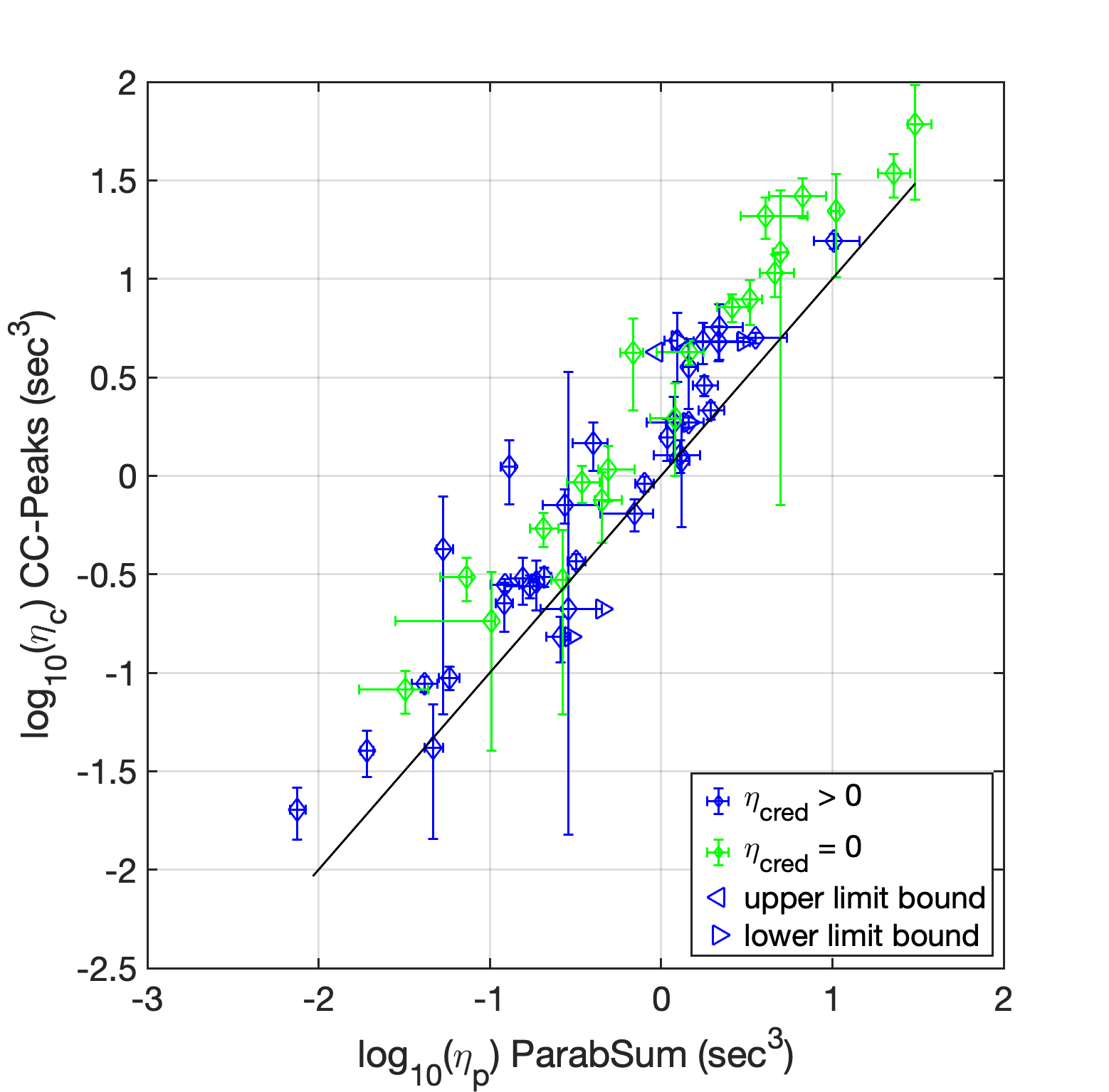}{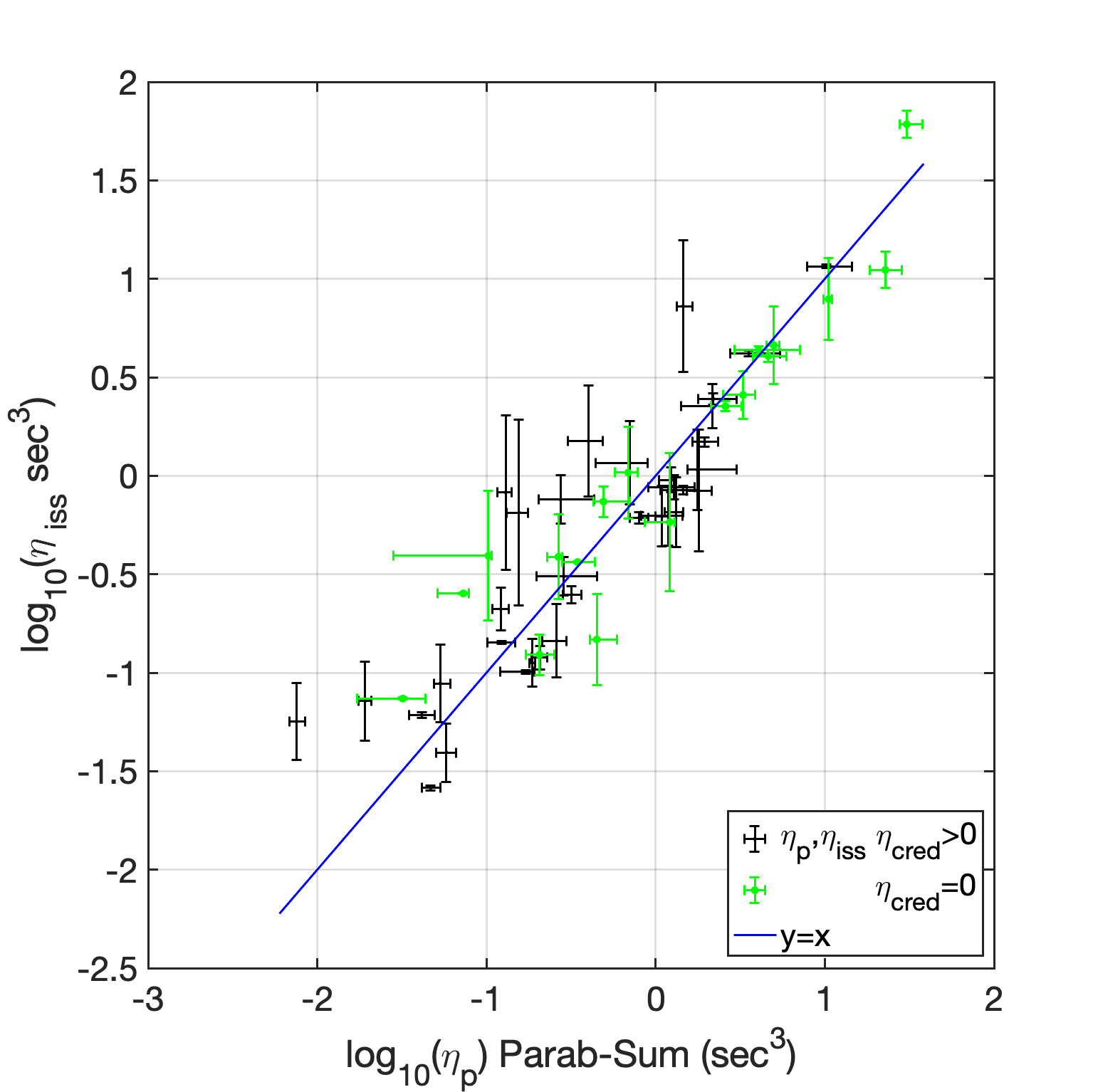}
\caption{Curvature.  \it Left: \rm Comparison of the two methods (maxima in cross cuts and parabolic summation) for estimating curvature described in the text, averaged from positive and negative Doppler frequencies. As discussed in \S\ref{sec:basicarc} we have created a subjective credibility index (0,1,2) for the $\eta_p$ estimate and plot green error bars to indicate low credibility index.
\it Right: \rm The pseudo-curvature $\eta_{iss}$, which is defined below in equation \ref{eq:etaiss}, plotted versus the observed  $\eta_p$ for data with credibility index greater than zero and valid results for the corrected $\dnuiss$ and $\dtiss$. 
}  
\label{fig:etac_etap} 
\end{figure}

We now compare the two methods of estimating curvature.
The estimates of curvature from  positive and negative Doppler are averaged for each method giving an overall $\eta_c$ and $\eta_p$ for each observation. These are included in Table~\ref{tab:analysis} and compared in the left panel of Figure~\ref{fig:etac_etap}.   There is a satisfactory agreement between the two methods. 
However, since the cross-cut method relies on finding the single highest peak at each delay, it can have quite large errors, and in what follows we focus on $\eta_p$ as our curvature estimator.
Note that the weighted $\eta_{p}$ can give an apparently reliable measure of curvature, even in the absence of a {\em{visible}} parabolic arc in $S_2$. 
As an example, the secondary spectrum in the left panel of Figure~\ref{fig:SS2217} exhibits no arc-like features, but there is a broad peak in the weighted parabola summation defining a specific curvature that is not seen in the unweighted summation. 

We include an estimate of the error in curvature,  calculated from the upper $\eta_{p,u}$ and lower $\eta_{p,l}$ range for which the parabola summation is above 95\% of its peak. 
This is illustrated in the lower right panels of Figure~\ref{fig:SS2217}.  Note however, that it is not a formal error estimate as we do not have a statistical model for the systematic variations in $S_2$.  
We arbitrarily choose 95\% reduction since the arcs represent only a small fraction of the total power in $S_2$ (equal to the variance in the DS), and so also only a small fraction of the parabolic sum.  Another estimate could be made at say 50\%, which would include a much wider range especially with double peaks and curves that saturate at the search limit.
The limits are included with each estimated $\eta_p$ in Table~\ref{tab:analysis}. 
These limits are also useful in characterizing the width of the arc as elaborated on in \S\ref{sec:arcwidth}. 

\subsubsection{How Curvature Is Related to the Basic ISS Parameters}
\label{sec:eta_basicISS}

Under the same single screen anisotropic
scattering assumptions made in \S\ref{sec:theoryCurv}, 
we examine the relationship to be expected between the curvature and the basic ISS parameters.
The characteristic time and frequency scales are related to the 
characteristic angular width on the long axis of the scattered brightness at the observer ($\theta_d$)
as follows:
\bea
\nuiss     &=&   c/(\pi D_{\rm eff} \theta_d^2) \\
\tiss &=&  c/(2 \pi \nu \theta_d  \; \cos\psi  V_{\rm eff} )    \label{eq:arceq}
\eea
We eliminate the dependence on $\theta_d$ in the following combination and obtain a quantity 
proportional to the curvature in equation \ref{eq:eta}.   
\be
\eta_{iss}  = 2\pi \tiss^2 /\nuiss =  c D_{\rm eff} / (2 \nu^2  V_{\rm eff}^2)  = \eta
\label{eq:etaiss}
\ee

In the right panel of Figure~\ref{fig:etac_etap}  we plot $\eta_{\rm iss}$ as defined in equation \ref{eq:etaiss}, against the curvature estimated from the parabola summation $\eta_p$. 
The points follow this relation and so confirm our basic assumptions.\footnote{It should be noted that our method of estimating $\eta_p$ involves visual selection of a rectangle in delay/Doppler, which might contribute to such a trend, because the search range for curvature is based on the apparent width of $S_2$ in delay and Doppler, which will be inverse-correlated with \nuiss and \tiss, respectively.}

Note that this result was already implied in the analysis of arcs by \citealp{crsc06}.  
They introduced scaled variables
$p$ and $q$ as:
\be
p=2 \pi \nuiss \tau  ; \;\; q=2 \pi \tiss \fD \; , 
\label{eq:pqdef}
\ee
which are related by the basic arc equation $p = q^2$. In \S~\ref{sec:pqAnalysis} we explore what additional insights are obtained from presenting SS in normalized ($pq$) coordinates.

In the next section we examine how to use the curvature \etap\ or \etaiss\  to estimate screen distance.  However, we note that they both suffer from the same problem: the screen location $s$ and the angle $\psi$, which appear in $D_{\rm eff}$ and $V_{\rm eff}$, are not separable in a single observation.
While there are cases where the orbital motion of the Earth or of a pulsar in a binary system can be used to break this degeneracy 
\citep[e.g.][]{shr05,rcb+20, mzsc22}
we do not have a sequence of observations necessary to pursue this further.

\subsubsection{Single Screen Model -- Estimating Screen Distance}
\begin{figure}[t]
\begin{tabular}{ll}
\includegraphics[trim = 10 0 90 0,clip,angle=0,width=8cm]{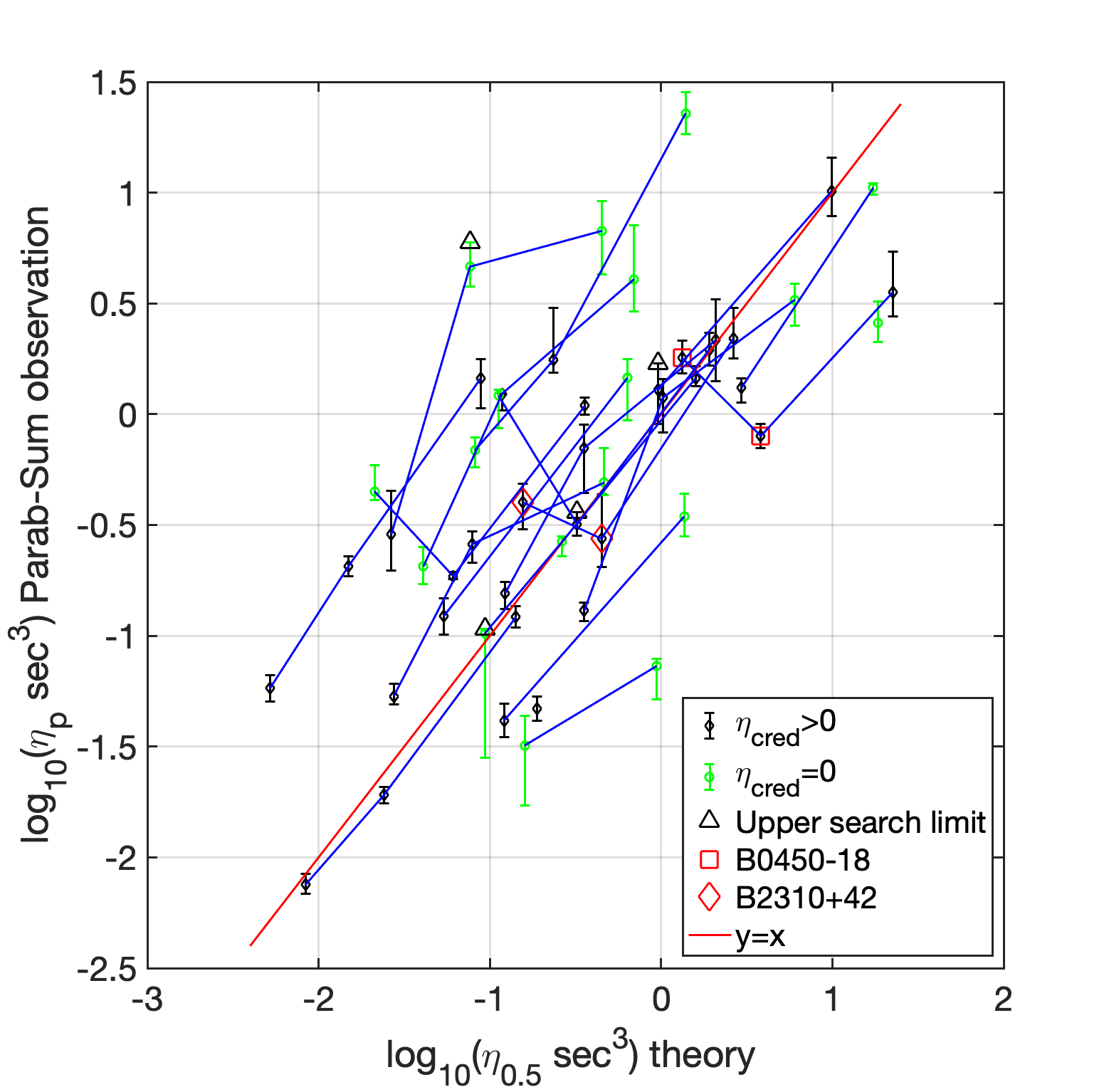} &
\includegraphics[trim = 10 0 90 0,clip,angle=0,width=8cm]{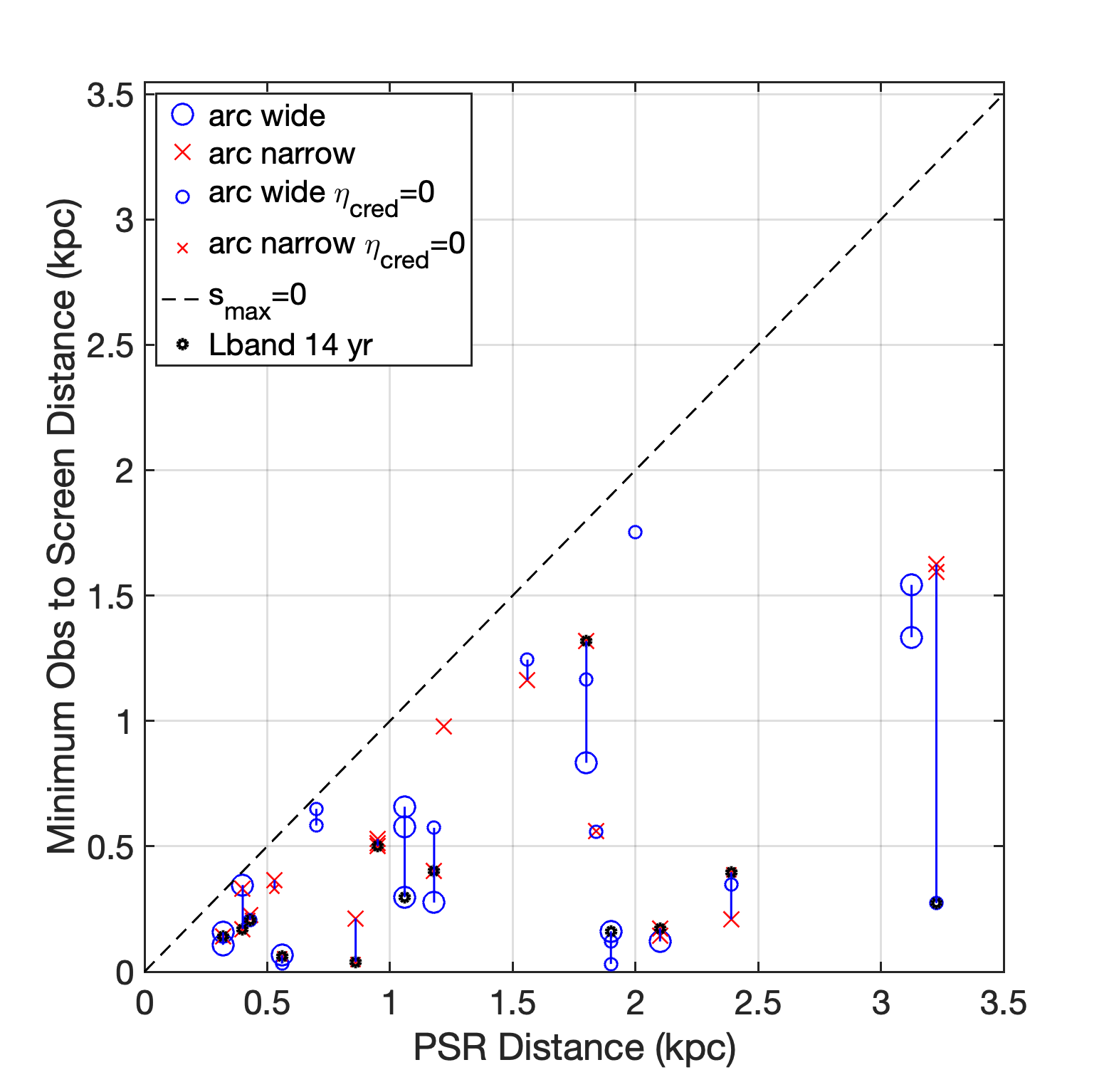}  \\
\end{tabular}
\caption{\it Left: \rm  Curvature $\eta_p$ versus single screen theory $\eta_{0.5}$;  points without error bars represent estimates with zero credibility index.
Points connected by a blue line are from the same pulsar at 2 or 3 frequencies. Two pulsars with discrepant frequency scaling between 1400 and 825 MHz are highlighted.
\it Right: \rm Single screen hypothesis:  Distance to the screen ($D_{psr}(1-s_{\rm max})$) versus distance to each pulsar, assuming isotropic scattering. If the scattering were anisotropic this becomes a minimum screen distance. Observations of the same pulsar are joined by a blue line, and marked by a black circle for 1400 MHz observations in 2020.   The small symbols mark curvature estimates classified as low credibility.   Narrow arcs are defined by $\Delta \log\eta < 0.2$ as estimated from the parabola summation plots. }  
\label{fig:screen}
\end{figure}

How consistent is the assumption of a single screen (or, more generally, a single ``dominant" screen) with the results of the survey? 
We start from the definition of the theoretical curvature for a mid-placed screen:
\be
\eta_{0.5} \equiv \frac{c\; D_{\rm psr}}{2\, \nu^2\, V_{\rm psr}^2 } =\,\, (0.462\,{\rm s}^3)\, 
\frac{D_{\rm psr,kpc}}{\nu_{\rm GHz}^2\, V_{\rm psr,100}^2}\\  \label{eq:eta_half},
\ee
where the pulsar velocity and distance and the observing frequency are expressed in convenient units ($ V_{\rm psr,100} = V_{\rm psr}/10^5~{\rm m\, s^{-1}}$).
We can then write equation \ref{eq:eta} as
\be
\eta  = \eta_{0.5}\; \frac{s}{\cos^2\psi\, (1-s)}   .  
\label{eq:eta_half2}
\ee
The theoretical quantity $\eta_{0.5}$ can be evaluated for each observation, using the published values for the pulsar distance and velocity, 
obtained from {\sc psrcat}
and listed in Table 1, for 20 of the 22 pulsars.  The left panel of Figure~\ref{fig:screen} shows the measured curvature (average of $\eta_{p}$ from positive and negative \fD) plotted against $\eta_{0.5}$.  In the plot a solid blue line joins observations of the same pulsar at 2 or 3 frequencies.  If the values were consistent with each other the line should have unit slope (parallel to the red line), corresponding to $\lambda^2$ scaling for $\eta$.   While many show reasonable agreement there are several discrepancies.   There are two pulsars in particular that stand out, B0450--18 and B2310+42, which are highlighted.   They are notable because $\eta$ is larger at 1400 than at 825 MHz, and so we examine them in detail in \S\ref{sec:Discrepant}.
 
As can be seen in the figure and from equation \ref{eq:eta_half2}, the actual value of $\eta$ can be above or below $\eta_{0.5}$, depending on the values for $s$ and $\psi$.   However, since $\cos^2\psi \le 1$ we can constrain
\be
s \le s_{\rm max} = \eta/(\eta+\eta_{0.5}).   \label{eq:smax}
\ee  
(This is the same estimate for $s$ used in \citealt{ps06a}.)

For isotropic scattering in a single thin screen, the equality can yield a direct estimate of 
screen distance $D_{\rm scr} = D_{psr}(1-s_{\rm max})$.    In the right panel of Figure~\ref{fig:screen} we plot such estimates against the distance to each pulsar. 
Assuming an
isotropic single screen scattering model, the 1--3 frequencies observed would yield the same $D_{\rm scr}$; in many cases the points, joined by a vertical line, do form a cluster.  We might expect that low $DM$ pulsars can be better modeled by discrete screens causing single or multiple well-defined arcs.
Hence we flag the narrow arcs by a red cross, but they show only a weak preponderance for small $D_{psr}$.  Another consideration here is the range in observing dates and the substantial proper motion of the pulsars.  In particular, the Green Bank observations at 1400 MHz were 14 years later than the observations at lower frequencies, and are flagged separately. 

\subsubsection{Apparently Discrepant Frequency-Scaling of the Curvature}
\label{sec:Discrepant}

As noted above B0450--18 and B2310+42 both show discrepant frequency scaling in the estimated curvature $\eta_p$.    Figure \ref{fig:SS2310} displayed the observations for B2310+42 at three frequencies.  In each case the curvature estimation comes from the lower right panel.  The major discrepancy is that $\eta_p$ is a factor 2.5 higher at 1400 MHz than at 825 MHz, but it should be a factor 0.35 smaller.   While the observing dates at 340 and 825 MHz differ by only 5 days, the 1400 MHz observations were 14.3 years later.  Similarly for B0450--18 the 1400 MHz observation were 14.4 years later than those at the lower frequencies.  Pulsar B0450--18 moved a transverse distance of 79 AU and B2310+42 moved 376 AU in the 14 years.   Many previous arc observations have shown evidence for significant structure in the interstellar plasma on AU scales, implying that interstellar scattering is due to a very patchy distribution of plasma.  Thus we interpret the discrepancies in the frequency scaling as due to changes in the plasma columns over the 14 years.

These changes in $\eta_p$ imply localized plasma concentrations at differing distances (unless the scattering were highly anisotropic with a change in orientation to the pulsar velocity), which is also illustrated by the widely differing values of $s_{max}$ in the right hand panel. For B2310+42 the 825 MHz result shows a well defined boundary arc whose curvature is 0.2 sec$^3$; scaling this to 1400 MHz predicts 0.07 sec$^3$.  However, the value estimated is about 0.3 sec$^3$, but with substantial differences between positive and negative $\fD$.  The SS at 1400 MHz has a poorly defined boundary arc at positive $\fD$.  
In a close inspection of the parabolic summation curve one can see this as a sharp rise in the summation at $\log_{10} \eta_p \sim  -1.1$ ($\eta_p \sim 0.08$ sec$^3$), and so might be due to scattering at the same distance as the boundary arc at 825 MHz seen 14 years earlier.

Now consider the results for B0450--18 shown in Figure~\psrC\ (and already published by Rickett et al., 2021).  At 825 MHz the strong forward arc has $\eta_p \sim 0.7 \pm 0.3$  sec$^3$ and is modulated by prominent reverse arclets.  However, at 1400 MHz there is a narrow forward arc with curvature $\eta_p \sim 1.8$ sec$^3$, in stark disagreement with the expected scaling from 825 MHz $\eta_p = 0.24$ sec$^3$.   Note that in the right panel $s_{max} \sim 0.05$ estimated from the earlier observations of the pulsar at 340 and 825 MHz implies a screen near the pulsar.  Even though our earlier analysis found anisotropic scattering, a low value of $s$ still holds since $\cos^2\psi \le 1$.  Thus the later value of $s_{max} \sim 0.2$ must be due to a new scattering screen substantially farther from the pulsar.

\subsection{Arc Width}
\label{sec:arcwidth}

\begin{figure}
\plottwo{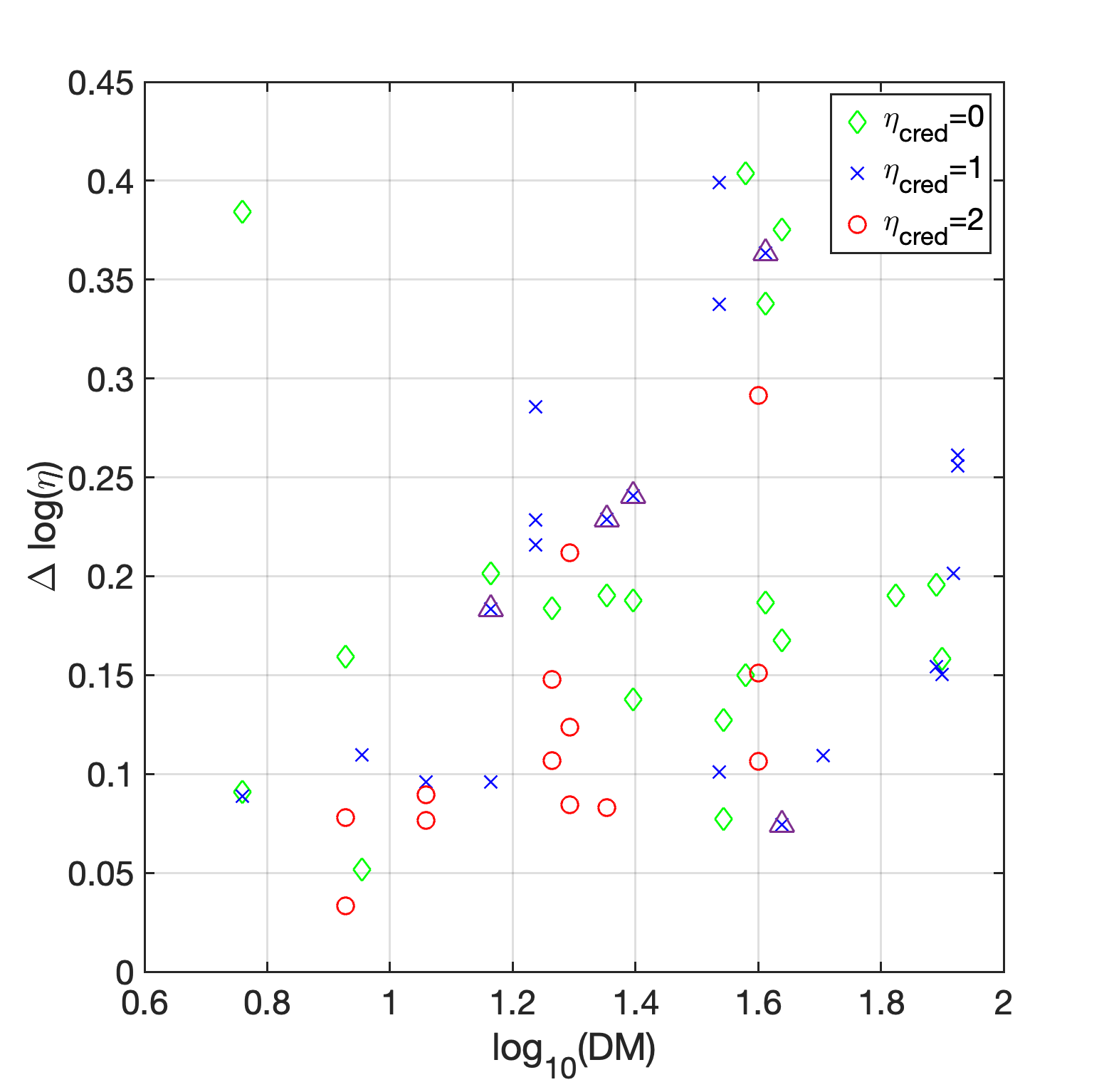}{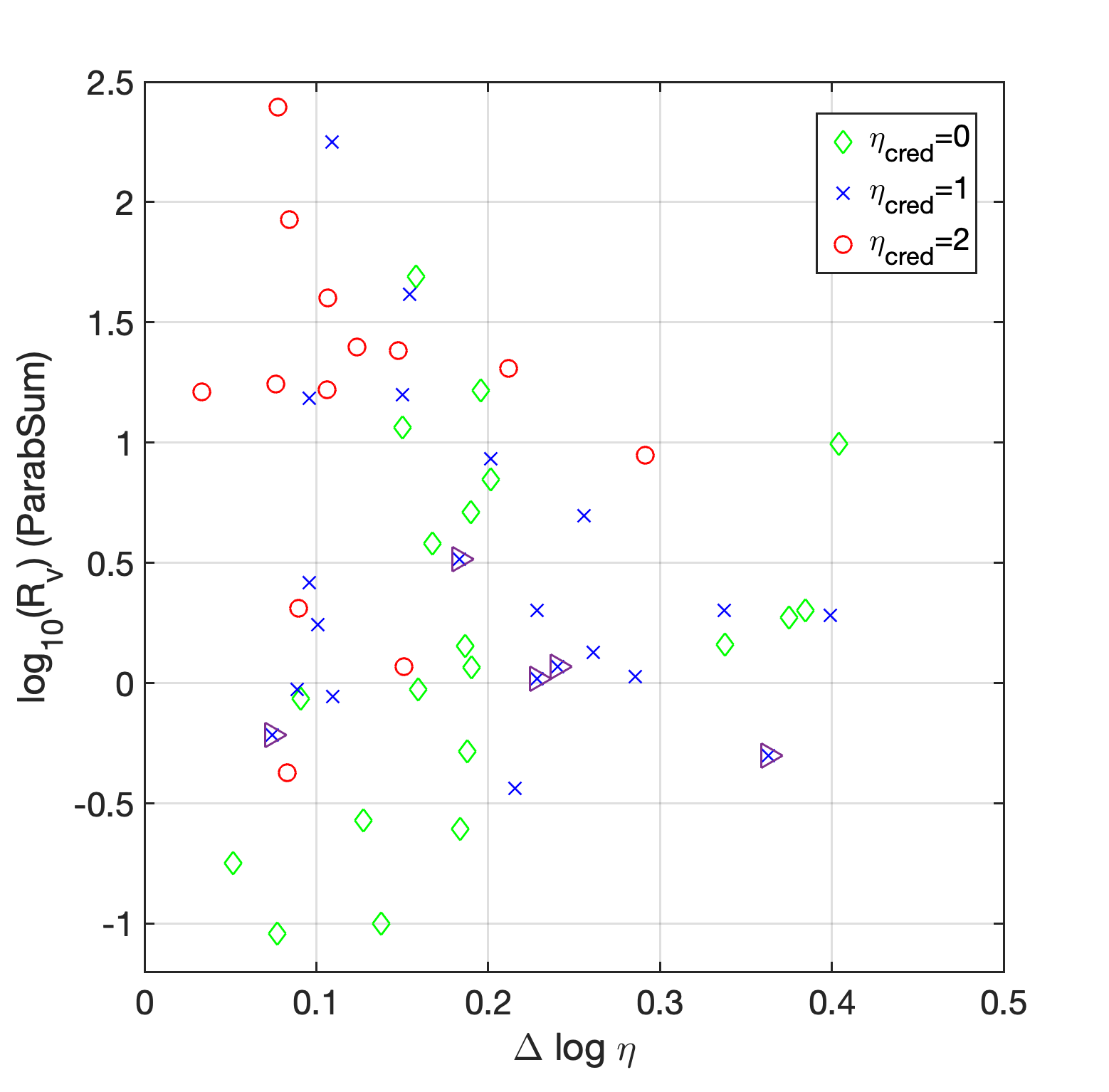}
\caption{\it Left: \rm  Arc width versus DM, flagged by curvature credibility index.
\it Right: \rm The depth of the valley in the SS, as a ratio $R_{\rm v}$, defined by parabola summation at the estimated curvature divided by a summation of SS near the delay axis.  In both left and right panels points are flagged by the curvature credibility index.
}
\label{fig:arcwidth} 
\end{figure}

We now characterize the relative prominence of arcs in the secondary spectrum.   In particular, we attempt to parameterize sharpness of an arc by its width and the depth of the valley along the delay axis.  
Using the analysis from \S\ref{sec:methodology}, we define a relative width of the arc:
\be
 \Delta \eta = \log_{10}[\eta_{p,u}/\eta_{p,l}].   \label{eq:Deleta}
 \ee
The left panel of Figure \ref{fig:arcwidth} plots $\Delta \eta$ against \DM\  and shows that arcs at low \DM\  are typically narrow, and at larger \DM\  the arcs usually widen, and $\Delta \eta$ covers a wide range.   

 We also use the parabola summation curves in an attempt to quantify the relative depth of any valley in SS near the delay axis.  We divide the peak in the summation curve by the summation of SS parallel to the delay axis 
(at $\fD=1$ resolution increment in \fD) over the same range in delay as used in the parabola summation; the result is a ratio $R_{\rm v}$ 
between typical SS amplitude along an arc and its value near the delay axis.  We avoid $\fD=0$ which is
influenced by the bandpass normalization.

In the right panel of Figure~\ref{fig:arcwidth} we investigate how $R_{\rm v}$ is related to the arc width $\Delta\eta$, defined by equation \ref{eq:Deleta}.   In the plot we flag the points by their credibility index.  It illustrates how the narrow arcs with deep valleys are often classified as $\eta_{\rm cred}=2$.   Note that in some cases $R_{\rm v} < 1$, which signifies a ridge along the delay axis rather than a valley, disrupting the curvature estimation.
The right panel of Figure~\ref{fig:arcwidth} shows that narrower arcs are associated with deeper valleys; thus the general increase in $\Delta \eta$ with \DM\  corresponds to a decrease in valley depth $R_v$ with \DM.   Figure~\ref{fig:arcwidth} provides  observational evidence that narrow arcs with deeper valleys are mostly seen at low dispersion measure.  Such a trend is expected since narrow arcs imply localized scattering from a thin region, and at larger distances (or DMs) it becomes more likely that the pulsar signal is scattered in multiple regions making arcs broader and less distinct.

\subsection{Theoretical Conditions for Arcs}
\label{sec:conditions}

In section \ref{sec:theoryCurv} we gave the theoretical relations for the curvature of parabolic arcs due to a single localized scattering screen.   Here we describe the form of the secondary spectrum for scattering by random irregularities in electron density, under some specific statistical assumptions, concerning their underlying spatial spectrum and their distribution along each LoS.  Consider, first, a thin region modeled as a phase-changing screen (phase screen) at a particular distance along the path from a pulsar.   Further, assume transverse variations in phase that follow the Kolmogorov spatial spectrum.  

The ISS observed in pulsars has narrow bandwidth, characterized by $\delta \nu_{iss}$ which is typically much less than the central frequency in the observations.  Thus the scintillations are strong in the sense that the rms variation of flux density is comparable to the mean flux density (see,\ \eg Rickett 1990).  Under strong scintillations there is negligible flux density from un-scattered waves.  However, the refractive index in the plasma varies as frequency$^{-2}$, and at frequencies above about 10 GHz, typically the ISS becomes weak (rms less than mean) and $\delta \nu_{iss}$ increases becoming comparable to the central frequency.    Such conditions give rise to a narrow forward arc in the SS caused by the interference of the unscattered wave with an angular spectrum of scattered waves. This forward parabola acts as an outer boundary, below which the SS is zero and above which $S_2$ is related by a simple expression to the angular spectrum in brightness.  At the other extreme, asymptotically strong scattering is due to the mutual interference between all possible pairs of scattered waves, as in the double integral equation (8) of \citet{crsc06}.

\begin{figure*}[h]
\begin{tabular}{lll}
\includegraphics[angle=0,width=5.3cm]{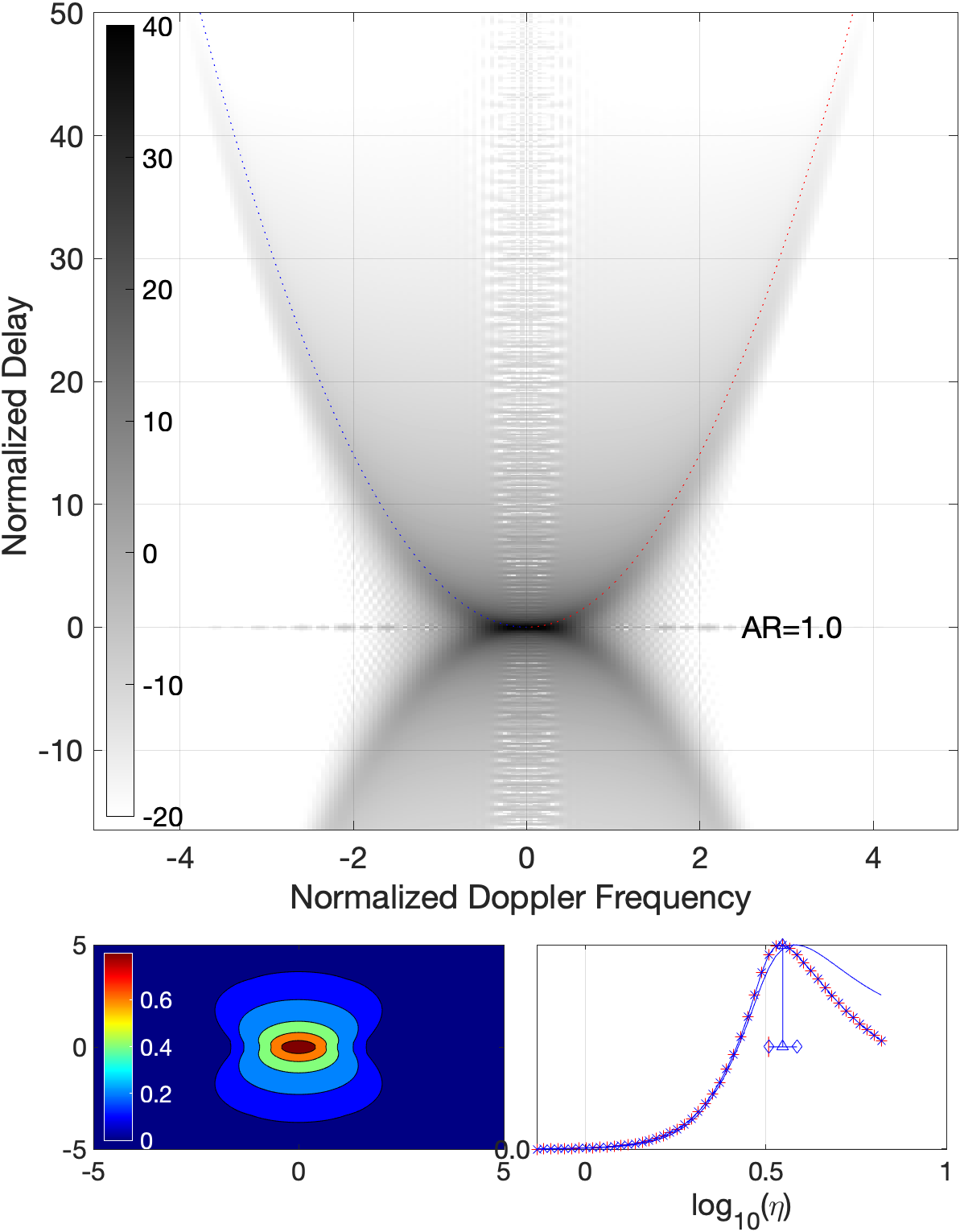}  &
\includegraphics[angle=0,width=5.3cm]{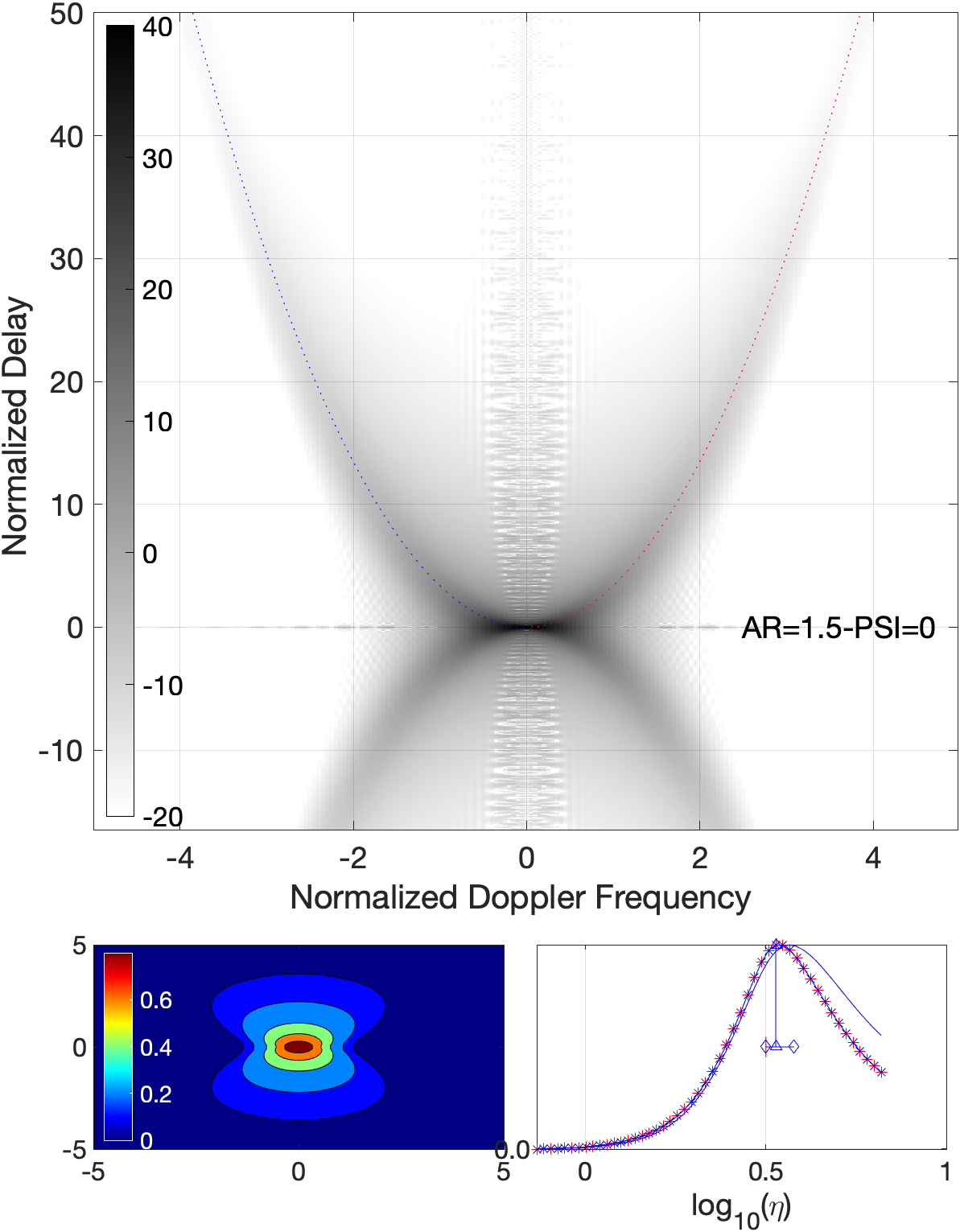} &
\includegraphics[angle=0,width=5.3cm]{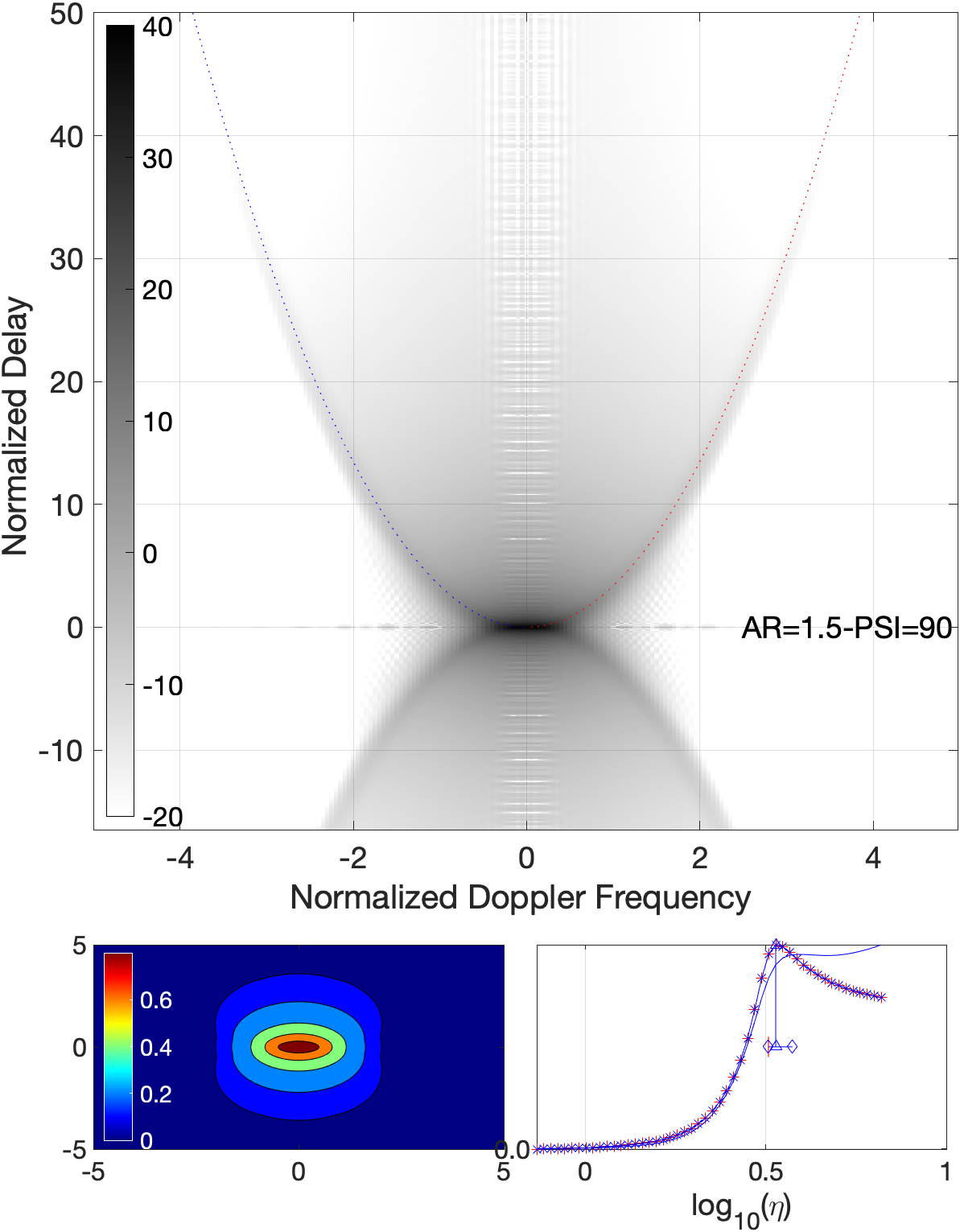}  \\
\end{tabular}
\caption{Theoretical secondary spectra for strong scattering in a screen with a Kolmogorov phase spectrum.  Format as for the observations.  \it Left: \rm Isotropic (axial ratio = 1).  \it Center: \rm Anisotropy axial ratio=1.5, velocity along major spatial axis ($\psi=0$)     \it  Right: \rm  axial ratio=1.5 velocity perpendicular to major spatial axis ($\psi=90$) .  The lower left panel is the autocorrelation function $R(\nu,t)$ versus normalized frequency lag (vertical) and normalized spatial lag (horizontal);  the lower right panel is an estimation of parabolic curvature from the secondary spectrum.}
\label{fig:KolmoSS} 
\end{figure*}

In Figure~\ref{fig:KolmoSS} we show the SS predicted in asymptotic strong scattering for a screen with a Kolmogorov phase spectrum.   The scattering is isotropic in the left panel;  it is slightly anisotropic in the center and right panels, with axial ratio $AR=1.5$ and orientation angles $\psi = 0, 90 \deg$, respectively.  The SS is calculated numerically by Fourier transforming the expressions for the frequency-time correlation function given by Lambert and Rickett, (1999).  (Note that the low level ripples in the SS near the delay axis are due to insufficient dynamic range in the computation.) See Figures~9 and 10 in \citet{rcb+20} for similar computations, which
also exhibit boundary arcs in the secondary spectra.

In all three panels of Figure~\ref{fig:KolmoSS} the parabola summation curves show significant peaks and  yield estimates for both the curvature and the arc width parameter $\Delta\eta$.   All three SS plots also exhibit a boundary arc, and demonstrate that boundary arcs do not require anisotropic scattering.
They can be seen even with modest anisotropy (axial ratio $AR=1.5$) and when the orientation angle $\psi=90^\circ$.
The boundary arc is caused by the interference of slightly scattered waves at very small angles with waves scattered at relatively large angles, similar to weak scintillation.   It is a property of the isotropic Kolmogorov spectrum that there is a bright compact core in the angular 
spectrum and also a tail of brightness at larger angles falling as angle$^{-11/3}$.  
The power law nature of this tail causes $S_2$ to decay slowly with delay, making the arc visible out to delays that are many times larger than the characteristic scatter-broadening time.

The center panel ($\psi=0,AR=1.5$) has the lowest interior SS levels (deepest valley) and the right hand panel ($\psi=90,AR=1.5$) has the highest interior SS levels. These differences can also seen from the parabola summation curves where the peak summation is greater than the summation at the maximum $\eta$, where the parabola lies close to the delay axis.

We do not have a full theory for the form of the SS when the scattering is distributed all along the path from the pulsar.   Under such conditions the tight quadratic connection between delay and Doppler frequency breaks down, which will certainly broaden any arcs and broaden any sharp boundary.  As a first approximation the SS can be considered as the superposition of the SS from multiple discrete screens, ignoring the effect of second (or higher) order scattering.  This approximation superimposes parabolic arcs of differing curvatures arising at differing distances and of differing velocities, and any anisotropy would likely be randomized in angle.   Thus the overall SS would exhibit few distinct parabolic arcs, but more likely become quite fuzzy and broadened curves of parabola summation with increased $\Delta\eta$.  Note, however, \citet{spmb19a}
describe a precise theory for SS scattered by two discrete screens. 

\subsection{Arc Width Versus Frequency}
\label{sec:WidthVsFreq}

As discussed throughout \S\ref{sec:comments}, a striking aspect of the observed SS is the systematic broadening of the arcs at the lower frequencies as in Figures~\ref{fig:SS0628} and \ref{fig:SS2310}, for example.  We interpret this as the widening of the scattered brightness distribution as the scintillations become stronger.  
Here we discuss how the width of the arcs changes for plasma scattering in a single screen with a Kolmogorov spectrum, for which we have a complete theory.

\begin{figure*}[t]
\begin{tabular}{ll}
\includegraphics[angle=0,width=7cm]{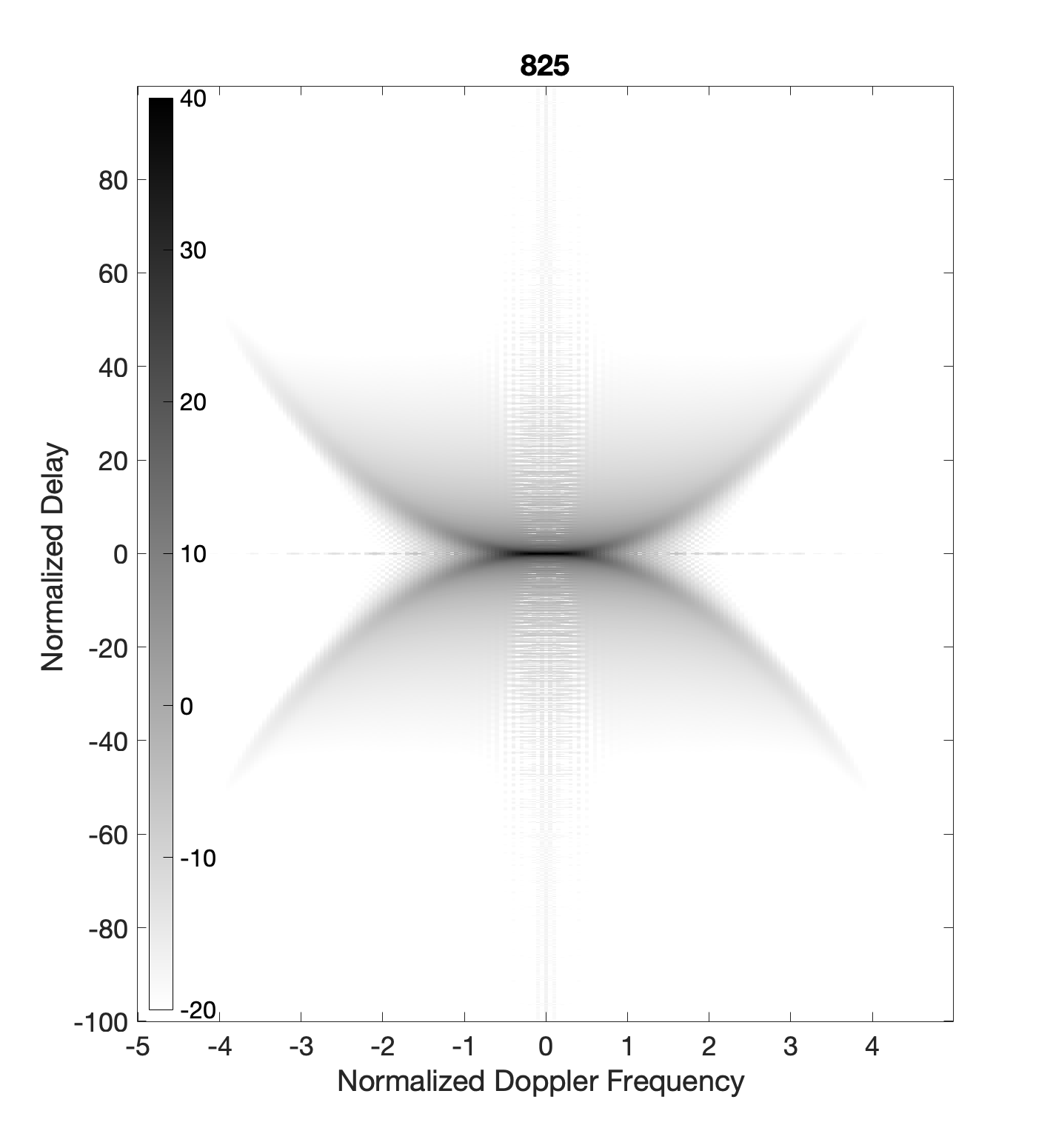}  &
\includegraphics[angle=0,width=7cm]{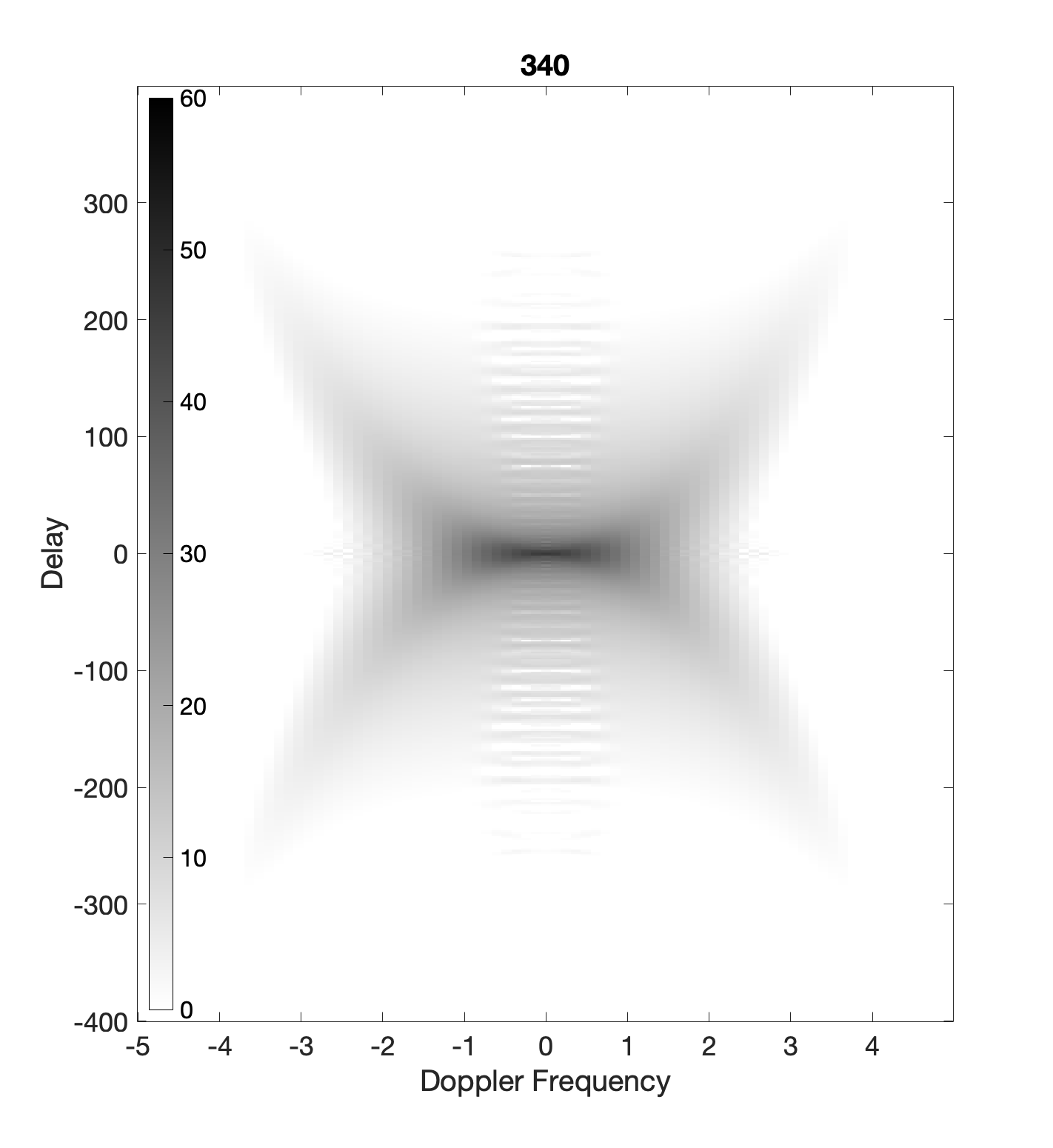} \\
\end{tabular}
\caption{Theoretical secondary spectra for strong scattering in a screen with an isotropic  Kolmogorov phase spectrum.  \it Left: \rm 
Normalized variables set at 825 MHz.    \it  Right: \rm  Stretched to 340 MHz: times $825/340 = 2.4$ in Doppler; and times $(825/340)^4 = 35$ in delay.  They are displayed over equal ranges in Doppler and 4 times larger in delay;  dynamic range in grayscale is set to 60 dB in both.}
\label{fig:Kolmo825-340} 
\end{figure*}

As noted above, a boundary arc is caused by the interference of a bright core of slightly scattered waves with those scattered at relatively large angles, which fall off in brightness as an inverse power law in the Kolmogorov spectrum. The power law is important in that the steeper decrease of a Gaussian spectrum suppresses the amplitude of the arc (see figures 5 and 7 of Cordes et al., 2006).  

Consider now the scaling versus frequency of the SS for an isotropic Kolmogorov spectrum displayed in the left panel of Figure~\ref{fig:KolmoSS}.  The key idea is that the angular width of the core in scattered brightness increases steeply with wavelength and so the boundary arc also widens with wavelength.  

As noted in \S \ref{sec:eta_basicISS},
the SS from a thin plasma screen can be expressed as a function of normalized delay $p$ and normalized Doppler $q$.
Hence, we can use $p,q$ variables to describe how arcs depend on the observing frequency.  Let the delay and Doppler at wavelength $\lambda_1$ be $\tau_1$ and $\fDi$.  Using  $p,q$ from equation \ref{eq:pqdef} we can scale them to the delay and Doppler at wavelength $\lambda_2$ as follows:
\bea
\tau_2 &=& p \: \Deff \theta_{o,2}^2/2c   \\
    &=&   \tau_1 \: (\theta_{o,2}/ \theta_{o,1})^2   \\   \nonumber
\fDii &=&  q \:  (\veff \theta_{o,2}/\lambda_2)  \\
      &=&  \fDi \: (\theta_{o,2}/\theta_{o,1})\; (\lambda_1/\lambda_2) \nonumber
\eea
$\theta_{o,1}$ is the characteristic width of the scattered brightness function at wavelength $\lambda_1$, similarly for wavelength $\lambda_2$.   For scattering in a
plasma $\theta_{o} \propto \lambda^2$  
Consequently  in scaling the calculated SS from a frequency $\nu_1$ to a lower frequency $\nu_2$,
the delay axis is stretched by a factor $(\nu_1/\nu_2)^4$ and the Doppler axis is stretched
by the lesser factor $(\nu_1/\nu_2)$.  
In Figure~\ref{fig:Kolmo825-340} an example is plotted in which the left and right panels represent 825 and 340 MHz, respectively.   The 825 panel is isotropic Kolmogorov calculation from the left panel of Figure~\ref{fig:KolmoSS}.  
The 340 panel is stretched 35 times in delay and 2.4 times in Doppler.  
Thus the 340 panel is an unequal zoom of the core of the 825 panel, adjusted to the same dynamic range.  
(If the Kolmogorov scaling exponents were used, the stretch factors would be somewhat larger, 49 and 2.9, in delay and Doppler, respectively).  
In practical observations the spectrometer channel width at 340 MHz is much finer than at 825, displaying the SS out to much greater delays, which is chosen here to be 4 times greater.

\subsection{Analysis of SS Using Normalized ({\em pq}) Coordinates}
\label{sec:pqAnalysis}
\begin{figure}    
\gridline{\fig{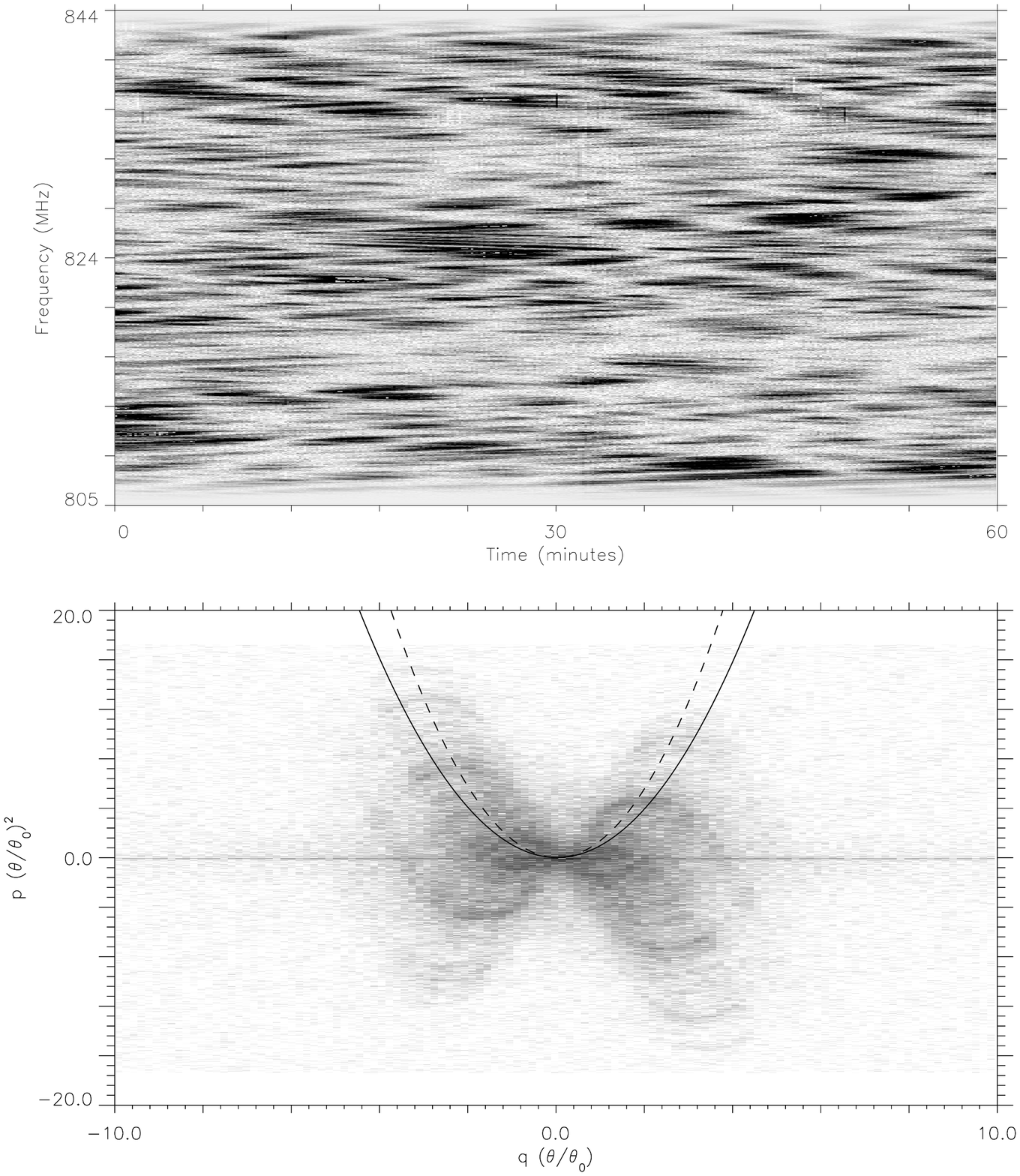}{0.3\textwidth}{(a)}
          \fig{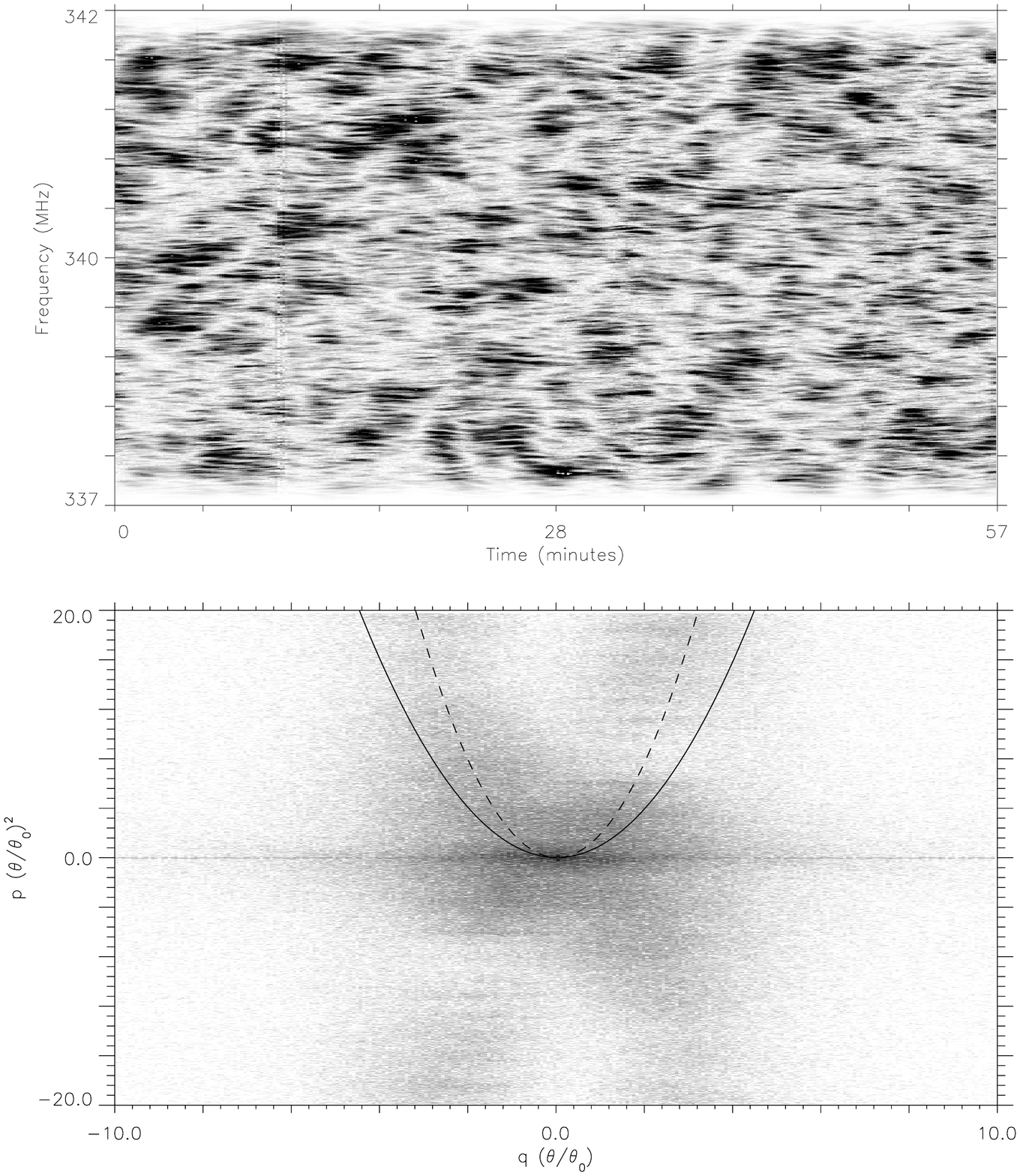}{0.3\textwidth}{(b)}
          \fig{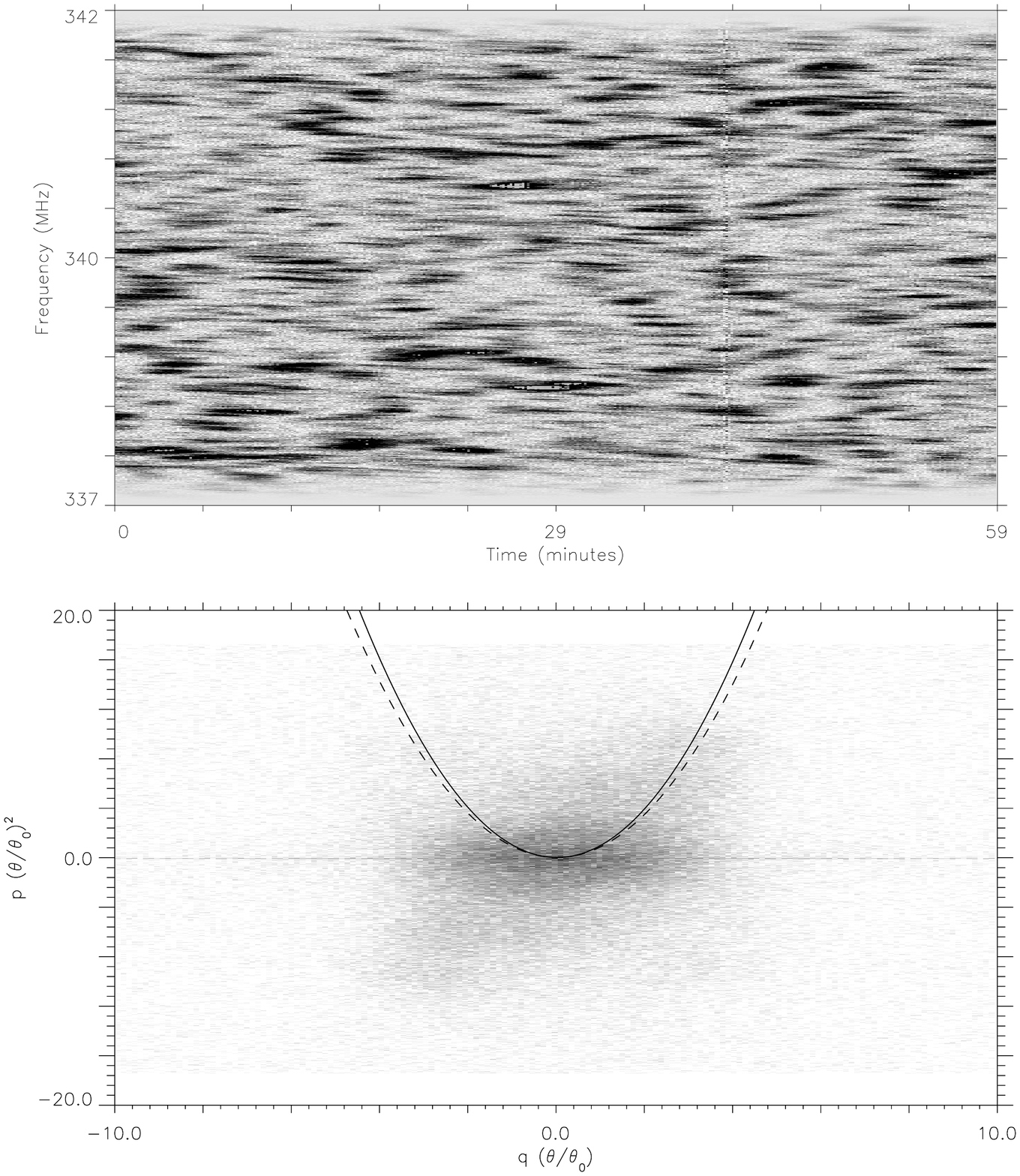}{0.3\textwidth}{(c)}}
\caption{Comparison of arc-based curvature estimate (\etap\ ) with ISS-parameter-based curvature (\etaiss\ ) for secondary spectra using normalized ($p, q$) coordinates.
All three panels are plotted to the same limits in $p$ and $q$ and hence can be directly compared. 
The dashed line follows the \etap\ parabola; the solid line indicates the \etaiss\ parabola.
{\em (a)} B0450--18 observed at 825~MHz (figure set~\psrC).
 {\em (b)} B1508+55 observed at 340~MHz (figure set~\psrK).
  {\em (c)} B2310+42 observed at 340~MHz (Figure~2 and figure set~\psrU).
  }
\label{fig:pq_overlay}  
\end{figure}

The strong correlation between \etaiss\ and \etap\ , described in \S~\ref{sec:eta_basicISS}, 
suggests that we explore the SS in terms of normalized ($pq$) coordinates, in which a scintillation arc has the simple form $p=q^2$. 
For convenience, we will call such parabolas {\em pq} arcs.
In Figure~\ref{fig:pq_overlay} we plot three examples of SS overlaid in {\em pq} coordinates, since there is good agreement between \etap\ and
\etaiss.

In panel~$a$ we have a case of strong inverted arclets and a deep valley along the $\tau$ axis, conditions indicative of a nearly 1D scattering profile made up of discrete local brightness peaks \citep{rszm21}.
It is simple in this case to fit the \etap\ parabola since it should coincide with the apexes of the inverted parabolas.
However, it is  
remarkable that the \etaiss\ parabola has nearly the same curvature since it is determined solely by the widths of the DS ACF --- which 
are (inversely) related to the widths near the origin of the SS ---
where there is little diffuse power. 

Considering Figure~\ref{fig:pq_overlay}($b$), the power distribution is much more diffuse.   Again, the \etap\ parabola is determined by the outlying features of the SS, although the precision of its curvature 
will be hampered by the blurriness of these features.
However, the \etaiss\ parabola, which is just determined by the central region of the DS ACF, matches the \etap\ curvature well.
Note the wide valley near the $\tau$~axis, indicative of anisotropic scattering in a thin screen.
This high-velocity pulsar exhibits several intriguing scintillation phenomena that are under current investigation.\footnote{See \eg\ the poster available at {\footnotesize\tt https://cloud.mpifr-bonn.mpg.de/index.php/s/YLR3M7YbGF4XzGG}, which is being developed into a paper by T.~Sprenger, O.~Wucknitz, and R.~Main (private communication).}

Finally, in Figure~\ref{fig:pq_overlay}($c$) 
(see also Figure ~\ref{fig:SS2310}) 
we have almost the opposite situation as in panel ($a$): it is clear how the DS ACF will yield a good measurement of \etaiss\ since there is power centered on the origin in the SS, but it is 
 surprising that a (weighted by $|\fD|$) parabolic summing of power along the SS plane results in such close agreement with the ACF-determined \etaiss value. 
 
\subsection{Scintillation Arcs Tend to Disappear at Low Frequency: Is This a Problem?}
\label{sec:lowfreq}
\begin{figure}    
\gridline{\fig{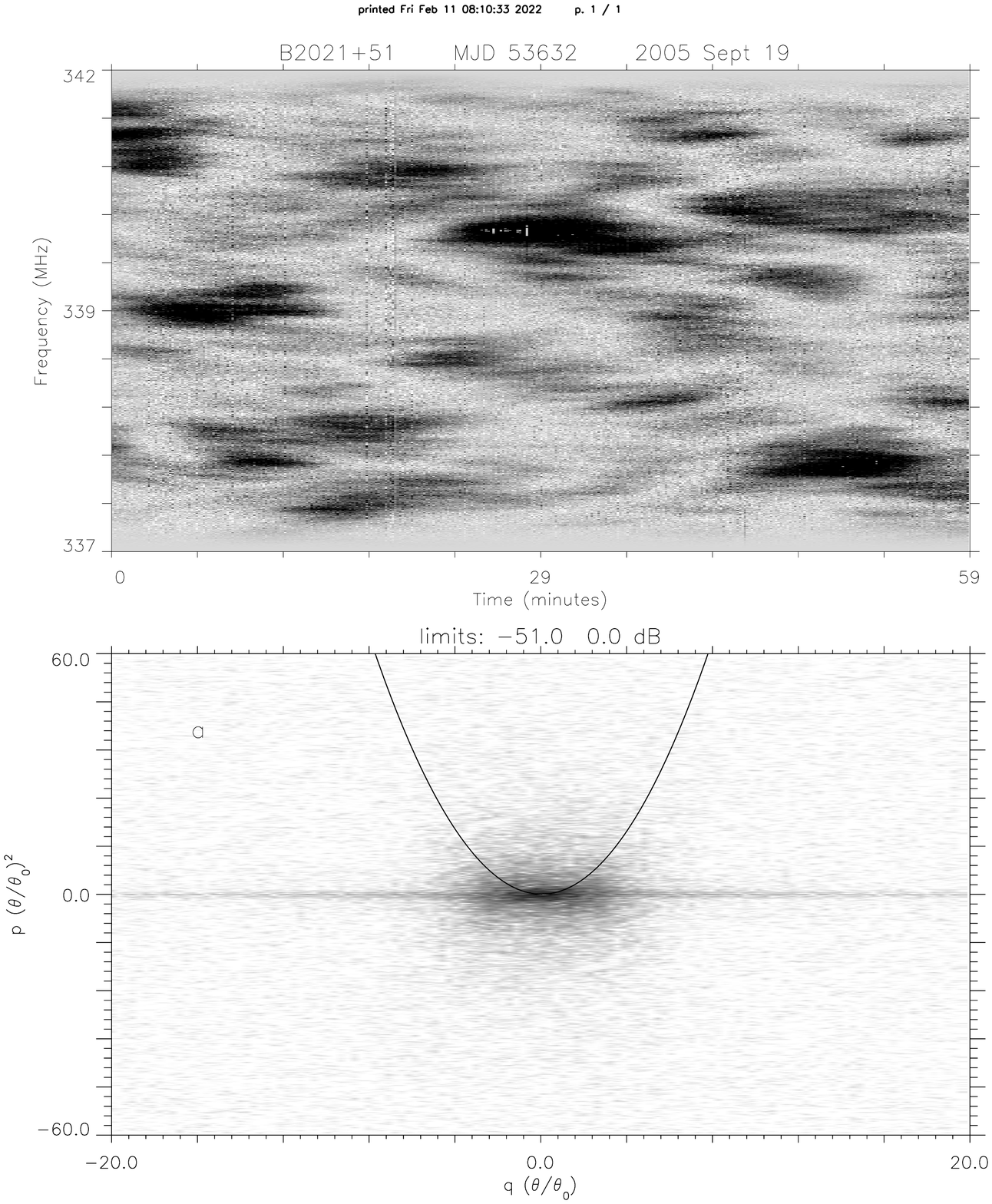}{0.3\textwidth}{(a)}
          \fig{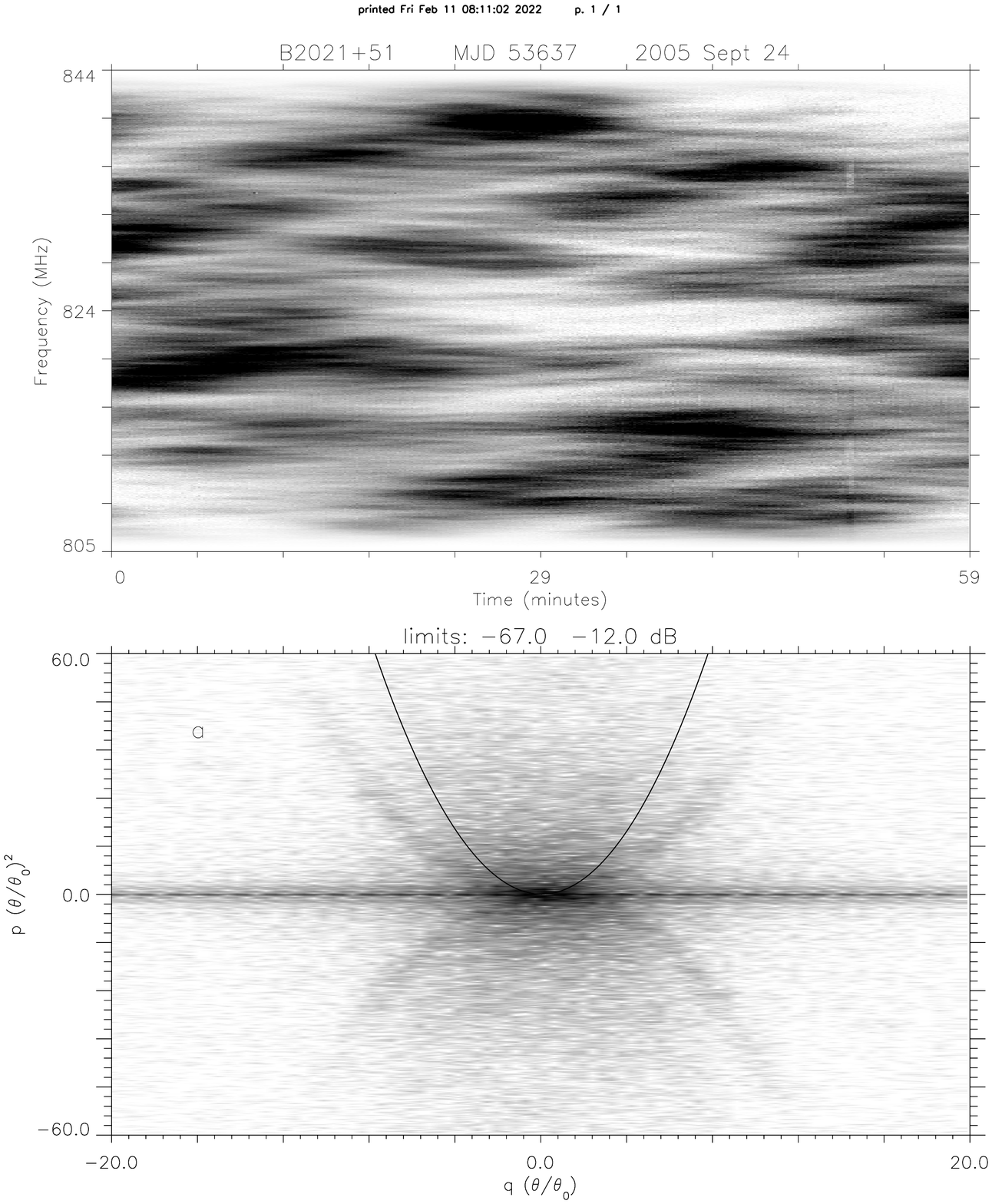}{0.3\textwidth}{(b)}
          \fig{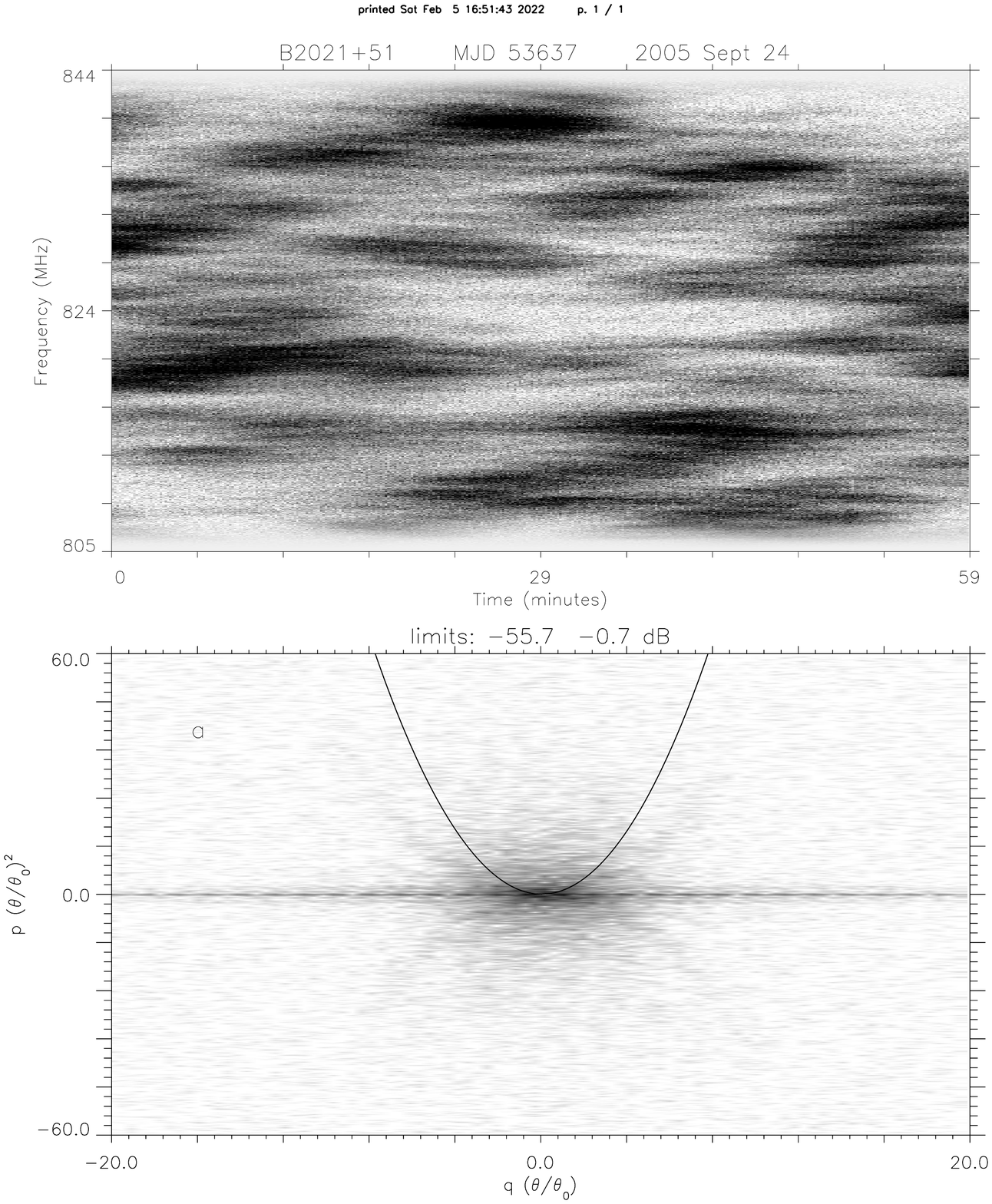}{0.3\textwidth}{(c)}}
\caption{Dual frequency comparison of secondary spectra using normalized ($p, q$) coordinates. {\em (a)} B2021+51 at 340~MHz. Display is a linear grayscale for the DS and logarithmic grayscale for the SS, covering about 50~dB from white to black. A pq~arc ($p = q^2$) is superposed on the SS. Since $p$ and $q$ are normalized by the angular width of the corresponding $B(\theta)$ curves, the axis stretch factors are accounted for automatically.
 {\em (b)} B2021+51 at 825~MHz observed 5 days later than the left panel. Display style is the same as in panel ($a$).
  {\em (c)} Same data as in ($b$), but with Gaussian white noise added to approximately match the S/N ratio of panel ($a$).}
\label{fig:2021pq}
\end{figure}

The multi-frequency aspect of this survey is a key asset, particularly when observations were made within a few days of each other as were the Green Bank 340~MHz and 825~MHz observations and observations of the three Arecibo pulsars with multi-frequency data.
Consider a direct comparison of the SS for pulsar B2021+51 at 340~MHz and 825~MHz as shown in the left two panels of Figure~\ref{fig:2021pq}.
 The $pq$~arc does not coincide with the boundary arc.
 However,
this could simply be due to a mis-estimation of the \dnuiss\ and \dtiss\ parameters, a more exaggerated form of the case shown in
Figure~\ref{fig:pq_overlay}(a).
The more important difference between panels ($a$) and ($b$)  is the 
absence of a boundary arc at the lower frequency. 
Is this surprising?
Displaying the SS using normalized coordinates should remove the issues with scaling of axes that were discussed in \S\ref{sec:WidthVsFreq}.
If the underlying assumptions hold, the low frequency SS should appear similar to that of the high frequency one in this normalized display.

The main additional assumptions are (reordered to match the list order):
an inhomogeneity spectrum in the screen that supports fluctuations at a small enough spatial scale to provide the high angle scattering needed to produce the arc;
 thin screen scattering with a screen that does not truncate the beam at lower frequency \citep{cl01,gkk+17};
and adequate S/N ratio to detect a scintillation arc.

 Here we explore these  possibilities:
 \begin{enumerate}
\item{{\em truncated screen}: that the scattered beam has become so large that it extends beyond the physical extent of the scattering material}
\item{{\em inner scale}: that there are no plasma fluctuations present at the small physical size needed to produce the halo power}
\item{{\em S/N inadequate}: that the observation at 340~MHz has insufficient sensitivity to reveal the low level scintillation arcs at high delay}
\end{enumerate}

Considering the first possibility, we use the information in Tables~\ref{tab:psrparms} and \ref{tab:analysis} to find that the screen must deflect a maximal ray at 340~MHz by about 2~mas.
The coherence scale $s_0$ in the screen necessary to do so can be found from $s_0 = \lambda/(2\pi\theta_{\rm scatt})$ (\eg Equation~2.4 in \citealt{ric90})
and is $s_0 \approx 1.5\times10^4$~km.
This is substantially larger than values of the inner scale of turbulence, which are in the range  $\sim 200 - 2000$~km 
\citep{sg90,mmrj95,bcc+04,rjtr09}.
Hence, it is unlikely that the absence of a scintillation arc at 340~MHz is caused by a deficit of irregularities at the  coherence scale.

Possibility 2 proceeds similarly. 
The width of the 340~MHz beam as it passes through the screen is $\approx 2$~au, a typical value for pulsars in this survey.
In order for this to explain the absence of a scintillation arc at 340~MHz
there would need to be a gap of this size in the medium producing the scattering. The material that produces the scintillation arc observed five days later at 825~MHz would need to comprise a small areal fraction of the $\sim 35$ times larger low-frequency beam and hence be diluted in its ability to produce a scintillation arc.
While not impossible, this seems unlikely.

Finally, we consider the option that the S/N of the 340~MHz observation is insufficient to allow detection of a scintillation arc.
As is evident from inspection of the figures and Table~\ref{tab:analysis}, the S/N at 340~MHz is substantially lower than for the 825~MHz observation. 
When we approximately match the S/N of the two observations by adding white Gaussian noise to the 825~MHz data, we obtain the result in Figure~\ref{fig:2021pq}(c).
Although there is a hint of a scintillation arc visible, slightly more additive noise would suppress the arc at 825~MHz entirely.
Hence, we consider this to be the most likely explanation:
the observation was simply not sensitive enough to detect the presence of the arc visible in the 825~MHz data.

\section{Discussion}
 \label{sec:discussion}

We draw a number of conclusions from the qualitative and quantitative analysis of the survey data in the preceding sections. We discuss those conclusions below. 

\subsection{Scintillation Arcs Are Prevalent}
\label{sec:prevalent}
Satisfactory S/N was obtained in 54 observations of 22 pulsars (at 1-3 frequencies) whose DMs range from 5.7 to 84 pc cm$^{-3}$.   Estimates of characteristic widths in frequency \nuiss\  and time \tiss\ were obtained from the ACFs of DS.   In all cases a curvature estimate was made by summing SS along forward parabolas, weighting $\propto |\fD |$.  In more than half the observations the summation exhibits a credible peak from which a curvature $\eta_{p}$ was estimated and a relative width parameter $\Delta \eta$ is defined.  These estimates are classified by a subjective credibility index (\etacred: 0, 1, 2: a compact maximum in the curve is rated 2; wide and double peaked curves are rated 1; cases where the peak is at the high or low limit in the search range or the secondary spectrum extends to the Nyquist delay are rated 0).   Of the 54 observations,  13 ranked as 2, 21 ranked as 1, and 20 ranked as 0.   Thus we have evidence for forward arcs in 34 of 54 observations.   In observing 22 pulsars, 19 exhibited an arc ranked $\etacred \ge 1$ at one frequency or more.

\subsection{Scintillation Arcs Are More Prominent at Higher Radio Frequencies}
\label{sec:prominent}
As discussed in \S\ref{sec:WidthVsFreq}, the much stronger scaling of the SS delay axis with frequency compared to the scaling of the Doppler axis results in the suppression of scintillation arcs at a radio frequency that depends upon the degree of scattering along the LoS.

\subsection{Scintillation Arcs Are Narrower in Low DM Pulsars}
\label{sec:narrowerLowDM}
Figure~\ref{fig:arcwidth} demonstrates that narrow well-defined arcs are common at low DM, but become rare at higher DM. 
It is well-established that a sharp scintillation arc can only be produced if the dominant scattering occurs in a relatively small fraction of the LoS, what is commonly referred to as a {\em thin screen}, although the transverse extent of the scattering region and its physical characteristics are left unspecified, \eg\ \citet{wmsz04,crsc06}.
Many of the narrower arcs at low DM are consistent with scattering from a localized plasma screen whose distance from the pulsar is no more than $s_{max} D_{psr}$.   
 
The trend toward narrow arcs at low DM follows naturally from a model for the ISM in which a pervasive but relatively low-scattering plasma is combined with  isolated regions or {\em clouds} of enhanced electron density $\langle n_e \rangle$, enhanced electron density variance $\langle n_e^2 \rangle$, or both. Such a trend is to be expected if the scattering is distributed along the LoS from each pulsar.  This decrease in arc definition with path length could be due to either multiple thin regions along the path or to a more general extended distribution in the scattering plasma. 

\subsection{Narrow Arcs Do Not Imply Anisotropy}
\label{sec:anisotropy}
Narrow arcs  do not, by themselves, imply an anisotropic plasma, but
they are consistent with a power-law spatial spectrum. 
The rich detail revealed in the reverse arclets reported for B0834+06 implies highly anisotropic plasma structures in the local ISM \citep{hsa+05,bmg+10}.  
The main forward arc with a deep valley along the delay axis in B1133+16 is also evidence for highly anisotropic local scattering \citep{sro19}.   
However a result of our analysis  is the recognition that boundary arcs do not necessarily imply anisotropic scattering \citep[see also][]{rcb+20}.    Under the conditions of strong scintillation that apply to our observations, relatively narrow boundary arcs can be caused by isotropic scattering when the underlying plasma density fluctuations follow some types of power law versus wavenumber.  While we have shown examples from screens that follow the Kolmogorov turbulence spectrum, other power law spectra can also cause forward arcs.  As analyzed earlier by \citet{crsc06}, forward arcs can be expected from spectra with a range of power law exponents; the simplest way to characterize them is via the structure function for phase perturbations that they impose on a propagating radio wave.  \citet{crsc06} show examples with phase structure functions that follow a power law versus spatial lag having exponents $\alpha \le 2$, where the corresponding exponent in a 3-dimensional wavenumber spectrum is $\alpha+2$.  The key point  is that no extended arcs are seen unless $\alpha < 2$.  
Media consisting of Gaussian-profiled density concentrations causing interstellar lenses are likely modeled by $\alpha = 2$, 
and so probably do not manifest parabolic arcs. 
Thus the defining property of the plasma density structures that cause arcs is the form of their high wavenumber spectrum rather than any anisotropy.  
Kolmogorov turbulence provides one possible physical origin for such fine scales in the plasma.

\subsection{Reverse Arclets and Power Asymmetries Indicate a Patchy Scattering Medium}
\label{sec:asymmetry}
Reverse arclets are seen in the SAS in  the following pulsars (see \S4 for more details):  B0450--18, B0525+21, B1540--06 and B2327--20, whose distances range from 0.4 to 1.2~kpc.   
In addition, B1508+55 ($D = 2.10$~kpc) shows discrete arclets, which appear to be flat (\ie\ low curvature).  
Overall the presence of arclets, and the relative frequency independence of their inferred angular locations \citep{hsa+05}, implies highly localized centers of scattering (or refraction) across the transverse dimension  and an anisotropic image on the sky \citep{wmsz04, crsc06, pl14, rcb+20}.

Power asymmetries in the SS along the scintillation arc and, in particular, between negative and positive \fD\ values, have been noted since early in the study of the phenomenon \citep{crsc06}.
In addition, discrete patches of power can be present, generally associated with reverse arclets. In a few cases it has been possible to track their motion from negative to positive \fD values along the arc \citep{hsa+05, whh+18}.
We see evidence of both phenomena in the survey data.
The occurrence of power asymmetry and discrete structure in scintillation arcs for the 22 pulsars in the survey is summarized in Table~\ref{tab:asymmetry}.

\begin{deluxetable}{c l c c}
\tabletypesize{\small}
\tablecaption{Scintillation Arcs: Power Asymmetries and Discrete Structure} 
\tablehead{
\colhead{     PSR      } &
\colhead{asymmetry\tablenotemark{a}}&
\colhead{discrete structure} &
\colhead{$l_{\rm 10mHz}$\tablenotemark{b}}\\
\colhead{}&
\colhead{}&
\colhead{}&
\colhead{(au)}
}
\tablenum{5}
\colnumbers
\startdata 
 B0450+55 & neg (340) & & 1.0 \\
 B0450--18	&	& (825) & 11.4 \\
 B0523+11	& neg (1450) & & 1.1 \\
 B0540+23	& neg (430, 1450) & & 1.8 \\
 B0626+24	&pos (430) & & 5.2 \\
 B1508+55	&	&(340, 825, 1400) & 0.3 \\
 B1540--06	& 	&(340, 825) & 3.2 \\
 B2310+42	&pos (340)	&(825) & 2.4 \\
 B2327--20	&	&(340) & 1.0 \\
\enddata
\tablenotetext{a}{Numbers in parentheses indicate data set frequency in MHz.
Negative (neg) asymmetry means stronger power for positive delay on the negative side of the \fD\ axis; conversely for positive (pos) asymmetry.}
\tablenotetext{b}{Approximate physical size (au) on the screen  with these assumptions: $s=0.5$, \fD = 10~mHz, $\nu_0 = 1$~GHz, velocity dominated by the pulsar.
Scalings: $l_{\rm 10mHz} \propto s \fD / (V \nu)$}
\label{tab:asymmetry}%
\end{deluxetable}%

Two explanations have been advanced for power asymmetry along the arc: 1) the presence of a refractive gradient across the image \citep{crsc06, crg+10, rcb+20} and 2) spatial variation of the properties of the scattering screen transverse to the LoS \citep[e.g.][] {hsa+05}.
These mechanisms are not mutually exclusive since a patchy medium, by which we mean variations in the scattering strength transverse to the LoS, will necessarily have substantial $n_e$ gradients.

\begin{figure}[h]   
\begin{center}
\includegraphics[width=0.6\textwidth]{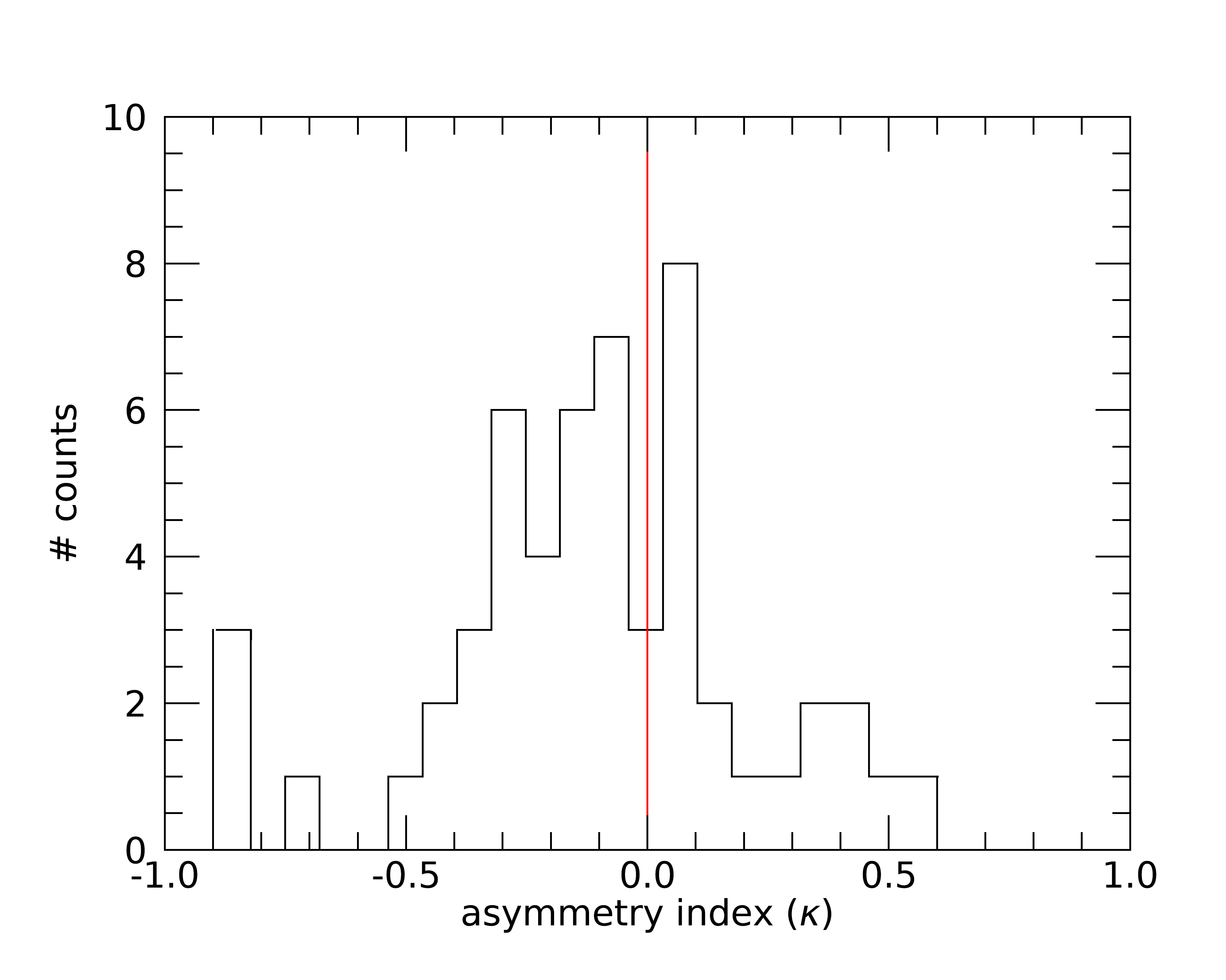} 
 \caption{\footnotesize 
 Histogram of arc power asymmetry index defined as $\kappa \equiv$~(R -- L)/(R + L), where R is the arc power for $\fD >0$ and L is the arc power for $\fD <0$.
 Negative values of $\kappa$ represent scattering material out in front of the projected path of the pulsar across the sky.
 }
\label{fig:kappa}
\end{center}
\end{figure}

In Table~\ref{tab:analysis} the asymmetry index $\kappa$ quantifies the power along the LHS (negative \fD; $\kappa < 0$) of the arc compared to along the RHS.
In Figure~\ref{fig:kappa} we note a {\em slight} tendency toward more instances of negative $\kappa$ values (33 vs. 21 positive), and the four largest values of $|\kappa|$ are all negative.
However, the sample is small, and this is likely a statistical fluctuation.
(The $\fD < 0$ side of the SS is the material that is out in front of the projected path of the pulsar across the sky \citep{hsa+05}.)

Kolmogorov density variations do not, in general, lead to substantial refractive shifts of the image.
In particular, it is rare to find refractive shifts as large as the width of the scattering disk size.
There is no way to adjust this fact for a Kolmogorov medium because it arises from the relative shallowness of the inhomogeneity power law.
During the 1980's this was a subject of substantial theoretical attention
 with a leading idea being inhomogeneity power laws with an index $\beta > 4$, where this index in the inhomogeneity power law has a value of $\beta=3.67$ for a Kolmogorov medium \citep{bn85,gn85,rnb86,grbn87}.
As an alternative to  explanations associated with power law density variations, the idea of discrete lenses was introduced  \citep{cfl98}.
More recently, highly inclined corrugated sheets viewed at nearly grazing incidence  \citep{pk12,pl14} and 
noodle-like models \citep{gwi19,gs19}
have been proposed.

As can be seen in Table~\ref{tab:asymmetry}, just under half of the pulsars in the survey show evidence for power asymmetry or discrete structures in the SS or both.
The last column in this table presents an approximate size scale, $l_{\rm 10mHz}$, probed by the scintillation arc observations.
Note that in the SS it is the separation of features in two coordinates, Doppler and delay, that allows discrete patches of power to be identified.
On the other hand, the occurrence of tilted scintles in the DS was noticed soon after the development of systematic scintillation observations (\citealt{hew80} and references therein).

The evidence for thin screen regions of scattering implies localization along the LoS direction.  
Together with the power asymmetries and reverse arclet structure in the SAS, these paint a picture of a very patchy distribution in the plasma responsible for the ISS within the $\sim$~3~kpc region sampled.

\subsection{The Galactic Distribution of Plasma Scattering}
\label{sec:galdist}

The  study by \citet{azg+20} revealed a coherent sheet-like structure that they refer to as the Radcliffe Wave, a 2.7-kpc-long filament of gas corresponding to the densest part of the Local Arm of the Milky Way. 
In addition, the understanding of the Local Bubble has improved markedly with the recent publication by \citet{zga+22}. There, using new spatial and dynamical constraints including recent Gaia data and carefully curated velocity information, they produce a three-dimensional map of dense gas and young stars within 200 pc of the Sun. They find evidence for stars preferentially concentrated near the edge of the Bubble at about 100 pc from the Sun.  
The boundaries are seen to be star-forming regions and are partially ionized by UV radiation from nearby stars. 
The structure and distribution of truly local interstellar clouds is also relevant.
\citet{lrt19} and \citet{lr21}  find partially ionized clouds on the scale of parsecs.

Our survey for scintillation arcs provides evidence for occasional localized plasma concentrations within about 1 kpc (sec 6.1 \& 6.3). The observations are consistent with the earlier more detailed observations of multiple discrete arcs in some nearby pulsars (\eg\ \citealt{ps06a};  see also \citealt{rcb+20}).  The reverse arclet phenomenon gives further evidence for discrete plasma concentrations down to au scales.  
However, the physical origin of  the clumps remains a mystery.

The evidence for the isolated concentrations of scattering plasma has to be reconciled with the strong evidence that the interstellar scatter broadening time for pulsars increases steeply with \DM, and so with pulsar distance. There is an absence of narrow arcs from pulsars beyond a few kpc in our survey, which is consistent with cumulative scattering along the LoS from many such concentrations at a wide range of distances (and so with differing arc curvature).  At present we lack a proper theory for the SS that would be observed through multiple plasma screens. While the SS from nearby pulsars can be understood by the superposition of arcs singly scattered by each screen, the SS due to successive scattering by multiple plasma screens have not been studied beyond the two screen analysis of \citet{spmb19a}. 

{\danx Pulsar dispersion and scattering studies over more than 40 years have resulted in a fairly consistent picture of the ionized gas within $\sim 5$~kpc of the Sun (\citealt{tc93}, \citealt{cl02} [NE2001], \citealt{ymw17}).}
{\danx Overall,} the geometry consists of relatively sparse regions of enhanced plasma scattering on scales smaller than 
1~kpc that are increasingly concentrated toward the Galactic plane and toward the center of the Galaxy.  
The models typically assume that the plasma density is ``turbulent,''  following a power-law  spectrum versus wavenumber over the micro-scales responsible for the ISS, but the strength of the turbulence varies widely over the much larger Galactic scales. Thus the turbulence level varies on scales ranging from parsecs to au and indeed down to the diffractive scale at $10^6 - 10^8$~m. In NE2001 some such concentrations are identified as known HII regions, but other clumps of denser scattering are added to model specific pulsars that exhibit extra scatter broadening. (See \citealt{mma+22} for an in-depth study of one such region associated with the pulsar J1643--1224 seen behind the HII region Sh~2--27.)

In order to model the arclet phenomenon and the multiple forward arcs in pulsar B1133+16 (see \citealt{mzsc22} and \citealt{ps06a}, for example), many more clumps are implied. A smaller scale is needed such that the mean free path for a pulsar sightline to intersect a clump is on the order of 100--500~pc.   
\citet{occd21} have proposed turbulence at stellar bow-shocks as the possible location of enhanced scattering. Scattering could even be caused in the plasma-spheres that surround hot stars, while \citet{wtb+17} proposed elongated plasma structures drawn out in the stellar winds from hot stars. 
Motivated by evidence of extreme 1D scattering images for some pulsars, \citet{pl14} proposed weak waves propagating along magnetic domain boundary current sheets as the origin of scintillation arcs.

The foregoing discussion suggests the big-picture hypothesis that the arc-causing clumps are so widely distributed {\danx {\em that they are the building blocks for all of interstellar scattering.}}
We suggest that the ISM contains multiple bubbles creating a foam-like structure with compressed regions of gas (neutral and plasma) at their interfaces, some of which cause observable arcs. 
The 50 -- 100 pc distance is comparable to the distance between the arc-causing clumps in the LoS to B1133+16 for which \citet{mzsc22} identified six discrete arc-causing screens along the 360~pc LoS.
If this path is typical of much longer paths through the Galaxy, it would explain the rarity of narrow arcs at higher \DM, since it would be  unlikely that a single arc would dominate the SS for pulsars beyond a kiloparsec or so.  Thus it becomes interesting to determine whether the arc-causing screens have a characteristic radius and scattering measure. Under the Kolmogorov scenario the radius might be identified with the outer scale of turbulence, as in  NE2001.

\section{Summary}
\label{sec:summary}
 We summarize our main results as follows:

\begin{itemize}
\item Scintillation arcs are prevalent 
\item Scintillation arcs are more prominent at higher radio frequencies
\item Scintillation arcs are narrower in low DM pulsars
\item Narrow arcs, especially sharp boundary arcs, do not imply anisotropy
\item Reverse arclets with deep valleys are seen in about 20\% of the survey pulsars
\item Power asymmetries in arcs indicate a patchy scattering medium on an au size scale
\end{itemize}

Overall, combining the SAS results with earlier ISS studies suggests 
that interstellar scattering is largely caused at the boundaries 
of structures in the plasma density similar to the Local Bubble.\\

We thank the developers and maintainers of the ATNF pulsar data base {\sc psrcat}
\citep{mhth05}, which we used extensively.
M.~McLaughlin, S.~Ransom, and D.~Stinebring acknowledge support from the NSF from a Physics Frontiers Center award (2020265) to NANOGrav.
In addition, D.~Stinebring thanks the NSF for support through grant 2009759.
S.~Ransom is a CIFAR Fellow.
S.~Ocker acknowledges support from the National Aeronautics
and Space Administration (NASA 80NSSC20K0784).
The Green Bank Observatory is a facility of the National Science Foundation operated under cooperative agreement by Associated Universities, Inc.
The Arecibo Observatory
is a facility of the NSF operated under cooperative
agreement (\#AST-1744119) by the University of Central Florida (UCF) in alliance with Universidad Ana G.\ 
Mendez (UAGM) and Yang Enterprises (YEI), Inc.
The National Radio Astronomy Observatory is a facility of the National Science Foundation operated
under cooperative agreement by Associated Universities, Inc.

\facilities{GBT, Arecibo}
\\
\\
{\em\large Objects:} 
\object{PSR B0138+59,}
\object{PSR B0450+55,}
\object{PSR B0450-18,}
\object{PSR B0523+11,}
\object{PSR B0525+21,}
\object{PSR B0540+23,}
\object{PSR B0626+24,}
\object{PSR B0628-28,}
\object{PSR B0809+74,}
\object{PSR B0818-13,}
\object{PSR B1508+55,}
\object{PSR B1540-06,}
\object{PSR B1706-16,}
\object{PSR B1821+05,}
\object{PSR B1857-26,}
\object{PSR B1907+03,}
\object{PSR B2021+51,}
\object{PSR B2045-16,}
\object{PSR J2145-0750,}
\object{PSR B2217+47,}
\object{PSR B2310+42,}
\object{PSR B2327-20.}


\end{document}